\documentclass[12pt,preprint]{aastex}

\usepackage{amsmath}
\usepackage{epsf}
\usepackage{cancel}

\usepackage[normalem]{ulem}

\usepackage{color}
\newcommand{\sfrac}[2]{\mathchoice
  {\kern0em\raise.5ex\hbox{\the\scriptfont0 #1}\kern-.15em/
   \kern-.15em\lower.25ex\hbox{\the\scriptfont0 #2}}
  {\kern0em\raise.5ex\hbox{\the\scriptfont0 #1}\kern-.15em/
   \kern-.15em\lower.25ex\hbox{\the\scriptfont0 #2}}
  {\kern0em\raise.5ex\hbox{\the\scriptscriptfont0 #1}\kern-.2em/
   \kern-.15em\lower.25ex\hbox{\the\scriptscriptfont0 #2}}
  {#1\!/#2}}

\newcommand{\myhalf}{\sfrac{1}{2}}
\newcommand{\thalf}{\sfrac{3}{2}}

\newcommand{\eb}{{\bf{e}}}
\newcommand{\Ub}{{\bf{U}}}
\newcommand{\Ubt}{\widetilde{\Ub}}

\newcommand{\xb}{{\bf{x}}}

\newcommand{\dr}{\Delta r}
\newcommand{\dt}{\Delta t}

\newcommand{\etarho}{\eta_\rho}
\newcommand{\gammaonebar}{\overline{\Gamma_1}}
\newcommand{\Hext}{H_{\rm ext}}
\newcommand{\Hnuc}{H_{\rm nuc}}
\newcommand{\omegadot}{\dot\omega}
\newcommand{\pred}{{\rm pred}}
\newcommand{\Sbar}{\overline{S}}

\newcommand{\inp}{\mathrm{in}}
\newcommand{\outp}{\mathrm{out}}
\newcommand{\nph}{{n+\myhalf}}
\newcommand{\nmh}{{n-\myhalf}}
\newcommand{\ow}{\overline{w_0}}
\newcommand{\dw}{\delta w_0}
\newcommand{\uadvone}{\Ubt^{\mathrm{ADV},\star}}
\newcommand{\uadvonedag}{\Ubt^{\mathrm{ADV},\dagger,\star}}
\newcommand{\uadvtwo}{\Ubt^{\mathrm{ADV}}}
\newcommand{\uadvtwodag}{\Ubt^{\mathrm{ADV},\dagger}}
\newcommand{\gcc}{\mathrm{g~cm^{-3} }}

\begin{document}

%==========================================================================
% Title
%==========================================================================
\title{{\tt MAESTRO}: An Adaptive Low Mach Number Hydrodynamics Algorithm for Stellar Flows}
\shorttitle{{\tt MAESTRO}: Low Mach Number Astrophysics}
\shortauthors{Nonaka et al.}

\author{A.~Nonaka\altaffilmark{1},
        A.~S.~Almgren\altaffilmark{1},
        J.~B.~Bell\altaffilmark{1},
        M.~J.~Lijewski\altaffilmark{1},
        C.~M.~Malone\altaffilmark{2},
        M.~Zingale\altaffilmark{2}}
        
\altaffiltext{1}{Center for Computational Science and Engineering,
                 Lawrence Berkeley National Laboratory,
                 Berkeley, CA 94720}

\altaffiltext{2}{Dept. of Physics \& Astronomy,
                 Stony Brook University,
		 Stony Brook, NY 11794-3800}

%==========================================================================
% Abstract and Keywords
%==========================================================================
\begin{abstract}
Many astrophysical phenomena are highly subsonic, requiring specialized 
numerical methods suitable for long-time integration.  In a series of
earlier papers we described the development of {\tt MAESTRO}, 
a low Mach number stellar hydrodynamics code that can be used to 
simulate long-time, low-speed flows that would be prohibitively expensive
to model using traditional compressible codes.  {\tt MAESTRO} is based
on an equation set derived using low Mach number asymptotics;
this equation set does not explicitly track acoustic waves and thus 
allows a significant increase in the time step.  
{\tt MAESTRO} is suitable for two- and three-dimensional local atmospheric flows 
as well as three-dimensional full-star flows.  Here, we continue the development of 
{\tt MAESTRO} by incorporating adaptive mesh refinement (AMR).
The primary difference between {\tt MAESTRO} and other structured grid AMR 
approaches for incompressible and low Mach number flows is the presence of the
time-dependent base state, whose evolution is coupled to the evolution of
the full solution.  We also describe how to incorporate the expansion
of the base state for full-star flows, which involves a novel mapping technique
between the one-dimensional base state and the Cartesian grid, as well as
a number of overall improvements to the algorithm.  We examine the efficiency and
accuracy of our adaptive code, and demonstrate that it is suitable for further
study of our initial scientific application, the convective phase of
Type Ia supernovae.
\end{abstract}
%==========================================================================
% Introduction
%==========================================================================
\section{Introduction}
Many astrophysical phenomena of interest occur in the low Mach number
regime, where the characteristic fluid velocity is small compared to
the speed of sound.  Some well-known examples are the convective phase
of Type Ia
supernovae (SN~Ia)~\citep{hoflichstein:2002,kuhlen-ignition:2005,ZABNW:IV},
classical novae~\citep{glasner:2007}, convection in
stars~\citep{meakin:2007}, and Type I X-ray bursts~\citep{Lin:2006}.
Such problems require a numerical approach capable of resolving
phenomena over time scales much longer than the characteristic time
required for an acoustic wave to propagate across the computational
domain.  In a series of papers (see \citet{ABRZ:I}---henceforth Paper I,
\citet{ABRZ:II}---henceforth Paper II, \citet{ABNZ:III}---henceforth
Paper III, and \citet{ZABNW:IV}---henceforth Paper IV), we have
described the initial development of {\tt MAESTRO}, a low Mach number 
hydrodynamics code for computing stellar flows using a time step
constraint based on the fluid velocity rather than the sound speed.
{\tt MAESTRO} is suitable for two- and three-dimensional local
atmospheric flows as well as three-dimensional full-star flows.  All
simulations are performed in a Cartesian grid framework, but rely
on the presence of a one-dimensional radial base state that describes the
average state of the star or atmosphere.  Starting
with the development of the low Mach number equation set (see Paper
I), we demonstrated how to capture the expansion of
the base state in a local atmospheric simulation
in response to large-scale heating (Paper II) and incorporate
reaction networks (Paper III).  In Paper IV, we presented the
initial application of {\tt MAESTRO}, following the last two hours of
convection inside a white dwarf leading up to the ignition of a SN Ia
using a three-dimensional, full-star simulation with a base state that is
constant in time.

In general, astrophysical flows are highly turbulent.  In the case of
the convective period preceding a SN Ia explosion, the
Reynolds number is $\mathcal{O}(10^{14})$ \citep{Woosley:2004},  
far larger than can be modeled on today's supercomputers.
Nevertheless, to understand the role of turbulence in these events,
we must use increasingly more accurate simulations.  
In this paper we describe how to incorporate adaptive mesh refinement (AMR),
in which we locally refine the Cartesian grid in regions of interest, to allow us to 
efficiently push to higher spatial resolutions and better capture the turbulent 
flow in critical regions of the simulation.  The primary difference between 
{\tt MAESTRO} and other structured grid AMR approaches for incompressible
and low Mach number flows is the presence of the time-dependent base state, whose 
evolution is coupled to the evolution of the full solution.
We also describe how to incorporate a time-dependent base
state for full-star problems, which involves a novel mapping technique between
the one-dimensional base state and the Cartesian grid.
This allows us to properly 
capture the effects of an expanding base state in full-star simulations.
We have also made a number of overall improvements to the algorithm, and 
all together, these enhancements will
allow us to compute more efficient and accurate solutions for our target
applications, including the convective phase of SNe Ia and Type I X-ray bursts.

This paper is divided into several sections, along with three detailed
appendices.  In \S \ref{Sec:Governing Equations}, we present the governing equations.
In \S \ref{Sec:Overview}, we give an overview of the methodology, referring the
reader to the appendices for full details.  In \S \ref{Sec:Mapping}, we describe
the new mapping procedure between one-dimensional and three-dimensional data structures
in full-star problems.  In \S \ref{Sec:AMR}, we discuss the extension of the
algorithm to include AMR.  In \S \ref{Sec:Test Problems}, we describe the results of 
our test problems.  We conclude with \S \ref{Sec:Conclusions}, which includes
future plans for scientific investigation.

%==========================================================================
% Equations
%==========================================================================
\section{Governing Equations}\label{Sec:Governing Equations}
Stellar flows are well characterized by
the compressible Euler equations (i.e., viscosity effects
are negligible).  These equations model all compressibility effects in
a fluid, and allow for the formation and propagation of shocks.  For
low speed convective flows in a hydrostatically stratified star or
atmosphere, we do not need to explicitly follow the propagation of
sound waves.  However, we do need to include large-scale
compressibility effects such as the expansion/contraction of a fluid
element as it changes altitude in the stratified background, and the
local changes to the density of the fluid element through heating
and compositional changes.  By reformulating the
equations of hydrodynamics to filter out sound waves but preserve the
correct large-scale fluid motions and hydrostatic balance, we can
retain the compressibility effects we desire while allowing for much
larger time steps than a corresponding compressible code.
The full derivation of the low Mach number hydrodynamics equations is 
given in papers I through III.  The resulting equations are:
\begin{eqnarray}
\frac{\partial(\rho X_k)}{\partial t} &=& -\nabla\cdot(\rho X_k\Ub) + \rho\omegadot_k,\label{eq:species}\\
\frac{\partial\Ub}{\partial t} &=& -\Ub\cdot\nabla\Ub  - \frac{1}{\rho}\nabla\pi - \frac{\rho-\rho_0}{\rho} g\eb_r,\label{eq:momentum}\\
\frac{\partial(\rho h)}{\partial t} &=& -\nabla\cdot(\rho h\Ub) + \frac{Dp_0}{Dt} + \rho\Hnuc + \rho\Hext,\label{eq:enthalpy}
\end{eqnarray}
where $\rho$, $\Ub$, and $h$ are the mass density,
velocity and specific enthalpy, respectively, and
$X_k$ are the mass fractions of species $k$ with associated
production rate $\omegadot_k$.  The species are constrained
such that $\sum_k X_k = 1$ giving $\rho = \sum_k (\rho X_k)$ and
\begin{equation}
\frac{\partial\rho}{\partial t} = -\nabla\cdot(\rho\Ub).\label{eq:mass}
\end{equation}
The source terms $\Hext$ and $\Hnuc$ are the external heating rate and nuclear energy 
generation rate per unit mass.  The pressure is decomposed into a hydrostatic base state
 pressure, $p_0 = p_0(r,t)$, and a dynamic pressure, $\pi = \pi(\xb,t)$, such that 
$|\pi|/p_0 = \mathcal{O}(M^2)$ (we use $\xb$ to represent the Cartesian coordinate 
directions of the full state and $r$ to represent the radial coordinate direction for 
the base state).  We also define a base state density, $\rho_0 = \rho_0(r,t)$, 
which is in hydrostatic equilibrium with $p_0$, i.e., 
$\nabla p_0 = -\rho_0 g\eb_r$, where $g=g(r,t)$ is
the magnitude of the gravitational acceleration and $\eb_r$ is the unit vector in the
outward radial direction. 

Mathematically, this system must still be closed by the equation of state which we
express as a divergence constraint on the velocity field (see Paper III),
\begin{equation}
\nabla\cdot(\beta_0\Ub) = \beta_0\left(S - \frac{1}{\gammaonebar p_0}\frac{\partial p_0}{\partial t}\right),\label{eq:U divergence}
\end{equation}
where $\beta_0$ is a density-like variable that carries background stratification, defined as
\begin{equation}
\beta_0(r,t) = \rho_0(0,t)\exp\left(\int_0^r\frac{1}{\gammaonebar p_0}\frac{\partial p_0}{\partial r'}dr'\right),
\end{equation}
and $\gammaonebar$ the lateral average (see \S \ref{Sec:Lateral Average}) of $\Gamma_1 = d(\log p)/d(\log\rho)$ at constant entropy.  The expansion term, 
$S$, incorporates local 
compressibility effects due to heat release from reactions, compositional changes, and 
external sources,
\begin{equation}
S = -\sigma\sum_k\xi_k\omegadot_k + \frac{1}{\rho p_\rho}\sum_k p_{X_k}\omegadot_k + \sigma\Hnuc + \sigma\Hext.\label{eq:S}
\end{equation}
where $p_{X_k} \equiv \left. \partial p / \partial X_k
\right|_{\rho,T,X_{j,j\ne k}}$, $\xi_k \equiv \left. \partial h /
\partial X_k \right |_{p,T,X_{j,j\ne k}},
p_\rho \equiv \left.
\partial p/\partial \rho \right |_{T, X_k}$, and $\sigma \equiv
p_T/(\rho c_p p_\rho)$, with $p_T \equiv \left. \partial p / \partial
T \right|_{\rho, X_k}$ and $c_p \equiv \left.  \partial h / \partial T
\right|_{p,X_k}$ is the specific heat at constant pressure.  

It is important to note that if the Mach number of the fluid in a
numerical simulation becomes $O(1)$, through large acceleration due to
buoyancy or nuclear energy generation, for example, the solution of
these equations would no longer be physically meaningful.  The low
Mach number equations do not enforce that the Mach number remain
small; rather, if the dynamics of the flow are such the Mach number
does remain small, then these equations are valid approximations for
the evolution of the flow.

As in Papers II and III, we decompose the full velocity field into a base 
state velocity, $w_0$, that governs the base state dynamics, and a 
local velocity, $\Ubt$, that governs the local dynamics, i.e.,
\begin{equation}
\Ub = w_0(r,t)\eb_r + \Ubt(\xb,t).
\end{equation}
with
$\overline{(\Ubt\cdot\eb_r)} = 0$ and
$w_0 = \overline{(\Ub\cdot\eb_r)}$.
The velocity evolution equations are then
\begin{eqnarray}
\frac{\partial w_0}{\partial t} &=& -w_0\frac{\partial w_0}{\partial r} - \frac{1}{\rho_0}\frac{\partial\pi_0}{\partial r},\label{eq:w0 evolution}\\
\frac{\partial\Ubt}{\partial t} &=& -\Ub\cdot\nabla\Ubt - \left(\Ubt\cdot\eb_r\right)\frac{\partial w_0}{\partial r}\eb_r - \frac{1}{\rho}\nabla\pi + \frac{1}{\rho_0}\frac{\partial\pi_0}{\partial r}\eb_r - \frac{\rho-\rho_0}{\rho}g\eb_r.\label{eq:utildeupd} 
\end{eqnarray}
where $\pi_0$ is the base state component of the perturbational pressure.
By laterally averaging to equation (\ref{eq:U divergence}), 
we obtain a divergence constraint for $w_0$:
\begin{equation}
\nabla\cdot(\beta_0 w_0 \eb_r) = \beta_0\left(\Sbar - \frac{1}{\gammaonebar p_0}\frac{\partial p_0}{\partial t}\right).\label{eq:w0 divergence}
\end{equation}
The divergence constraint for $\Ubt$ can be found by subtracting (\ref{eq:w0 divergence}) 
into (\ref{eq:U divergence}), resulting in
\begin{equation}
\nabla\cdot\left(\beta_0\Ubt\right) = \beta_0\left(S-\Sbar\right).\label{eq:utilde divergence}
\end{equation}

In the present paper, we revert back to the method introduced in
paper~II and define a base state enthalpy, $(\rho h)_0$.  We use
$\rho_0$ and $(\rho h)_0$ to define the perturbational quantities $\rho'
= \rho - \rho_0$ and $(\rho h)' = (\rho h) - (\rho h)_0$, which are
predicted to the Cartesian edges to compute fluxes for the
conservative updates of $\rho_0$ and $(\rho h)_0$.  Experience has
shown that by advancing perturbational quantities, the slope limiters
are more effective at reducing numerical oscillations since they are
being applied to a normalized signal, rather than a signal that spans
many orders of magnitude over a small number of cells.  This is a
departure from paper~III where we predicted temperature to the
Cartesian edges.  Evolution equations for $\rho_0$ and $(\rho h)_0$
are designed so that $\rho_0$ and $(\rho h)_0$ will remain the average
over a layer of constant radius of $\rho$ and $(\rho h)$.  The fluxes
for $(\rho X_k)$ are computed by first predicting $\rho_0, \rho'$, and
$X_k$ to time-centered Cartesian edges.  The flux for $(\rho h)$ is
computed by first predicting $(\rho h)_0$ and $(\rho h)'$ to time-centered
Cartesian edges.

We now derive the equations used to predict the time-centered Cartesian edge values in
the actual algorithm.  The species evolution equation is
found by combining equations (\ref{eq:species}) and (\ref{eq:mass}):
\begin{equation}
\frac{\partial X_k}{\partial t} = -\Ub\cdot\nabla X_k + \omegadot_k.
\label{eq:Primitive Species}
\end{equation} 
The base state evolution equations for density and enthalpy can be
found by averaging (\ref{eq:mass}) and (\ref{eq:enthalpy})
respectively over a layer of constant radius, resulting in
\begin{eqnarray}
\frac{\partial\rho_0}{\partial t} &=& -\nabla\cdot(\rho_0 w_0 \eb_r),\label{eq:Base State Density}\\
\frac{\partial(\rho h)_0}{\partial t} &=& -\nabla\cdot\left[(\rho h)_0w_0\eb_r\right] + \psi + \overline{\rho \Hnuc} + \overline{\rho \Hext}. \label{eq:Base State Enthalpy}
\end{eqnarray} 
where $\psi$ is the Lagrangian change in the base state pressure
defined as $\psi \equiv {D_0p_0}/{Dt} \equiv {\partial p_0}/{\partial t} + 
w_0 {\partial p_0}/{\partial r}$ and is related to the total pressure by
\begin{equation}
\frac{Dp_0}{Dt} = \psi + \Ubt\cdot\nabla p_0.
\end{equation}
Subtracting the base state evolution equations from the corresponding
full state equations yields
\begin{eqnarray}
\frac{\partial\rho'}{\partial t} &=& -\Ub\cdot\nabla\rho' - \rho'\nabla\cdot\Ub - \nabla\cdot\left(\rho_0\Ubt\right),\label{eq:Perturbational Density}  \\
\frac{\partial(\rho h)'}{\partial t} &=& -\Ub\cdot\nabla(\rho h)' - (\rho h)'\nabla\cdot\Ub - \nabla\cdot\left[(\rho h)_0\Ubt\right] + \Ubt\cdot\nabla p_0 \nonumber \\
   && + ( \rho\Hnuc - \overline{\rho \Hnuc}) + (\rho\Hext - \overline{\rho \Hext}) \label{eq:Perturbational Enthalpy}.
\end{eqnarray}
In our treatment of enthalpy, we split the reactions and external heating from 
the hydrodynamics, i.e., during the hydrodynamics step, we 
neglect the $\omegadot_k$, $\Hnuc$, and $\Hext$ terms.  Also, in our treatment
of species, we similarly split the reactions from the hydrodynamics.

While equation (\ref{eq:Base State Density}) properly captures the
change in $\rho_0$ due to atmospheric expansion caused by heating, it
neglects changes that can occur due to significant convective
overturning.  We impose the constraint that $\overline{\rho'}=0$
for all time.  In Paper III, we quantified the drift 
in $\overline{\rho'}$ by introducing $\etarho$ in the equation
\begin{equation}
\frac{\partial\overline{\rho'}}{\partial t} = -\nabla\cdot(\etarho\eb_r).
\end{equation}
However, we incorrectly derived $\etarho$ by assuming
$\overline{\nabla\cdot(\rho'w_0\eb_r)}=0$, when in general this is not true
since $\rho'$, when predicted to time-centered edges, does not in general
satisfy $\overline{\rho'}=0$.  Therefore, the correct expression is
$\etarho = \overline{\left(\rho'\Ub\cdot\eb_r\right)}$.
In practice, we correct the drift by simply setting $\rho_0 =
\overline{\rho}$ after the advective update of $\rho$.  However we still need to
explicitly compute $\etarho$ since it appears in other
equations.

%==========================================================================
% Numerical Methodology
%==========================================================================
\section{Overview of Numerical Methodology}\label{Sec:Overview}
\begin{figure}[tb]
\centering
\includegraphics[width=2.0in]{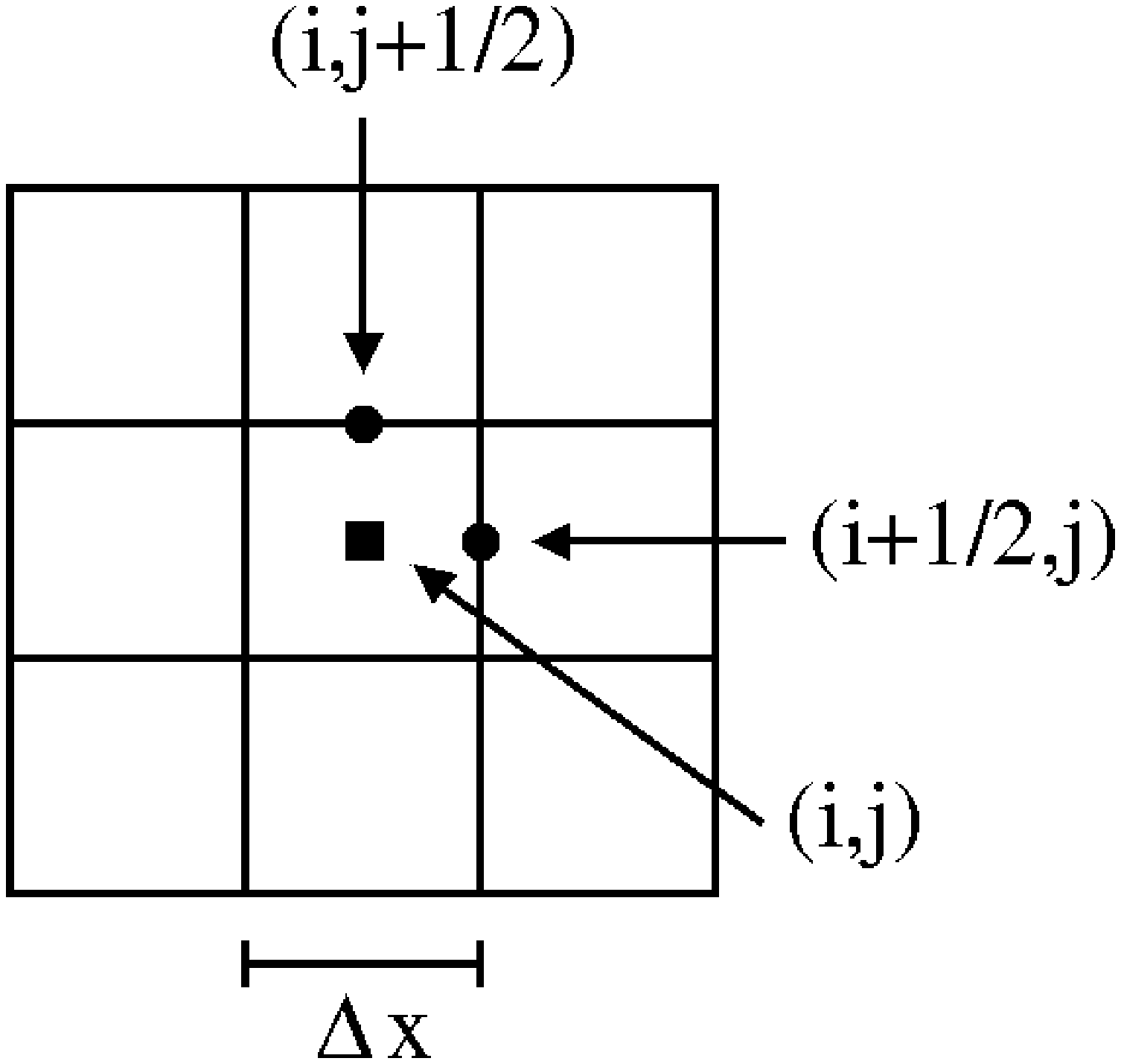}
\includegraphics[width=3.0in]{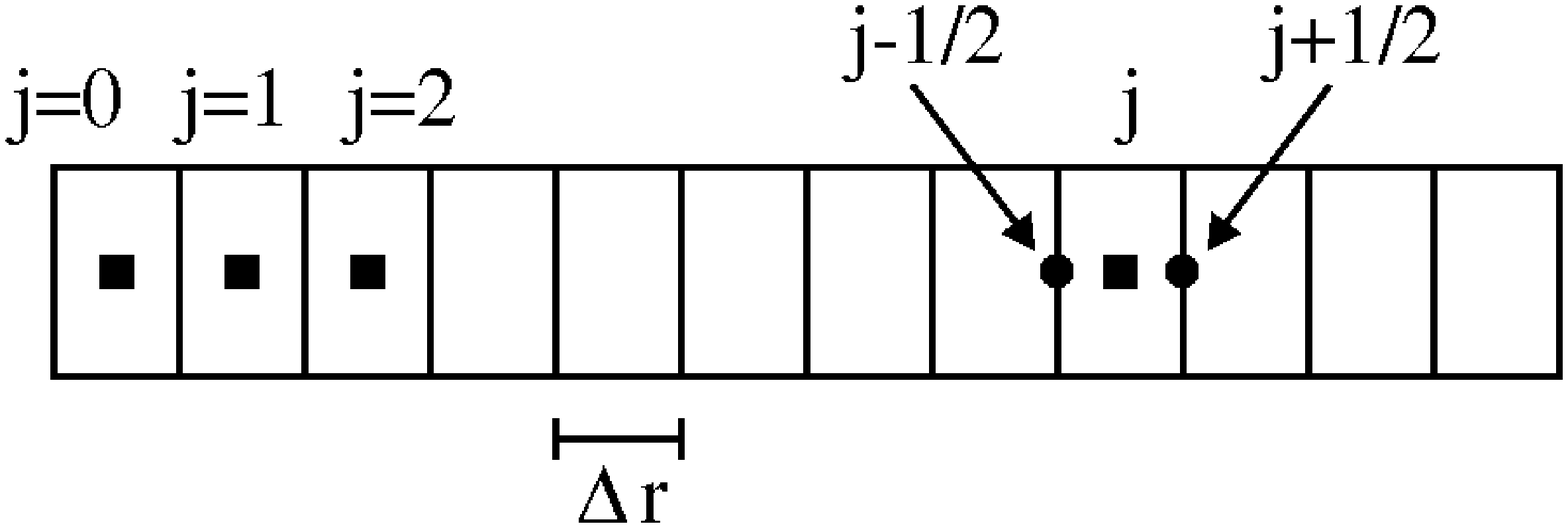}
\caption{\label{Fig:grid}
(Left) For data on the Cartesian grid (shown here in two dimensions), we use a
cell-centered convention with indices $i,j,k$ (in three dimensions).  Edges are denoted 
with a half-integer.  (Right)  The base state lives on a radial array and 
uses a cell-centered convention with index $j$.  Edges are denoted with a 
half-integer.}
\end{figure}

We shall refer to local atmospheric flows in two and three dimensions as problems 
in ``planar'' geometry, and full-star flows in three dimensions as problems in 
``spherical'' geometry.
The solution in both cases consists of the Cartesian grid solution
($\Ubt,\rho,h,X_k,T$) and the one-dimensional base state solution 
($w_0,\rho_0,(\rho h)_0,p_0$),
all of which are cell-centered except for $w_0$, which is edge-centered.
Edge-centered data is denoted by a half-integer subscript.  See Figure
\ref{Fig:grid} for a representation 
of each grid structure.  The time step index is denoted as a superscript.
%%%%%%%%%%%%%%%%%%%%%%%%%%%%%%%%%
%%%%%%%%%%%%%%%%%%%%%%%%%%%%%%%%%

\begin{figure}[tb]
\centering
\includegraphics[height=2.0in]{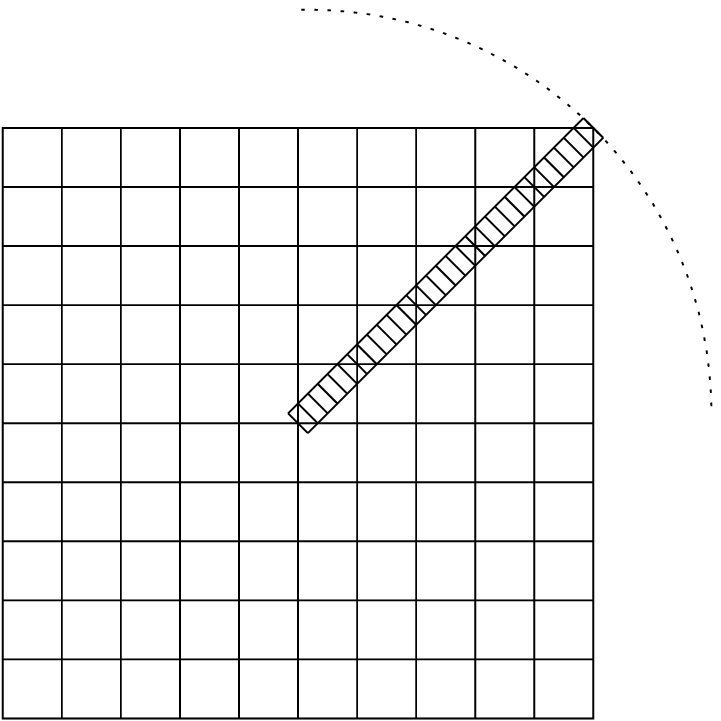} \hspace{0.5in}
\includegraphics[height=2.0in]{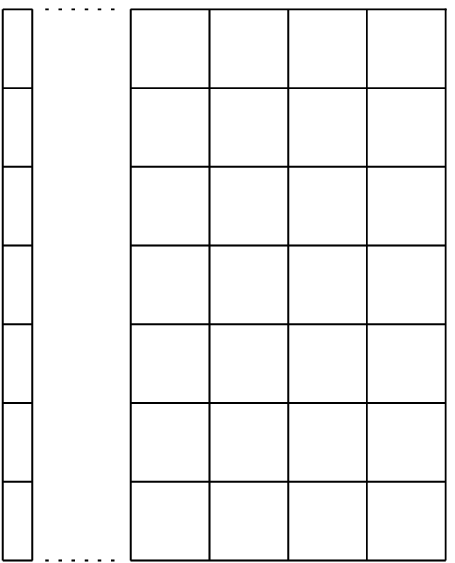}
\caption{\label{Fig:Base Grid Single}
(Left) For problems in spherical geometry, there is no direct alignment between the radial 
array cell centers and the Cartesian grid cell centers.  (Right) For problems in planar 
geometry, there is a direct alignment between the radial array cell centers and the 
Cartesian grid cell centers.}
\end{figure}

For planar problems, $\eb_r$ is in alignment with the Cartesian grid
unit vector in the outward radial direction, $\eb_y$ (in two dimensions) or $\eb_z$ 
(in three dimensions).  We choose $\Delta r = \Delta x$ so that there will be a 
simple, direct mapping between the radial array and the Cartesian grid.  
For spherical problems, $\eb_r$ is not in alignment with any Cartesian coordinate 
direction. Our choice of $\Delta r$ can be independent of $\Delta x$; as in 
Paper IV, we use $5\Delta r = \Delta x$.  Note that for spherical problems, we place the 
center of the star at the center of the computational domain, and therefore 
the center of the star lies at a corner where 8 Cartesian cells 
meet.  See Figure \ref{Fig:Base Grid Single} for an illustration of the relationships
between the radial array and the Cartesian grid for spherical and planar geometries.
%%%%%%%%%%%%%%%%%%%%%%%%%%%%%%%%%
%%%%%%%%%%%%%%%%%%%%%%%%%%%%%%%%%

The time-advancement algorithm uses a predictor-corrector formalism.  In the predictor step,
we compute an estimate of the expansion of the base state, then compute a preliminary
estimate of the state at the new time level.  In the corrector step, we use this
preliminary state to compute a new estimate of the expansion of the base state, and
then compute the final state at the new time level.  We incorporate reactions and 
external heating using Strang-splitting.
As in previous papers, our algorithm is second-order in space and time.

The full details of the algorithm are presented in Appendix 
\ref{Sec:Time Advancement Algorithm}.  The main algorithm description in Appendix
\ref{Sec:Main Algorithm Description} is similar
to the description in Paper III, but has been significantly updated to show how we 
incorporate the time-dependent spherical base state.  There are numerous other 
improvements we have made to the algorithm since papers III and IV, which are described 
in Appendix \ref{Sec:Changes}.  Note that these changes have also been incorporated into the
main algorithm description.  Overall:
\begin{itemize}
\item Appendix \ref{Sec:Changes} is a summary of algorithmic changes since papers III and IV.
\item Appendix \ref{Sec:Gravity} describes how we compute and discretize gravity.
\item Appendix \ref{Sec:Main Algorithm Notation} is a description of shorthand notation we
      use in describing the algorithm.
\item Appendix \ref{Sec:Main Algorithm Description} steps through the algorithm in detail.
\item Appendix \ref{Sec:Cutoffs} describes special treatment given to low density regions
      in the simulation.
\end{itemize}

\section{Mapping}\label{Sec:Mapping}
At many points in the algorithm, we need to map the full state on the
Cartesian grid onto a one-dimensional radial array, and vice-versa.
Since Paper IV, we have greatly increased the accuracy of the
numerical mapping to and from these data structures for spherical
problems, most notably the lateral average routine described below.
We refer to the procedure for mapping a cell-centered Cartesian field
to a cell-centered radial array as a ``lateral average'', and we refer
to the procedures for mapping an edge- or cell-centered radial array
to an edge- or cell-centered Cartesian grid as a ``fill''.
\subsection{Lateral Average}\label{Sec:Lateral Average}
\begin{figure}[t]
\centering
\includegraphics{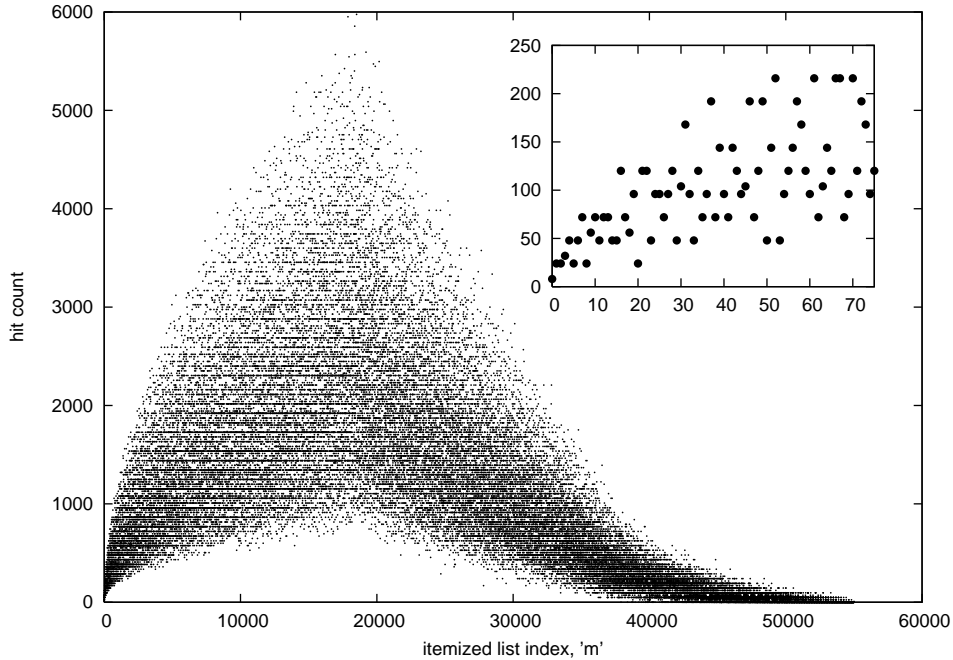}
\caption{\label{Fig:test_average3}
The number of Cartesian cells whose centers lie
at a radius $\hat r_m$ (i.e., the ``hit count'') for a 384$^3$ domain vs.\ the
itemized list index, $m$.  
Indices $m<18,432$ correspond to locations within half the width 
of the computational domain.  A non-zero set of Cartesian cell centers maps 
into the radius associated with every $m\le 37,912$, which corresponds to 
approximately 0.72 times the width of the computational domain.  The inset plot
is a zoom-in of the innermost 75 values of $m$.}
\end{figure}

For any Cartesian cell-centered field, $\phi$, we define 
$\overline{\phi} = ${\bf Avg}$(\phi)$ as the lateral average over a layer at 
constant radius $r$, $\Omega_H$, as
\begin{equation}
\overline{\phi(r)} = \frac{1}{A(\Omega_H)}\int_{\Omega_H}\phi(r,\xb)d\Omega; \qquad
A(\Omega_H) = \int_{\Omega_H}d\Omega.
\end{equation}
\begin{description}
\item[planar:] 
  This is a straightforward arithmetic average of cells at a particular
  height since the radial cell centers are in alignment with the Cartesian 
  grid cell centers.
\item[spherical:] It can be shown that any Cartesian cell center is a
  radius $\hat r_m = \Delta x\sqrt{\sfrac{3}{4} + 2m}$ from the center of
  the star, where $m \ge 0$ is an integer.  
  For example, the Cartesian cell with coordinates $(i,j,k) = (1,1,1)$ relative
  to the center of the star lies at a distance of $\Delta x\sqrt{\sfrac{3}{4}+6}$
  from the center of the star, corresponding to $m=3$.  The Cartesian cells
  with coordinates $(i,j,k) = (2,0,0), (0,2,0)$, or $(0,0,2)$ relative to
  the center of the star also lie at that same 
  distance.  For the 384$^3$ resolution
  examples in this paper, we have verified that a non-zero set of
  Cartesian cell centers map into each radius $\hat r_m$ until $m$ is 
  large enough to correspond to a radius larger
  than half the width of the computational domain (i.e., the edge of the
  domain, not the corner of the domain).  Figure 
  \ref{Fig:test_average3} shows the number of Cartesian cells that map into each 
  radius $\hat r_m$, which we refer to as the ``hit count'', for a 384$^3$ domain.  
  We use this mapping to help construct the lateral average, using the following steps:
\begin{enumerate}
\item Create an itemized list, $\hat\phi_m$, where each element is
  associated with a radius $\hat r_m = \Delta x\sqrt{\sfrac{3}{4} + 2m}$ from the 
  center of the star.
\item For each $\hat\phi_m$, compute the arithmetic average value of the Cartesian 
  cells whose centers lie at the associated radius.  As an additional 
  element in the itemized list, include the center of the star (corresponding
  to a radius of $r=0$).  Compute this additional value of $\hat\phi$
  at this location using quadratic interpolation with $\hat\phi_0$, $\hat\phi_1$,
  and a homogeneous Neumann condition at $r=0$ as the stencil points.  Note
  that for very large values of $m$, it is possible that no Cartesian cell centers
  exist at a radius $\hat r_m$ (i.e., the hit count is zero).  
  If so, we say that $\hat\phi_m$ has an 
  undefined/invalid value, and we ignore such values for the rest of this procedure.
\item To compute the lateral average, use quadratic interpolation using the 
  value in the itemized list with the closest associated radius, $\hat\phi_k$,
  and the nearest values above and below, $\hat\phi_{k_+}$ and $\hat\phi_{k_-}$,
  using divided differences:
  \begin{equation}
    \overline{\phi(r)} = \hat\phi_{k_-} + \frac{\hat\phi_k-\hat\phi_{k_-}}{\hat r_k-\hat r_{k_-}}(r-\hat r_{k_-}) + \frac{\frac{\hat\phi_{k_+}-\hat\phi_k}{\hat r_{k_+}-\hat r_k}-\frac{\hat\phi_k-\hat\phi_{k_-}}{\hat r_k-\hat r_{k_-}}}{\hat r_{k_+}-\hat r_{k_-}}(r-\hat r_{k_-})(r-\hat r_k)\nonumber,
  \end{equation}
  where $\hat r_{k_-}$, $\hat r_k$, and $\hat r_{k_+}$ are the three radii
  associated with $\hat\phi_{k_-}$, $\hat\phi_k$, and  $\hat\phi_{k_+}$.
  Finally, constrain $\overline{\phi(r)}$ to lie within the range of
  $\hat\phi_{k_-}$, $\hat\phi_k$, and $\hat\phi_{k_+}$ so as to not introduce
  any new maxima or minima.
\end{enumerate}
  In \S \ref{Sec:Mapping Test}, we show the improvement of this averaging procedure 
  over the Paper IV procedure.
\end{description}
%%%%%%%%%%%%%%%%%%%%%%%%%%%%%%%%%
%%%%%%%%%%%%%%%%%%%%%%%%%%%%%%%%%
\subsection{Fill}\label{Sec:Fill}
\begin{figure}[tb]
\centering
\includegraphics[width=1.75in]{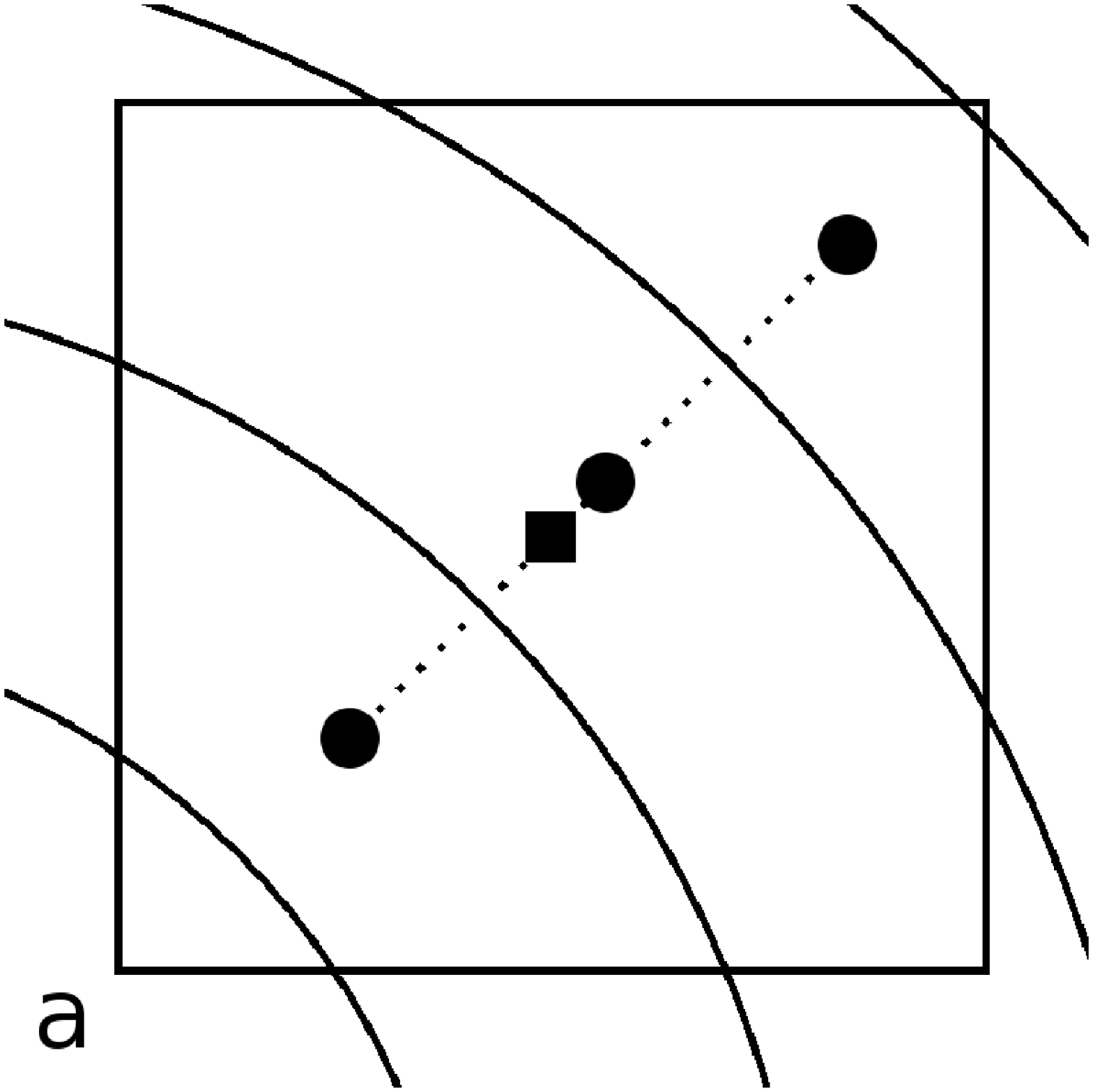}\hspace{0.1in}
\includegraphics[width=1.75in]{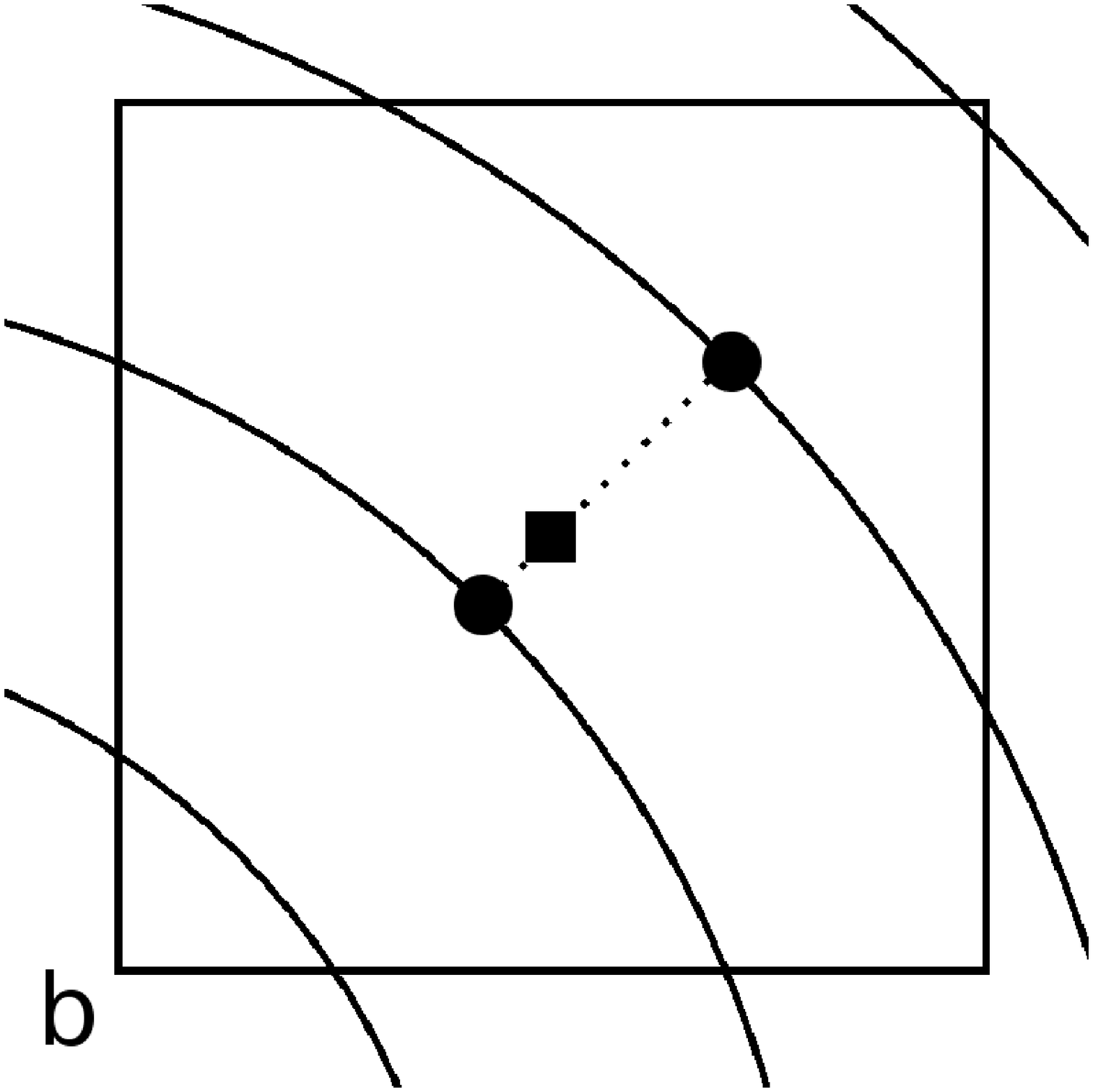}\hspace{0.1in}
\includegraphics[width=1.75in]{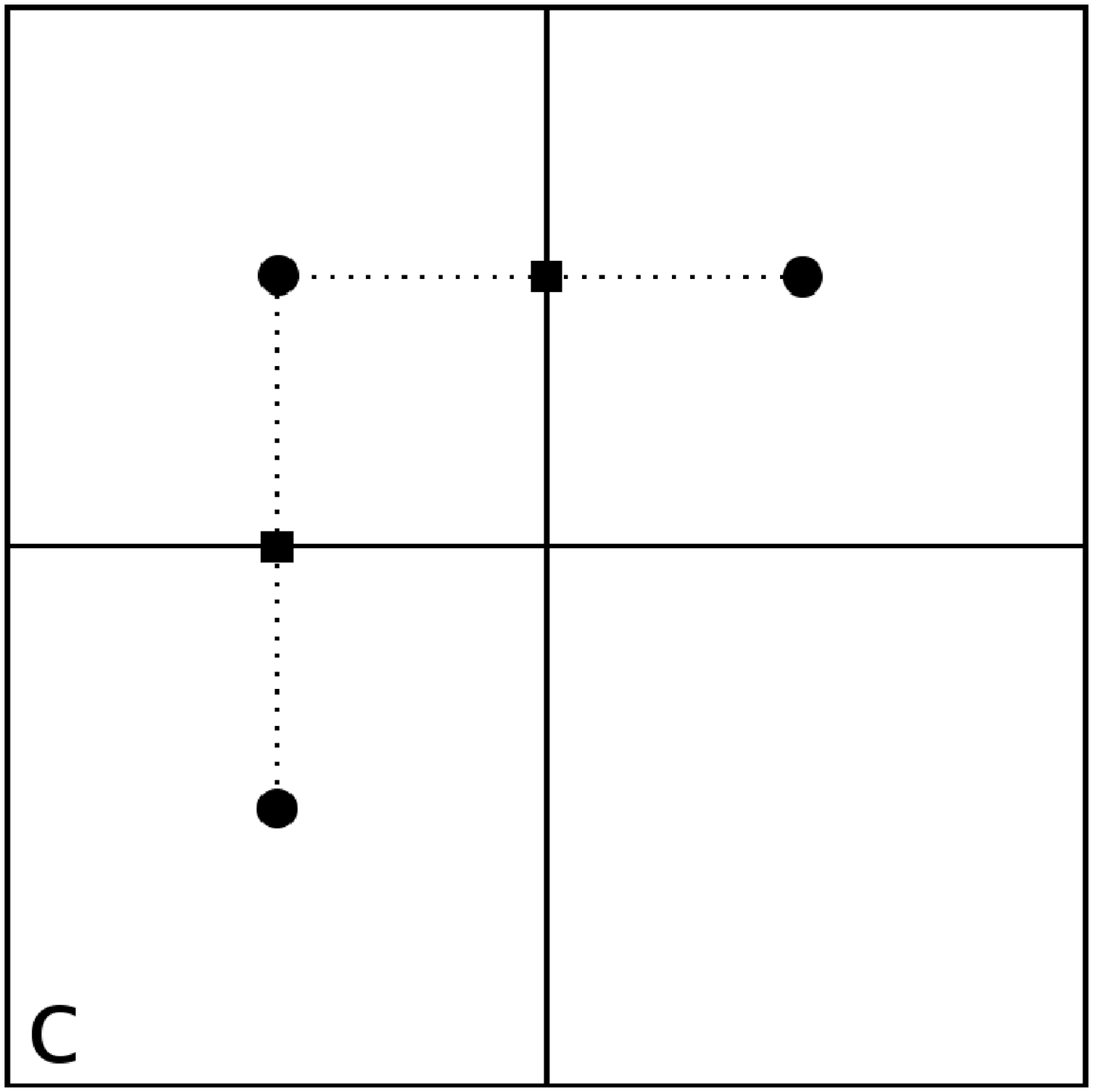}
\caption{\label{Fig:fill} Illustrations of the fill operation for
  spherical geometry.  (a) To fill the Cartesian cell center (fill
  type 1), represented by the square, from radial cell-centered data,
  represented by the circles, we use quadratic interpolation from the
  nearest three points.  (b) To fill the Cartesian cell center
  (fill type 2), represented by the square, from radial edge-centered
  data, represented by the circles, we use linear interpolation from
  the nearest two points. (c) To fill a Cartesian edge (fill
  types 3 and 4), represented by the squares, first fill the 
  Cartesian cell centers, represented by the circles, then average
  the two neighboring cell centers.}
\end{figure}

There are four different mappings from a one-dimensional radial array to the three-dimensional Cartesian grid;
below we describe the procedures for planar and spherical geometries separately.
\begin{description}
\item[planar:]
\end{description}
\begin{enumerate}
\item To map a cell-centered radial array onto Cartesian cell centers,
  we use direct-injection since the radial cell centers are in
  alignment with the Cartesian cell centers. 
\item To map an edge-centered radial array onto Cartesian 
  cell centers, we average the two nearest radial edge-centered values.
\item To map a cell-centered radial array onto Cartesian edges with normal in 
  the radial direction, we use 4th order spatial interpolation.  For example,
  in two dimensions,
\begin{equation}
\phi_{i,j+\myhalf} = \frac{7}{12}(\phi_{j}+\phi_{j+1}) - \frac{1}{12}(\phi_{j-1}+\phi_{j+2}).
\end{equation}
  We constrain $\phi_{i,j+\myhalf}$ to lie between the interpolated values, and 
  lower the order of interpolation near domain boundaries.  For the 
  Cartesian edges transverse to the base state direction, we use direct-injection 
  since the radial cell centers are in alignment with these 
  Cartesian edges.
\item To map an edge-centered radial array onto Cartesian edges, 
  we use direct-injection on Cartesian edges normal to the base state direction
  since the radial edges are in alignment with these Cartesian edges.  
  For the remaining Cartesian edges, we average the two nearest radial 
  edge-centered values.
\end{enumerate}
\begin{description}
\item[spherical:] 
\end{description}
\begin{enumerate}
\item To map a cell-centered radial array onto Cartesian cell centers,
      we use quadratic interpolation from the nearest 
      three radial cell centers (see Figure \ref{Fig:fill}a).
      This is a departure from Paper IV, in which we used 
      piecewise constant interpolation.  
\item To map an edge-centered radial array onto Cartesian cell centers, 
      we use linear interpolation from the nearest two points
      (see Figure \ref{Fig:fill}b).
\item To map a cell-centered radial array onto Cartesian edges,
      we first map the radial array onto Cartesian cell centers (see 1.), 
      then average the two neighboring centers to obtain the Cartesian edge values
      (see Figure \ref{Fig:fill}c).
\item To map an edge-centered radial array onto Cartesian edges,
      we first map the radial array onto Cartesian cell centers (see 2.), 
      then average the two neighboring centers to obtain the Cartesian edge values
      (see Figure \ref{Fig:fill}c).
\end{enumerate}

%%%%%%%%%%%%%%%%%%%%%%%%%%%%%%%%%
%%%%%%%%%%%%%%%%%%%%%%%%%%%%%%%%%

\section{Adaptive Mesh Refinement}\label{Sec:AMR}
Our approach to AMR uses a nested hierarchy
of logically rectangular grids with successively finer grids at higher levels.
This is based on the strategy introduced for gas dynamics by 
\citet{berger-colella}, extended to the incompressible 
Navier-Stokes equations by \citet{AlmBelColHowWel98},
and extended to low Mach number reacting flows by
\citet{pember-flame} and \citet{DayBell:2000}.
We refer the reader to these works for more details.  
The key difference between our method and these earlier methods stems from 
the presence of a one-dimensional base state whose time evolution is coupled to that of 
the full solution.  To the best of our knowledge, there are no existing AMR algorithms
for astrophysics or any other field,
for flows with a time-dependent base state coupled to the full solution.
For simplicity, we present a version of the algorithm with no subcycling in time,
i.e., the solution at all levels is advanced with the same time step.

We first summarize our AMR approach without the 
base state, then discuss how the base state is incorporated in 
both the planar and spherical cases.

\subsection{Creating and Managing the Grid Hierarchy}

At each time step the state data is defined on a nested hierarchy
of grids, ranging from the base level ($\ell = 1$), which covers the
entire computational domain, to the finest level ($\ell = {\ell}_\mathrm{max}$).  
At each level there is a union of non-intersecting rectangular grids with the 
same spatial resolution.  For simplicity, we require that the cells composing the 
grids be square ($\Delta x = \Delta y = \Delta z$), and that the refinement
ratio between levels be 2.  The grids in the interior of the
computational domain are required to be properly nested, i.e., the union of
grids at level $\ell+1$ is completely contained in the union of grids
at level $\ell$.  Additionally, in the interior, we require that each 
grid at level $\ell+1$ be a distance of at least two level $\ell$ cells 
from the boundary between level $\ell$ and level $\ell-1$ grids;
this allows us to always fill ``ghost cells'' at level $\ell+1$ from the
level $\ell$ data (or the physical boundary conditions, if appropriate).
We initialize the grid hierarchy and regrid following the procedure
outlined in \citet{Bell:1994}.  A user-specified error
estimation routine is used to tag cells where more resolution is
desired.  The tagged cells are grouped into rectangular patches
following \citet{bergerRigoutsos:1991}, and
subsequently refined to create new grids at next level.  
Refinement continues until the maximum level is reached.

During {\bf Step 0},\footnote{The ``{\bf Step}'' notation is used in describing the
full details of the algorithm in Appendix \ref{Sec:Main Algorithm Description}}
grids at all levels are filled
directly from the initial data.  As the simulation progresses, we
periodically check our refinement criteria and regrid as necessary.
This regridding takes places during {\bf Step 12}, before computing
the next time step.  Newly created grids are filled by using data from
previous grids at the same refinement level (if available) or by
interpolating from underlying coarser grids.

\subsection{Communication between levels}

Since we use the same time step to advance the solution at all levels, 
much of the complication associated with synchronization of data between levels in a 
subcycling algorithm (see \citealt{AlmBelColHowWel98}) is eliminated.  
The MAC projections in {\bf Step 3} and {\bf 7} enforce that 
$\uadvone$ and $\uadvtwo,$ respectively, on any coarse edge underlying fine
edges are the average of the values on the fine edges.  Similarly, the nodal projection in
{\bf Step 12} enforces that at any coarse node underlying a fine node,
the value of $\phi$ on the coarse node is identical to the value on the fine
node above it.  The additional communication of data between levels occurs as follows:
\begin{itemize}

\item Before any explicit operation at level $\ell>1,$ data in ghost cells at that level
      are filled by interpolating from level $\ell-1$, or imposing physical boundary 
      conditions, as appropriate.

\item Edge-based fluxes at level $\ell < \ell_{max}$ that
      underlie edges at level $\ell+1$ are defined to be the average of the fluxes  
      on level $\ell+1$ at that edge.  This enforces conservation.

\item After any update to the solution, data at finer levels is conservatively 
      averaged onto the underlying coarse grid cells, starting at the finest level.

\end{itemize}

\subsection{AMR with a Time-Dependent Base State}
Our specific treatment of AMR is guided by our initial scientific applications,
including Type I X-ray bursts and the convective phase of SNe Ia, as
well as numerical concerns, most notably the presence of the one-dimensional base state.
Our treatment of the base state in an AMR framework differs for planar vs.\ 
spherical problems.

\begin{figure}[tb]
\centering
\includegraphics[height=2.0in]{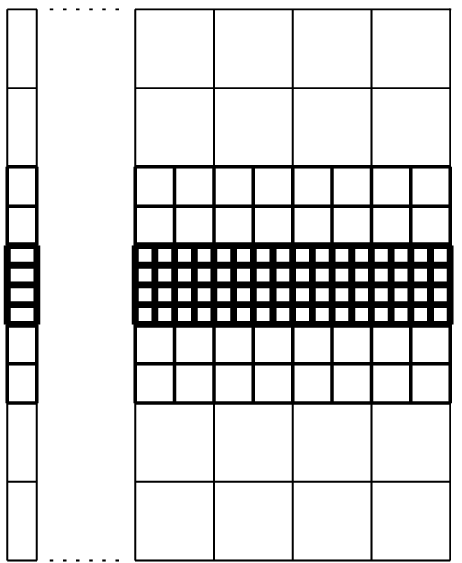} \hspace{0.5in}
\includegraphics[height=2.0in]{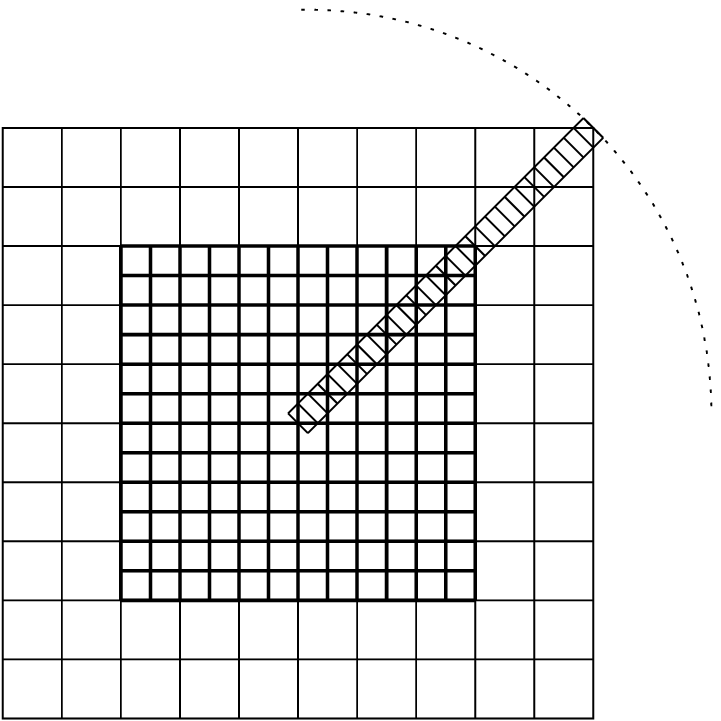}
\caption{\label{Fig:Base Grid}  
(Left) For multi-level problems in planar geometry, we force a direct alignment
between the radial array cell centers and the Cartesian grid cell centers by 
allowing the radial base state spacing to change with space and time.
(Right) For multi-level problems in spherical geometry, since there is no direct alignment
between the radial array cell centers and the Cartesian grid cell centers, we choose to fix
the radial base state spacing across levels.}
\end{figure}

For planar problems, our approach is to define a
radial base state array with variable mesh spacing.  A general localized
fine Cartesian grid would require either a base state that exists at multiple resolutions
at a particular height, or an interpolation algorithm to obtain the base
state value at a particular height if $\Delta r \ne \Delta x$.
Both of these methods pose problems, as they generate
oscillations in perturbational quantities (such as $\rho', (\rho h)'$ and the 
$S-\Sbar$ term on the right hand side of the divergence constraint)
since the lateral average routine is only defined 
when the base state is aligned with the Cartesian grid across the width of
the domain.  Any attempt at interpolation will cause oscillations in
the perturbational quantities directly related to the interpolation
error.  We have found that such oscillations can be detrimental to the
results.  With these issues in mind, we choose to only allow fine
grids to exist that span the width of the domain.  This way, the base
state exists as a single seamless entity with multiple resolutions
depending on height (see Figure \ref{Fig:Base Grid}).  We will take advantage
of this type of grid structure in our studies of Type I X-ray bursts.
%%%%%%%%%%%%%%%%%%%%%%%%%%%%%%%%%
%%%%%%%%%%%%%%%%%%%%%%%%%%%%%%%%%

Next, we define ghost cell values for the finer base state 
levels, and fill these values by interpolating coarser data.
This makes the algorithm directly compatible with the one-dimensional time-centered
edge state calculation used in {\bf Advect Base Density}
\footnote{The boldface notation refers to numerical modules we have described in
Appendix \ref{Sec:Main Algorithm Notation}.}
and {\bf Advect Base Enthalpy}.  In particular, the slopes
can be used with a consistent stencil at each 
level, that is not dependent on the data from any other level once the ghost
cells are set.

Finally, whenever we regrid the Cartesian grid data, 
we regrid the base state to match the grid structure of the Cartesian grid.
Then, we set $\rho_0=\overline{\rho}$ and compute $p_0$ using {\bf Enforce HSE}.  
To compute $\psi$ and $w_0$ on the new base state array, we use piecewise linear
interpolation of the coarser data to fill any new fine radial cells/edges.

For spherical geometry, we first note that even in the single-level case 
the radial base state is not aligned with the Cartesian grid.
Therefore, we use a base state with a fixed $\Delta r$ for all levels (see Figure
\ref{Fig:Base Grid}).  As in the single-level algorithm, we
choose $\Delta r = \Delta x/5$, but here, $\Delta x$ corresponds to
resolution of the Cartesian grid at the finest level.

Our next consideration is defining the radial average.  First, 
we first create an itemized list associated with each level of refinement using
only Cartesian cells that are not covered by cells at a finer level.
At this point one option would
be to merge the lists and proceed as in the single-level algorithm; this 
was tested and found to be problematic.  Instead, we choose the list from
a chosen particular level and define the average using quadratic interpolation 
with only this list, as in the single-level case.
To decide which list to use, we first examine the three points that would
be used by quadratic interpolation at each level.   The guiding principle is to avoid 
using interpolation points with low hit counts.  Thus, at each level we find the minimum 
hit count of the three points; the level which has the largest minimum hit count is the 
level whose list we use for interpolation.  We note that this multi-level average
works particularly well when the center of the star is fully refined.  This is the case
since near the center of the star, there are relatively few Cartesian cells that
contribute to each radial bin, so by fully refining the center of the star, we ensure
that the multi-level averaging procedure retains the accuracy of a single-level spherical
average near the center.  For our studies of SNe Ia, we will take advantage of this fact
by always refining the center of the star, which is our region of interest
(see Figure \ref{Fig:wdconvect_grid} in \S \ref{Sec:Full-Star AMR}, as an example).
In \S \ref{Sec:Mapping Test}, we present numerical tests of the new multi-level 
averaging procedure for spherical geometry.

For both planar and spherical problems, after regridding the Cartesian grid, we
make the state thermodynamically consistent by computing $T=T(\rho,h,X_k)$ 
(for planar problems) or $T=T(\rho,p_0,X_k)$ (for spherical problems).
Then, we recompute $\gammaonebar$ and $\beta_0$ as described in {\bf Steps 10} 
and {\bf 11}.

\subsection{Parallel Implementation}

We parallelize the algorithm by distributing the grids on each level 
across processors.  Each grid carries a perimeter of ghost cells that
are filled from neighboring grids at the same level or interpolated from
coarser grids as needed.  This allows the data on each grid to be updated 
independently of the other grids.  A typical grid is large enough (e.g., $32^3$ cells) 
that for explicit operations the cost of computation within each grid greatly exceeds the 
cost of communication between grids.   The linear solves necessary for the
MAC projection and approximation nodal projection have higher communication
costs, but we still obtain good parallel efficiency for the overall algorithm.
A scaling study for {\tt MAESTRO} can be found in \citet{maestro-scidac2007}.

Since the one-dimensional base state arrays are so much smaller than the three-dimensional arrays
holding the full solution,  each processor owns a copy of the entire one-dimensional base
state arrays.  Operations such as averaging to define base state quantities
require a collection operation among grids, followed by a distribution of the
average state to each processor.

%==========================================================================
% Results
%==========================================================================
\section{Test Problems}\label{Sec:Test Problems}
We have developed a suite of test problems in order to test various aspects of our
code.
\begin{itemize}
\item In \S \ref{Sec:Mapping Test}, we show that our new mapping procedure from 
      \S \ref{Sec:Mapping} is much more accurate than the mapping procedure
      from Paper IV.
\item In \S \ref{Sec:Spherical Base State}, we show that we are able to properly
      capture the expansion of the base state in a three-dimensional full star simulation
      due to heating at the center of the star.
\item In \S \ref{Sec:Convergence Test}, we show that our multi-level algorithm
      is second-order accurate in space and time by tracking a hot bubble rising in a 
      white-dwarf environment.
\item In \S \ref{Sec:Bubble Rise}, we show that an adaptive algorithm in
      three-dimensional planar geometry can properly track a hot bubble rising
      in a white-dwarf environment.
\item In \S \ref{Sec:Forced Convection}, we demonstrate that a multi-level, 
      two-dimensional planar simulation will properly
      capture the expansion of the base state due to a heating layer, and also
      that a multi-level simulation is able to capture the same fine-scale structure
      as a single-level simulation at the same effective resolution over a short time.
\item In \S \ref{Sec:Full-Star AMR}, we demonstrate that a full-star simulation
      with AMR can be used to study the dynamics of convection in white dwarfs.
\end{itemize}

For test problems \S \ref{Sec:Spherical Base State}-\S \ref{Sec:Full-Star AMR},
we compute flows in which the density spans at least four orders of magnitude.
The large drop in density in the upper atmosphere results in high velocities 
due to conservation of momentum.  This region should not affect the dynamics 
below the surface in the convecting regions of the star.  However, because 
the time step in the low Mach number code is limited by the highest
velocity in the computational domain, the efficiency gains of the 
low Mach number algorithm are reduced if those velocities persist.
We employ a sponging technique to damp such velocities.  Damping techniques 
are commonly used in
modeling atmospheric convection (see, for example \citet{Durran_mono}).
In Paper IV, for full-star convection, we explored the effects of sponging the velocity
beginning at two different heights to demonstrate that the dynamics in the upper
atmosphere do not affect the convecting regions of the star.

Full details for the sponge implementation can be found in papers III and IV, 
but in summary, we add a forcing term to the velocity
update before the final projection.  We use the parameters $r_{\rm sp}, r_{\rm md}$, 
and $\kappa$ to describe the sponge.  The sponge forcing turns on at radius 
$r_{\rm sp}$ and reaches half of its peak strength at radius $r_{\rm md}$.  
We can control the strength of the forcing with the parameter, $\kappa$.  For
full-star problems, we also use an outer sponge which prevents the velocities
near the domain boundaries from becoming too large.

For all of these tests, we use a publicly available, general stellar equation 
of state \citep{flash,timmes_swesty:2000}, with contributions from ions, 
radiation, and degenerate/relativistic electrons.

\subsection{Mapping}\label{Sec:Mapping Test}
To test the new spherical fill and lateral average routines from \S \ref{Sec:Mapping}, 
we first create a unit cube with $384^3$ resolution and no refinement.
We create a radial array with $\Delta r = \Delta x/5$, and initialize the radial array cell 
centers with the Gaussian profile $\phi_{\rm exact}(r) = e^{-10r^2}$.  We map 
$\phi_{\rm exact}$ to Cartesian cell centers with the fill operation, then 
compute the lateral average of this Cartesian field, ${\bf Avg}(\phi)$.
We repeat this process by choosing the grid with two levels of refinement
used in the full-star simulation 
test in \S \ref{Sec:Full-Star AMR}, shown in Figure \ref{Fig:wdconvect_grid}.

\begin{figure}[tb]
\centering
\includegraphics{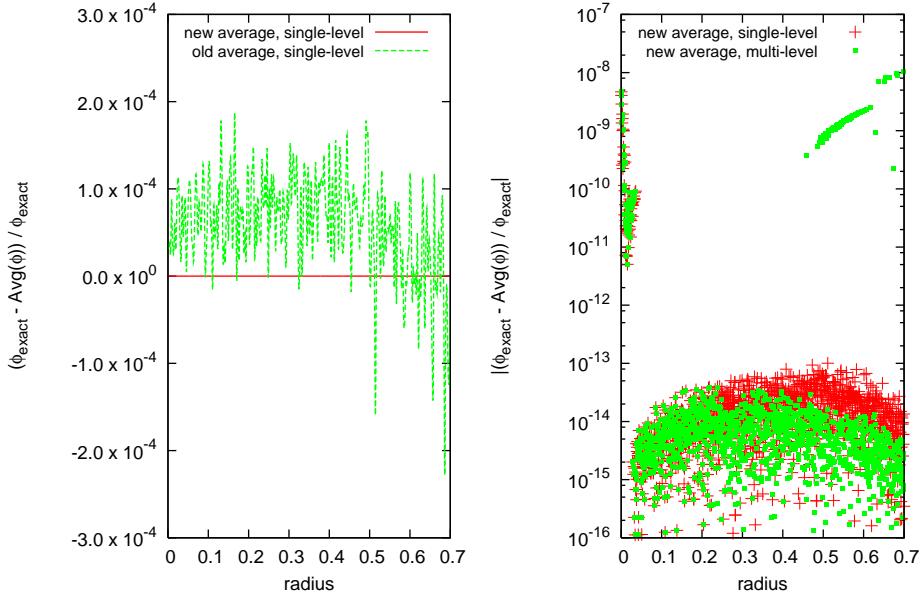}
\caption{\label{Fig:test_average1}
(Left) Relative error between $\phi_{\rm exact}$ and ${\bf Avg}(\phi$) from averaging
a mapped Gaussian profile centered on a unit cube with 384$^3$ resolution and no refinement
for the current mapping (solid line) and the mapping from Paper IV (dashed line).
The new mapping procedure greatly decreases the relative error.
(Right) A zoom-in of the relative error for the new mapping.  
The red markers correspond to the test with no refinement.  
The relative error is $\mathcal{O}(10^{-8})$ for $r\in[0,0.036]$ and is 
$\mathcal{O}(10^{-13})$ for $r\in[0.036,0.7]$.  The
green dots shows the relative error for a test with two levels of refinement
using the grid structure shown in Figure \ref{Fig:wdconvect_grid}.
The relative error is $\mathcal{O}(10^{-8})$ for $r\in[0,0.7]$, which is still
a vast improvement when compared to the Paper IV mapping applied to a single-level
simulation.}
\end{figure}

Figure \ref{Fig:test_average1} (left) shows the relative error between $\phi_{\rm exact}$ 
and ${\bf Avg}(\phi$) for the new mapping procedures and the
mapping procedures from Paper IV for the single-level test.
The new mapping procedure greatly decreases the relative error.  
Figure \ref{Fig:test_average1} (right)
is a zoom-in of the relative error for the new mapping.  For the single-level grid,
the relative error is $\mathcal{O}(10^{-8})$ for $r\in[0,0.036]$ and is 
$\mathcal{O}(10^{-13})$ for $r\in[0.036,0.7]$.  For the test with two levels
of refinement,
the relative error is $\mathcal{O}(10^{-8})$ for $r\in[0,0.7]$, which is still
a vast improvement when compared to the Paper IV mapping applied to a single-level
simulation.
%%%%%%%%%%%%%%%%%%%%%%%%%%%%%%%%%
%%%%%%%%%%%%%%%%%%%%%%%%%%%%%%%%%

\subsection{Spherical Base State}\label{Sec:Spherical Base State}
To test the base state expansion for spherical geometry, we perform a
series of tests similar to those in Paper II which tested the base
state expansion in planar problems.  We run the same test using three
codes--- a one-dimensional version of the compressible code, 
{\tt CASTRO} \citep{castro:2010}, in spherical
coordinates; a one-dimensional version of {\tt MAESTRO} in spherical coordinates, 
and a full three-dimensional spherical
star in {\tt MAESTRO}.

Our initial model is generated by specifying a core density
($2.6\times 10^9~\gcc$), temperature ($6\times
10^8$~K), and a uniform composition ($X(^{12}\mathrm{C}) = 0.3$,
$X(^{16}\mathrm{O}) = 0.7$) and integrating the equation of
hydrostatic equilibrium outward while constraining the specific entropy, $s$, to be
constant. In discrete form, we solve:
\begin{eqnarray}
p_{0,j+1} &=& p_{0,j} + \frac{1}{2} \Delta r ( \rho_{0,j} +
\rho_{0,j+1} ) g_{j+\myhalf}. 
\label{eq:hse} \\
s_{0,j+1} &=& s_{0,j} \label{eq:entropy_constraint}
\end{eqnarray}
We begin with a guess of $\rho_{0,j+1}$ and $T_{0,j+1}$ and use the equation of state
and Newton-Raphson iterations to find the values that satisfy our system.
Since this is a spherical, self-gravitating star, the gravitation
acceleration, $g_{j+\myhalf}$, is updated each iteration based on the
current value of the density. Once the temperature falls below $10^7$~K, we
keep the temperature constant, and continue determining the density via
hydrostatic equilibrium.  This uniquely determines the initial model.

For the one-dimensional simulations, we map the inner $5\times 10^8$~cm of the model onto a
one-dimensional array with 1280 elements with $\Delta r = 3.90625\times 10^5$~cm.
For the full-star three-dimensional simulation, we map the model onto a 
$5\times 10^8$~cm$^3$ domain with 256$^3$ Cartesian grid cells with 
$\Delta x=5\Delta r=19.53125\times 10^5$~cm.  
For the one and three-dimensional {\tt MAESTRO} calculations, 
we use cutoff densities (see Appendix \ref{Sec:Cutoffs}) of $\rho_{\rm cutoff} = 10^5~\gcc$ 
and $\rho_{\rm anelastic} = 10^6~\gcc$, corresponding to radii of approximately
$1.8\times 10^8$~cm and $1.9\times 10^8$~cm, so the star easily fits within the
computational domain for each problem.  For the full-star three-dimensional simulation, 
we use an inner sponge with $r_{\rm sp}$ equal to the radius where $\rho_0=10^7~\gcc$,
$r_{\rm md}$ equal to the radius where $\rho_0=3\times 10^6~\gcc$,
and $\kappa =10$~s$^{-1}$.  We use the same outer sponge as in Paper IV.
All boundary conditions are outflow, except for the center of the one-dimensional
simulations, which uses a symmetry boundary condition.  We run each simulation
using a CFL number of 0.5.

\begin{figure}[tb]
\centering
\includegraphics{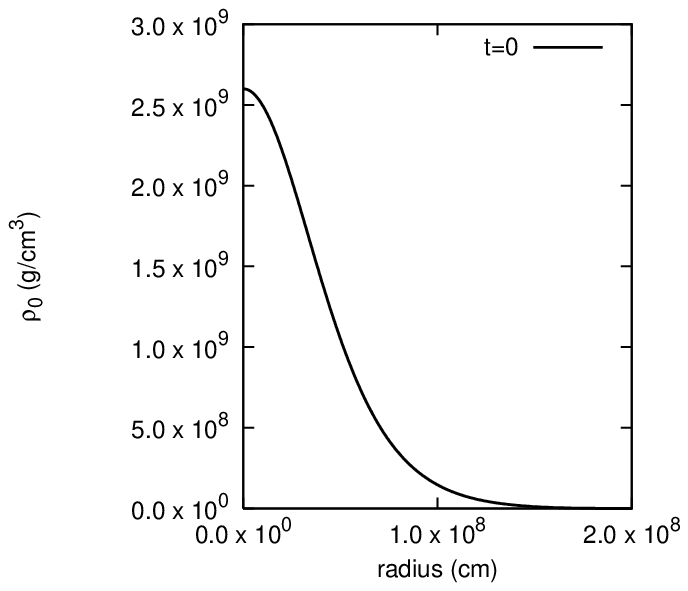}
\includegraphics{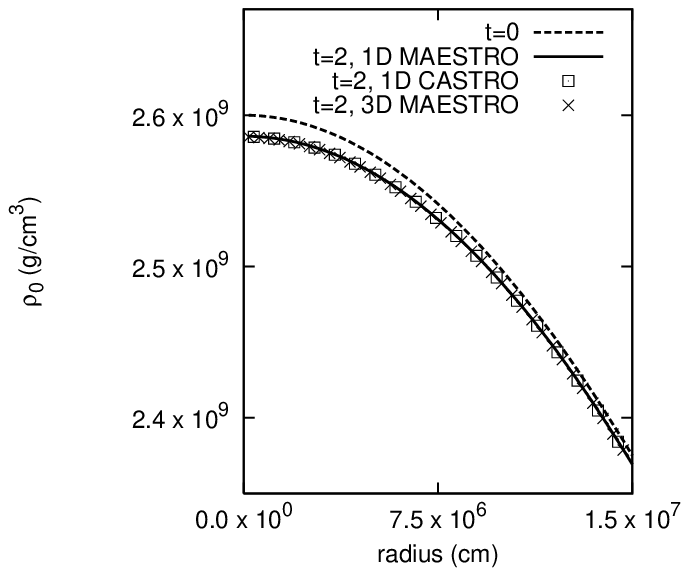} \\
\includegraphics{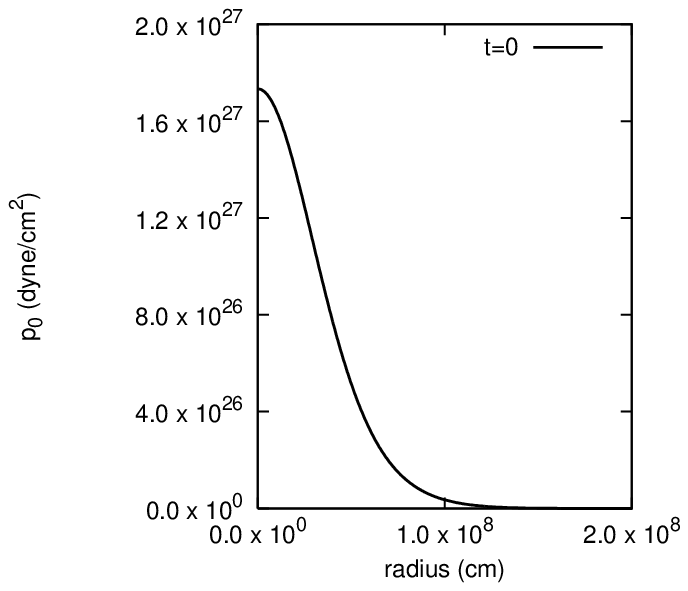}
\includegraphics{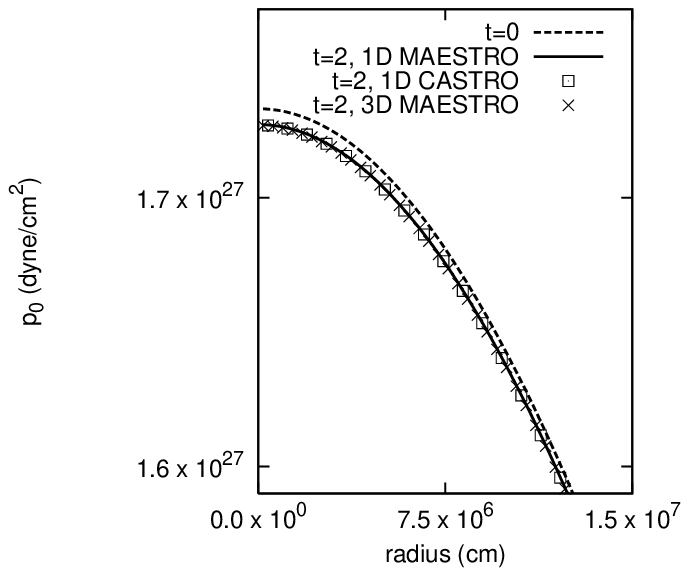} \\
\includegraphics{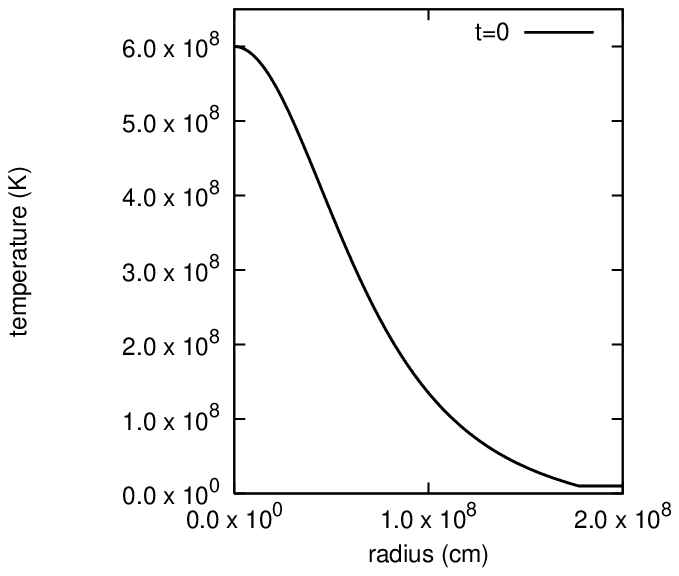}
\includegraphics{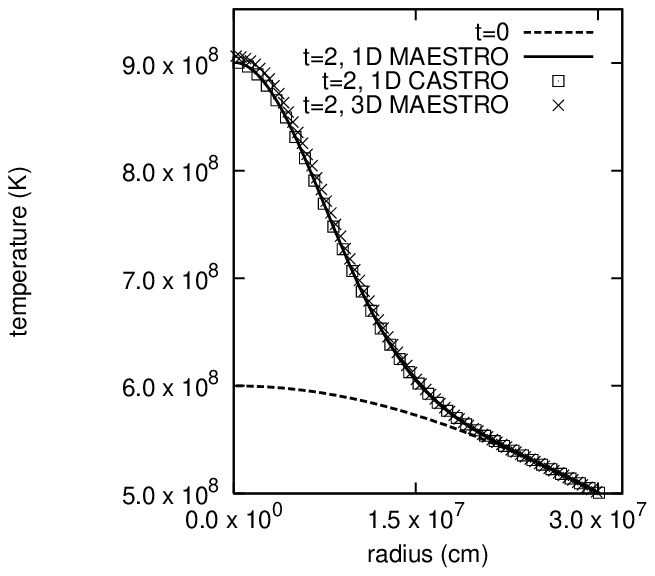}
\caption{\label{Fig:spherical_heat1} Plots of $\rho_0$ (Top), $p_0$ (Middle), 
  and $\overline{T}$ (Bottom) vs.\ radius for a
  white dwarf star subject to heating.  The initial profiles are on
  the left.  Close-up views of the initial profiles and final solutions are on the 
  right.  We use three test codes: a one-dimensional version of
  {\tt MAESTRO} in spherical coordinates, a one-dimensional version of the compressible code, 
  {\tt CASTRO}, in spherical coordinates, and a full-star three-dimensional version of {\tt MAESTRO}.}
\end{figure}

We heat the center of the star for 0.5 seconds and look at the
solution at 2.0 seconds (after the compressible solution has had time to 
re-equilibrate).  We use $H_{\rm ext} = H_0
e^{-(r/10^7~\mathrm{cm})^2}$, with $H_0 =
10^{16}~\mathrm{erg~g^{-1}~s^{-1}}$ chosen to be much larger than the
nuclear energy generation rate during the convective phase of SN~Ia, 
in order to see a measurable effect over a few seconds.
A comparison of $\rho_0,p_0$, and $\overline{T}$ at $t=0$ and $t=2$~s
for each code is shown in Figure \ref{Fig:spherical_heat1}.
There is excellent agreement between each of the simulations.
%%%%%%%%%%%%%%%%%%%%%%%%%%%%%%%%%
%%%%%%%%%%%%%%%%%%%%%%%%%%%%%%%%%

\subsection{Convergence Test}\label{Sec:Convergence Test}
In Paper III, we demonstrated that our single-level algorithm is second-order in space
and time by tracking a hot bubble rising in a white-dwarf environment using
two-dimensional planar geometry.  Here, we perform
the same test to show that our algorithm, with a level of refinement, is second-order
in space and time.

We choose a domain size of $7.2\times 10^7$~cm by 
$2.88\times 10^8$~cm and generate a high resolution initial model with 
$\dr = 7.03125\times 10^4$~cm (which is equal to $\Delta x$ for our highest 
resolution, ``exact'' solution) using the method described in 
Appendix \ref{Sec:Test Problem Initial Model} with $r_\mathrm{base} = 0$.
For each of the remaining, lower resolution 
simulations for this test, we generate an initial model using $\dr$ equal to 
the effective $\Delta x$ of each simulation by linearly interpolating values from 
the high resolution model.  Next, we add a temperature perturbation of the form:
\begin{equation}
T_{ij} = T_{0,j} + 0.3\left[1 + \tanh\left(2-\frac{d_{ij}}{\sigma}\right)\right],
\label{eq:temperature pert}
\end{equation}
where $\sigma = 2.5\times 10^6~$cm and $d_{ij}$ is the physical distance between the
cell center corresponding to cell $(i,j)$ and the location 
$(3.6\times 10^7~$cm$,3.2\times 10^7~$cm).  Then, we call the equation of state to 
compute a consistent $\rho,h = \rho,h(p_0,T,X_k)$ everywhere.
We use the reaction network described in \S~4.2 in Paper III.
We use cutoff densities of $\rho_{\rm cutoff}=\rho_{\rm anelastic}=3\times 10^6~\gcc$, 
and a sponge with $r_{\rm sp}$ equal to the radius where $\rho_0 = 10\rho_{\rm cutoff}$,
$r_{\rm md}$ equal to the radius where $\rho_0 = \rho_{\rm cutoff}$, and $\kappa=10$~s$^{-1}$.
We specify periodic boundary conditions on the side walls, outflow at the top, and a solid
wall at the bottom of the domain.

Since we do not have an exact analytical solution, we consider a single-level simulation
run with $1024\times 4096$ cells and $\dt = 3.125\times 10^{-3}$~s to be the exact 
solution.  We perform three single-level
simulations using resolutions of $64\times 256, 128\times 512$, and $256\times 1024$
grid cells using fixed time steps of $\dt = 0.05$~s, $0.025$~s, and $0.0125$~s, 
respectively.  We also perform two simulations with a single level of refinement with
effective resolutions of $128\times 512$ and $256\times 1024$ grid cells
with fixed time steps of $\dt = 0.025$~s and $0.0125$~s, respectively.
These fixed time steps correspond to a CFL of 0.9.
We refine all cells in the range $r\in [1.8\times 10^7,5.4\times 10^7]$ cm, so 
effectively we have refined 1/8$^{\rm th}$ of the domain, making sure the hot spot 
is contained within the refined region.  We run each simulation to $t=1$~s.
For this test, whenever we call {\bf Enforce HSE} to compute $p_0$ from $\rho_0$, we
use $r = 1.8\times 10^7$~cm as the starting point for 
integration rather than the location of the cutoff density to ensure that 
numerical errors due to integrating the equation of hydrostatic equilibrium
across simulations with different resolutions are minimized.

In order to compute the $L_1$ error norm for each simulation, we average the data from 
the exact solution onto a grid with corresponding resolution.  We measure the $L_1$ error norm
in the physical space corresponding to the refined region using
\begin{equation}
L_1 = \frac{1}{n_{\rm cell}} \sum_{i,j} |\phi_{ij} - \phi_{ij,{\rm exact}}|,
\end{equation}
where $n_{\rm cell}$ is the number of cells we sum over.  This form of the $L_1$ error norm
gives us the average error per cell.  We compute the convergence rate, $p$, between
a coarser and finer simulation using
\begin{equation}
p = \log_2\left(\frac{L_{1,{\rm coarser}}}{L_{1,{\rm finer}}}\right).
\end{equation}
Tables \ref{table:conv_single} and \ref{table:conv_multi} show the $L_1$ error norms and 
convergence rates for the single-level and multi-level solutions, respectively.
The convergence rates correspond to the two columns on either side of the reported value.
We note second order convergence in each variable.  Additionally, the magnitude of
the $L_1$ error norms for the multi-level simulations is comparable to the corresponding
resolution error norms for the single-level simulations.  This means that the multi-level
simulations are accurately capturing the finer-scale features, as compared to the
single-level simulations, i.e., the presence of coarse grid data and/or
coarse-fine interfaces is not harming the solution in the refined region.
\begin{table}
\begin{center}\caption{\label{table:conv_single} $L_1$ error norms and convergence rates
for the single-level simulations.\newline}
\begin{tabular}{cccccc}
\tableline
\tableline
& \multicolumn{1}{c}{$64\times 256$ Error} & \multicolumn{1}{c}{Rate, $p$} & \multicolumn{1}{c}{$128\times 512$ Error} & \multicolumn{1}{c}{Rate, $p$} & \multicolumn{1}{c}{$256\times 1024$ Error} \\
\tableline
$\rho$                & $2.23\times 10^{4}$ & 2.02 & $5.51\times 10^{3}$ & 2.30 & $1.12\times 10^{3}$ \\
$u$                   & $1.40\times 10^{4}$ & 2.02 & $3.44\times 10^{3}$ & 2.13 & $7.90\times 10^{2}$ \\
$v$                   & $1.82\times 10^{4}$ & 2.03 & $4.45\times 10^{3}$ & 2.24 & $9.40\times 10^{2}$ \\
$h$                   & $3.14\times 10^{13}$ & 1.97 & $8.03\times 10^{12}$ & 2.09 & $1.89\times 10^{12}$ \\
$X(^{24}\mathrm{Mg})$ & $5.06\times 10^{-9}$ & 2.14 & $1.15\times 10^{-9}$ & 2.01 & $2.86\times 10^{-10}$ \\
$T$                   & $1.38\times 10^{6}$ & 1.94 & $3.59\times 10^{5}$ & 2.04 & $8.72\times 10^{4}$ \\
\tableline
\end{tabular}
\end{center}
\end{table}
\begin{table}
\begin{center}\caption{\label{table:conv_multi} $L_1$ error norms and convergence rates 
for the multi-level simulations.\newline}
\begin{tabular}{cccc}
\tableline
\tableline
& \multicolumn{1}{c}{$128\times 512$ Error} & \multicolumn{1}{c}{Rate, $p$} & \multicolumn{1}{c}{$256\times 1024$ Error} \\
\tableline
$\rho$                & $5.83\times 10^{3}$ & 2.25 & $1.23\times 10^{3}$ \\
$u$                   & $4.30\times 10^{3}$ & 2.09 & $1.01\times 10^{3}$ \\
$v$                   & $4.99\times 10^{3}$ & 2.20 & $1.09\times 10^{3}$ \\
$h$                   & $8.20\times 10^{12}$ & 2.08 & $1.94\times 10^{12}$ \\
$X(^{24}\mathrm{Mg})$ & $1.15\times 10^{-9}$ & 2.01 & $2.86\times 10^{-10}$ \\
$T$                   & $3.66\times 10^{5}$ & 2.04 & $8.03\times 10^{4}$ \\
\tableline
\end{tabular}
\end{center}
\end{table}

\subsection{Adaptive Bubble Rise}\label{Sec:Bubble Rise}
To test the ability for an adaptive, three-dimensional planar simulation 
to track a localized feature, we examine a hot bubble rising in a white-dwarf 
environment.  The problem setup is exactly the same as in \S \ref{Sec:Convergence Test},
except that we now compute in three dimensions and allow the grid structure to
change with time.  We choose a domain size of $7.2\times 10^7$~cm by $7.2\times 10^7$~cm
$2.88\times 10^8$~cm and for each simulation, we generate an initial model with 
$\dr = 5.625\times 10^5$~cm (which is equal to the effective $\Delta x$ for both of the 
simulations in this test) using the method described in Appendix 
\ref{Sec:Test Problem Initial Model} with $r_\mathrm{base} = 0$.
We add a temperature perturbation of the form
given in equation (\ref{eq:temperature pert}), but in three dimensions with
the hot spot centered at location 
$(3.6\times 10^7~$cm$,3.6\times 10^7~$cm$,3.2\times 10^7~$cm). 
We will show that the adaptive 
simulation captures the same dynamics as the single-level simulation in a more 
computationally efficient manner.

We compute a single-level simulation with $128\times 128\times 512$ grid cells,
and an adaptive simulation with two levels of refinement at the same effective
resolution.  For each cell, if the $T-\overline{T} > 3\times 10^7$~K, 
we tag all cells at that height
to ensure they are computed at the finest refinement level.  We run each
simulation to $t=2.5$~s using a CFL number of 0.9.

\begin{figure}[tb]
\centering
\includegraphics[width=0.75in]{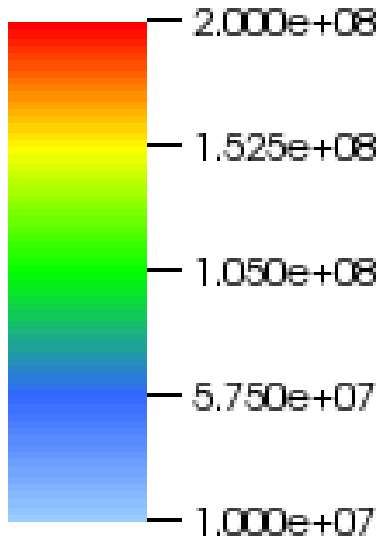}
\includegraphics[width=2in]{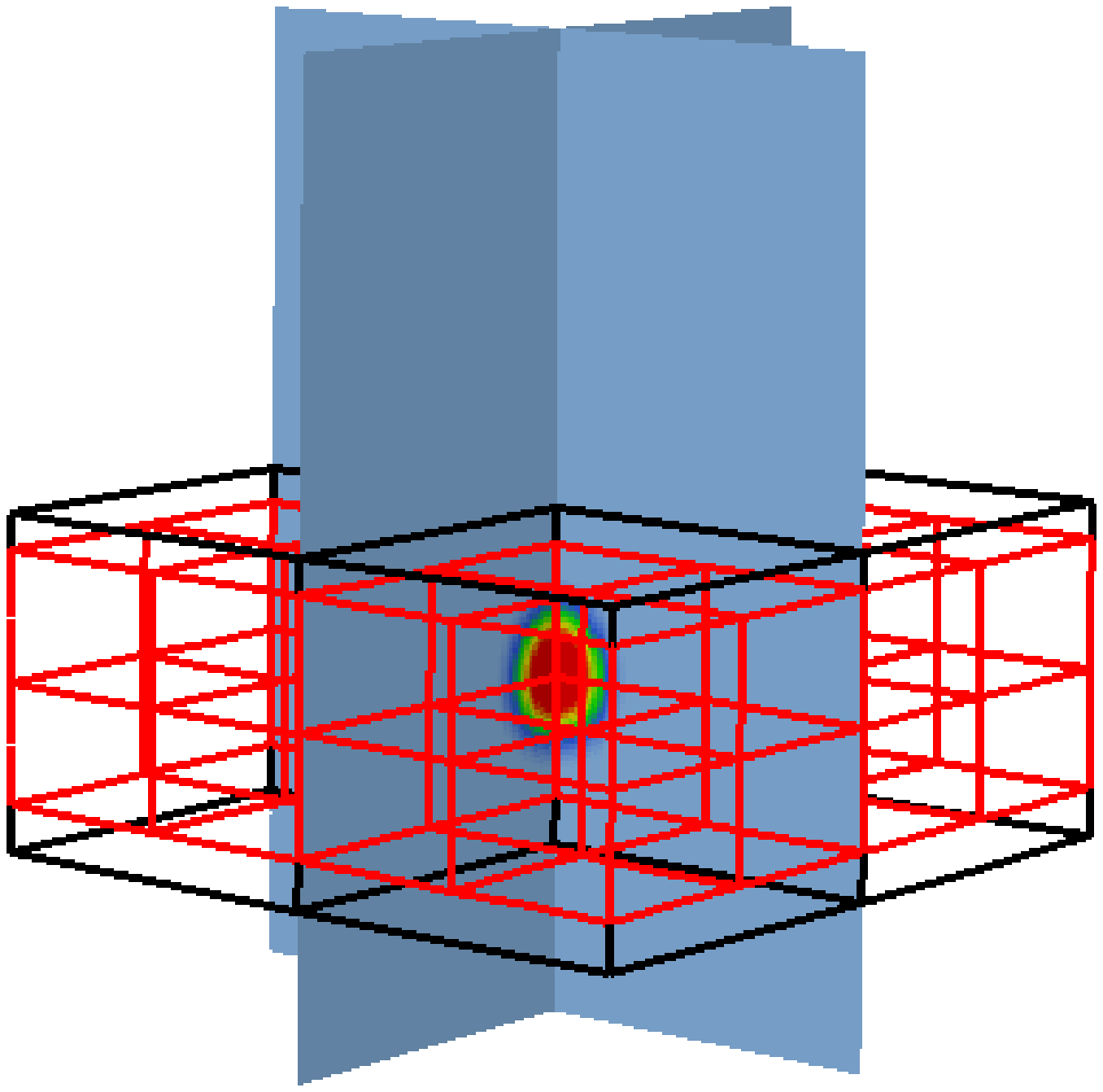}
\caption{\label{Fig:test2_3d}
Profile of $T-\overline{T}$ for a hot bubble in a white dwarf environment.  
The black and red lines represent grids of increasing refinement.
The vertical distance shown is from $z=0$ to $9.2\times 10^7$~cm.}
\end{figure}

\begin{figure}[tb]
\centering
\includegraphics[width=1in]{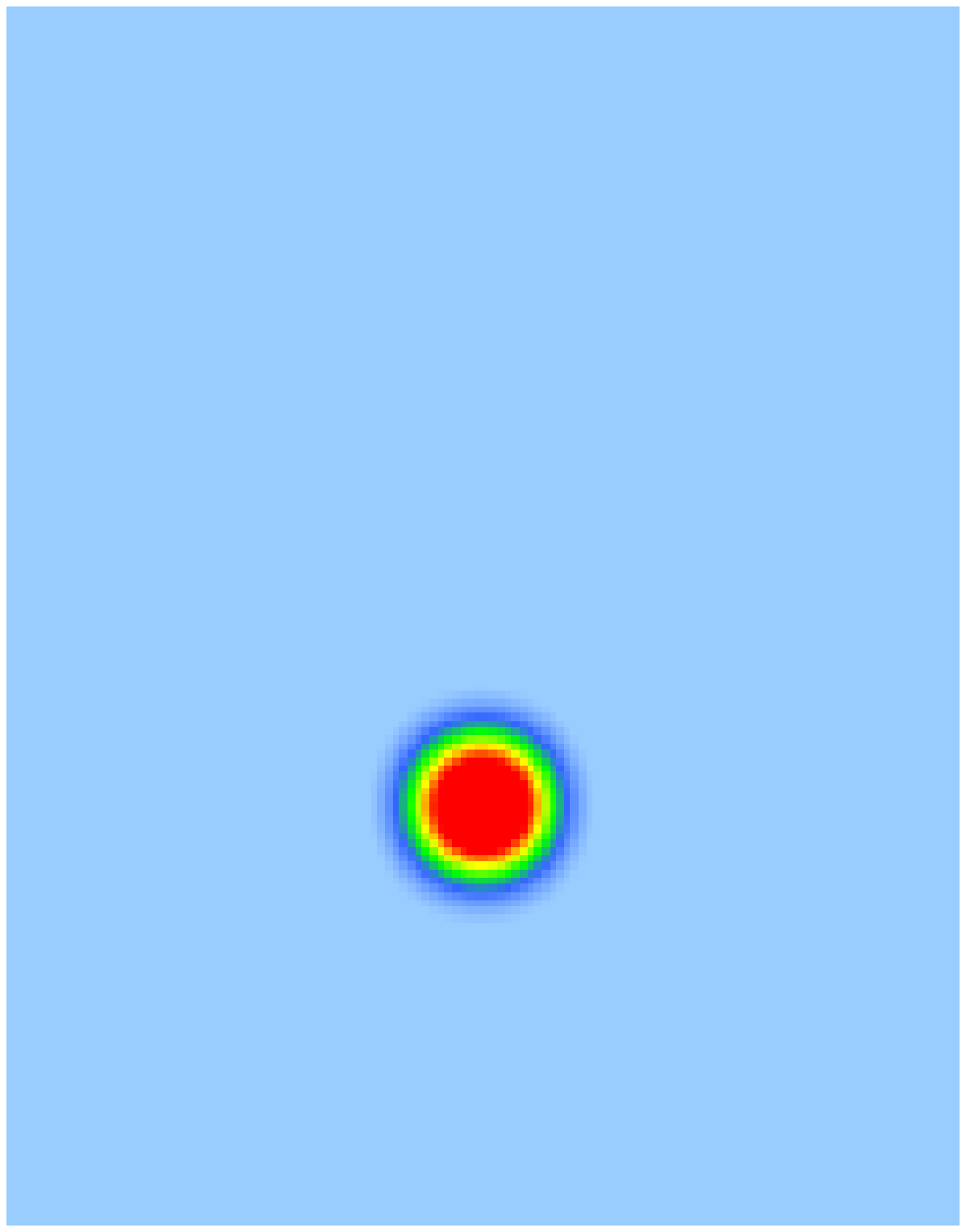}~
\includegraphics[width=1in]{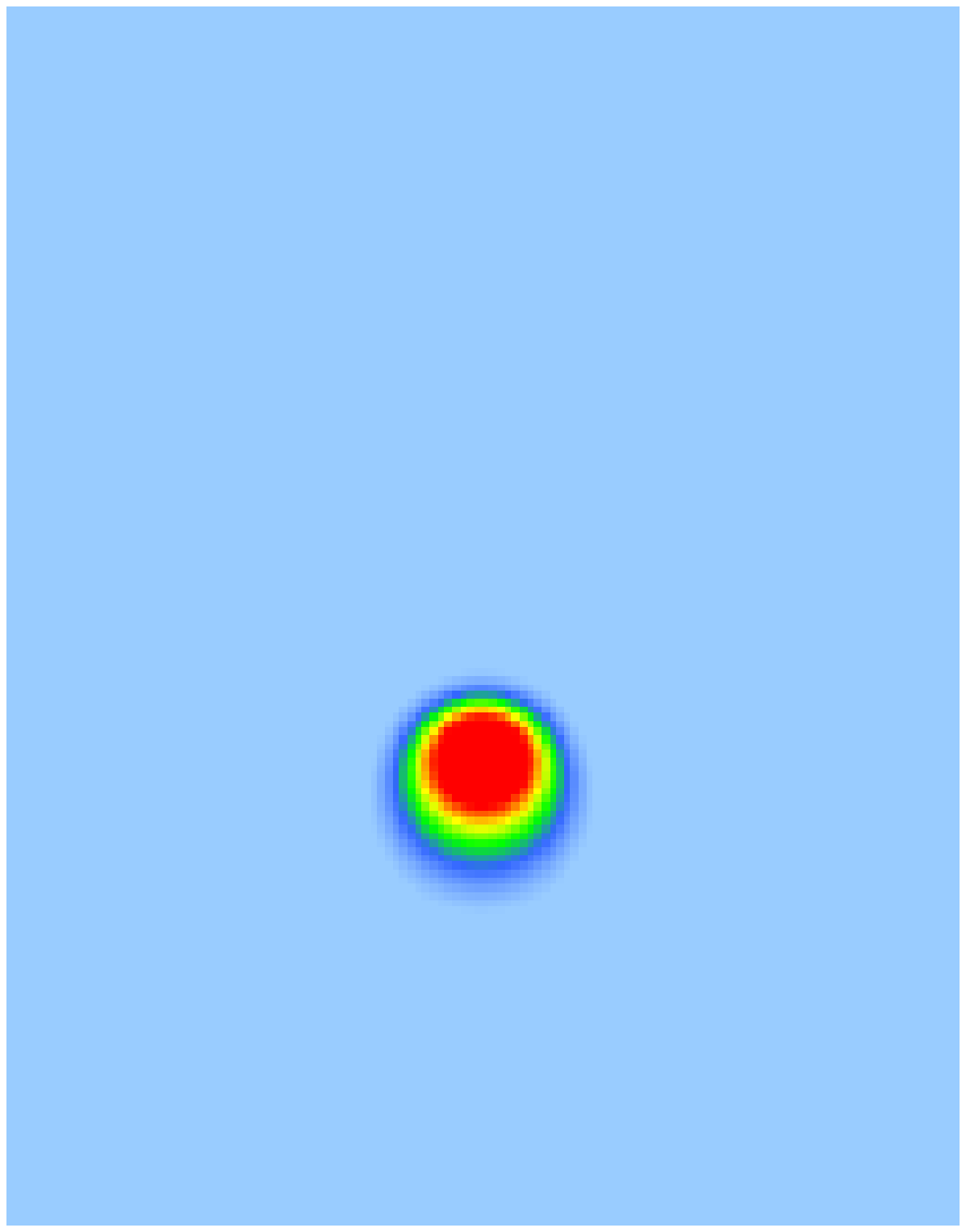}~
\includegraphics[width=1in]{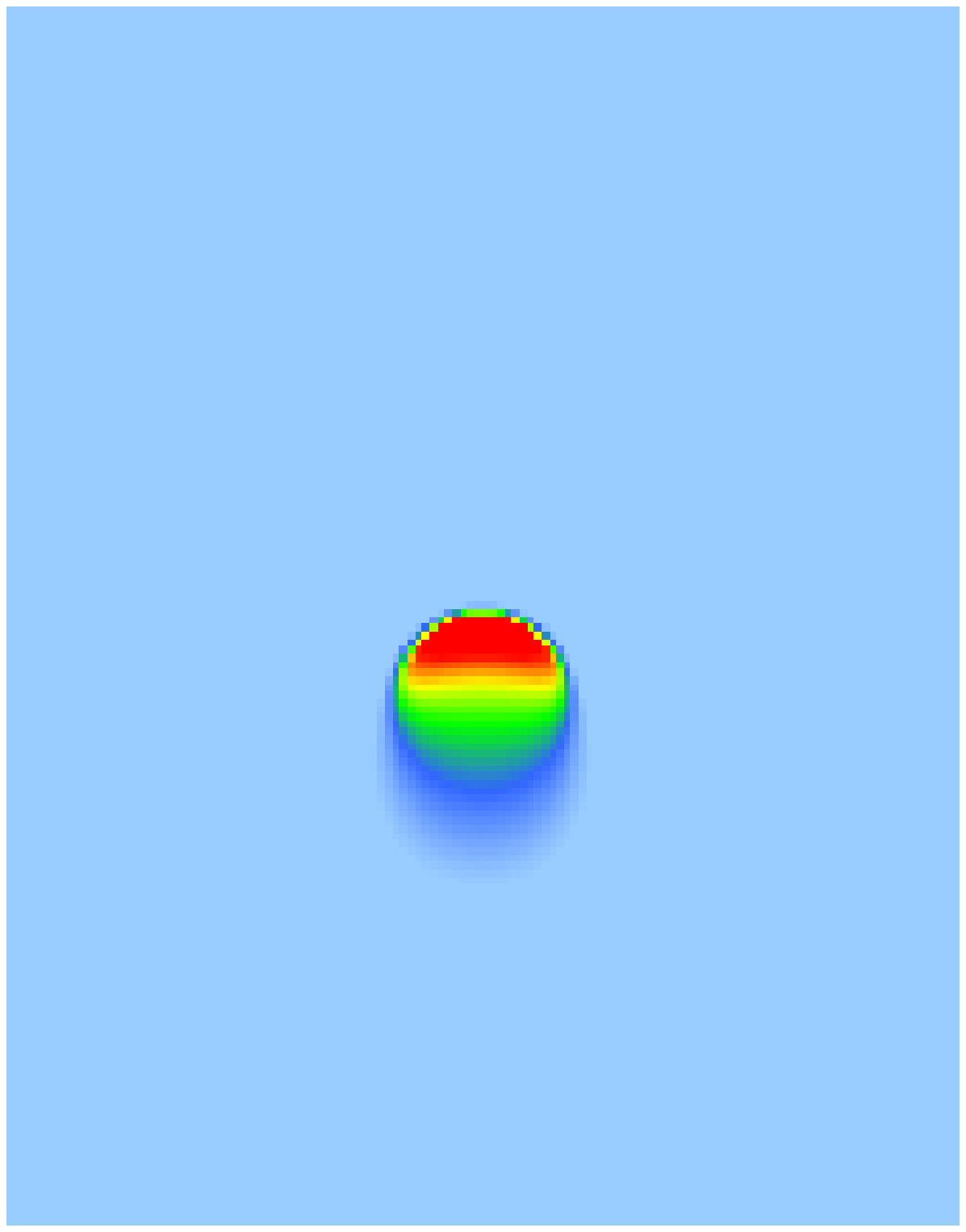}~
\includegraphics[width=1in]{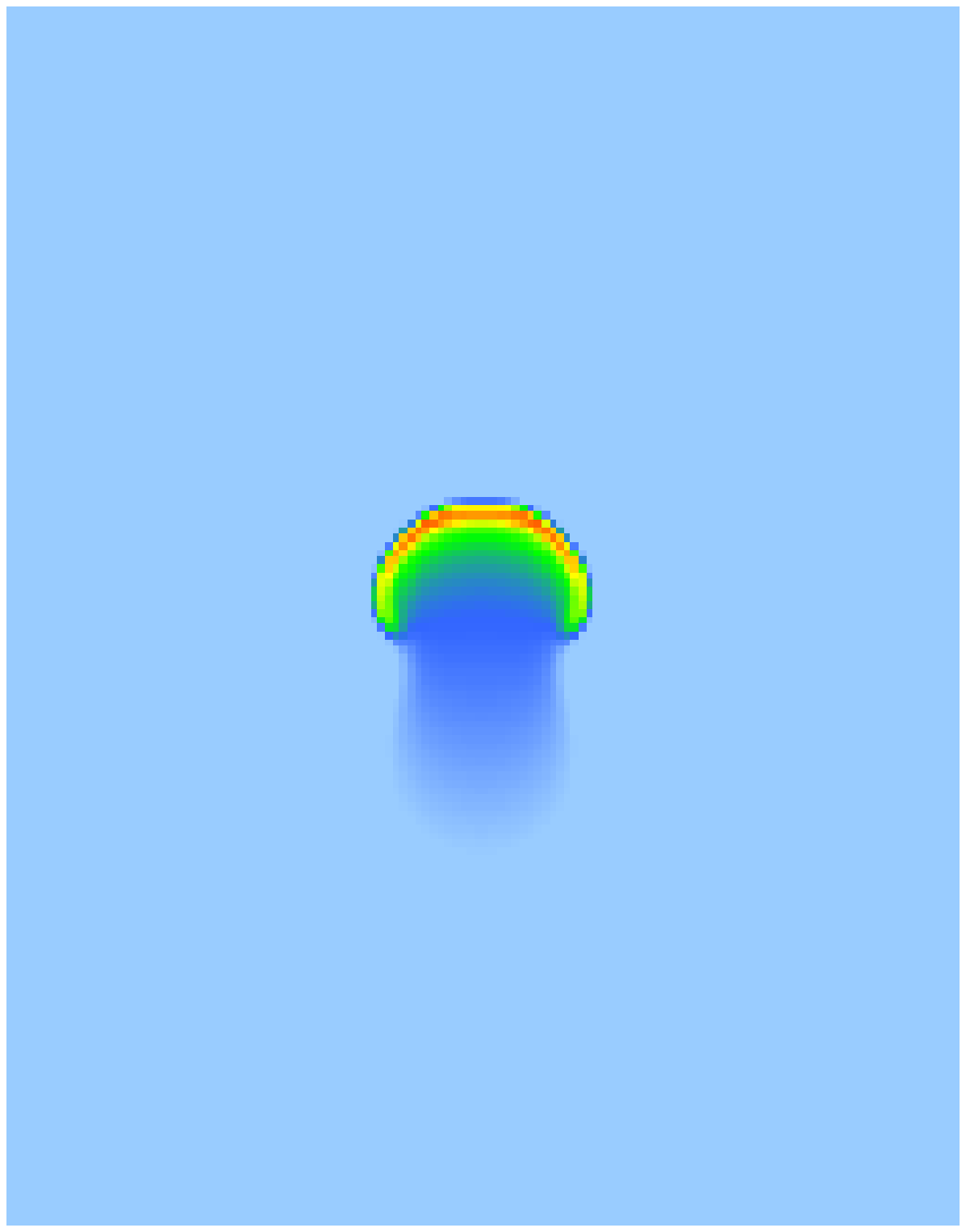}~
\includegraphics[width=1in]{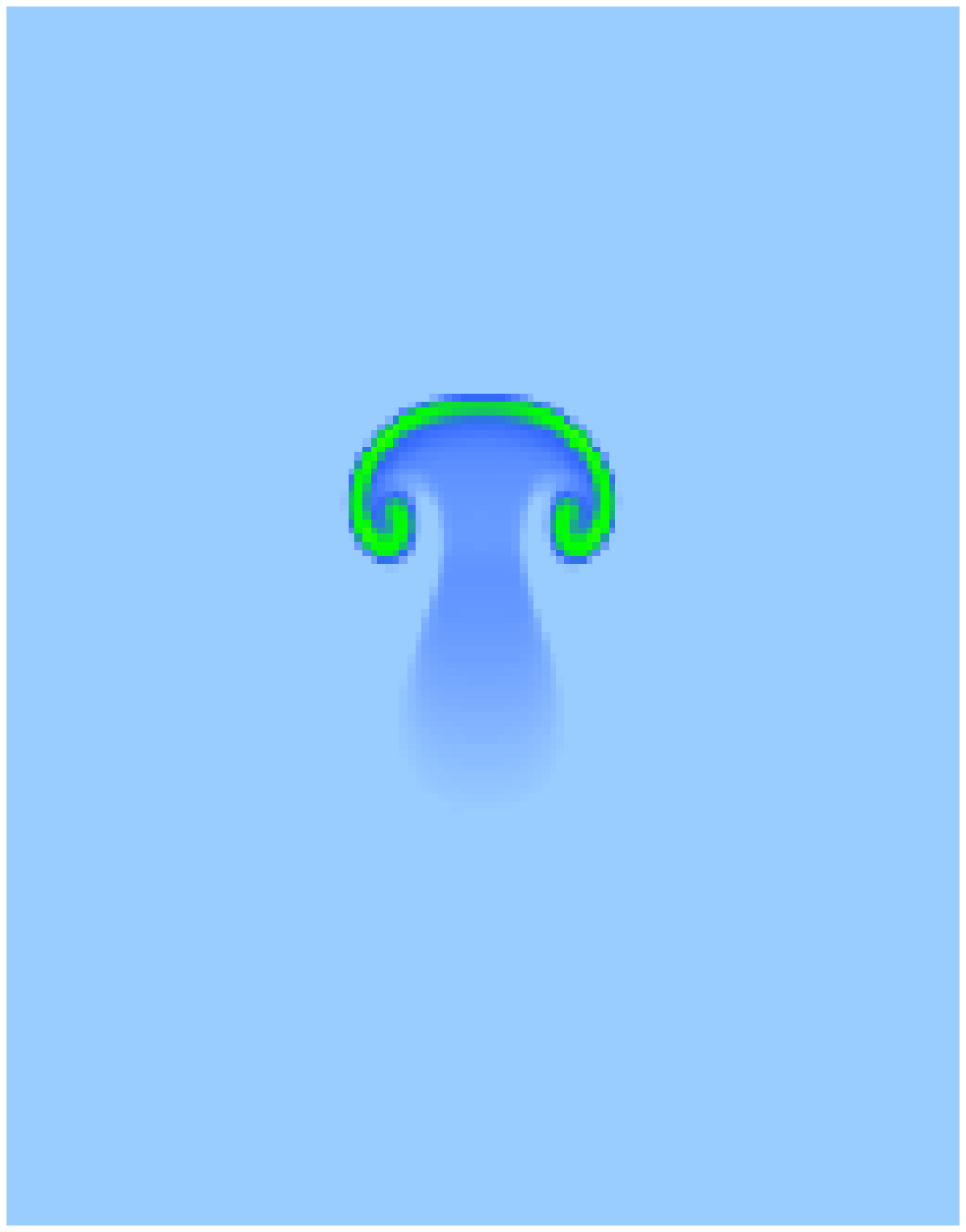}~
\includegraphics[width=1in]{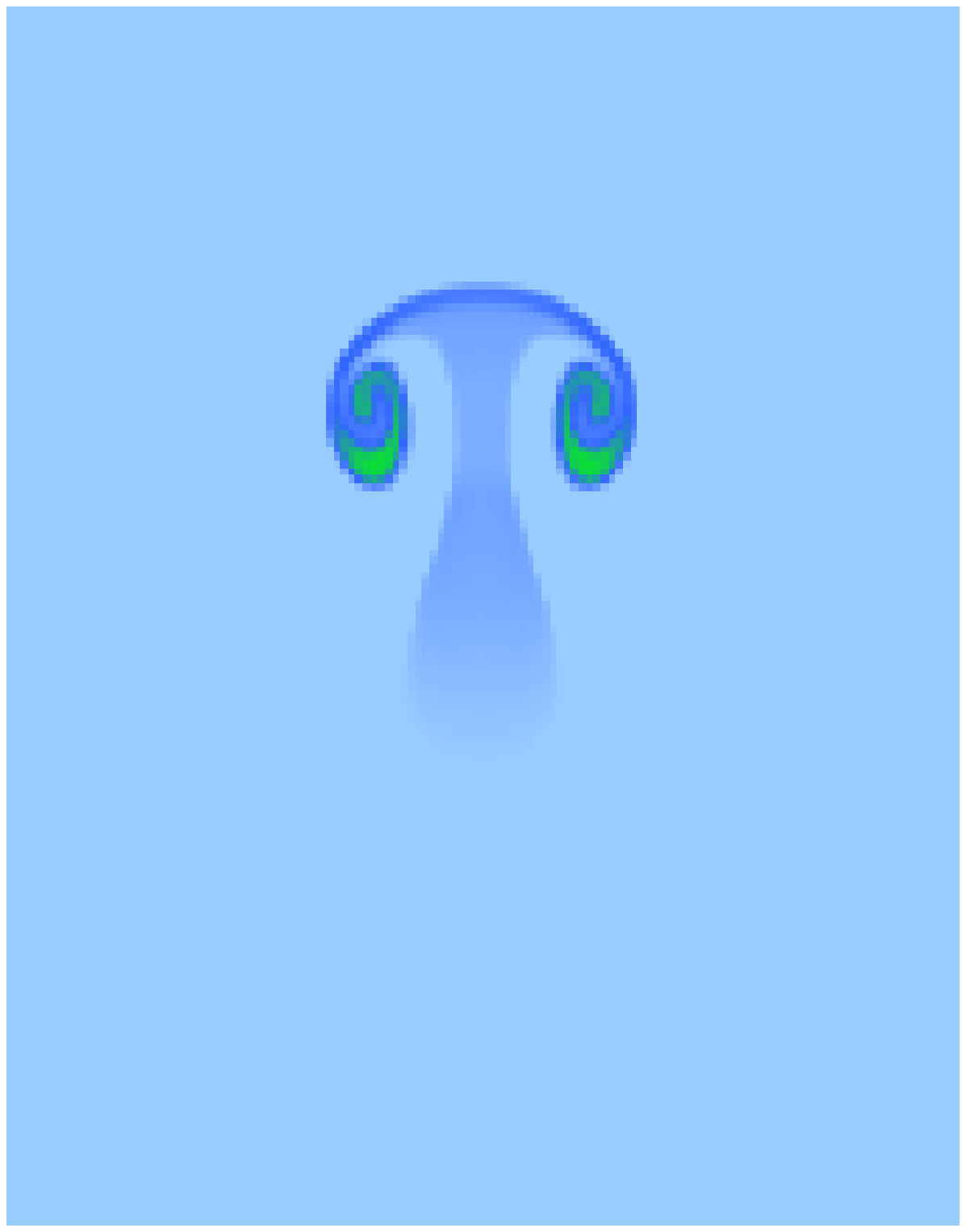}
\\
\includegraphics[width=1in]{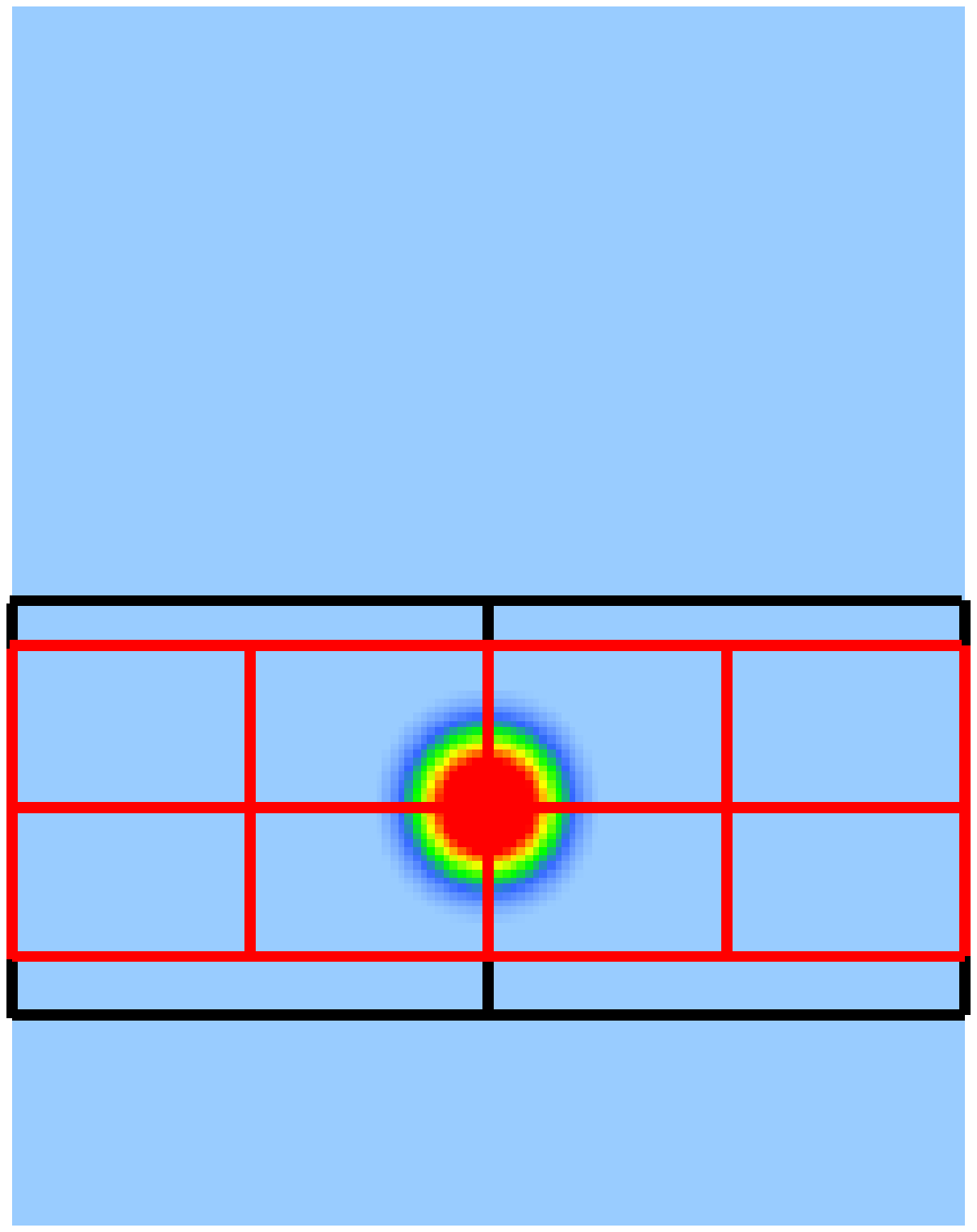}~
\includegraphics[width=1in]{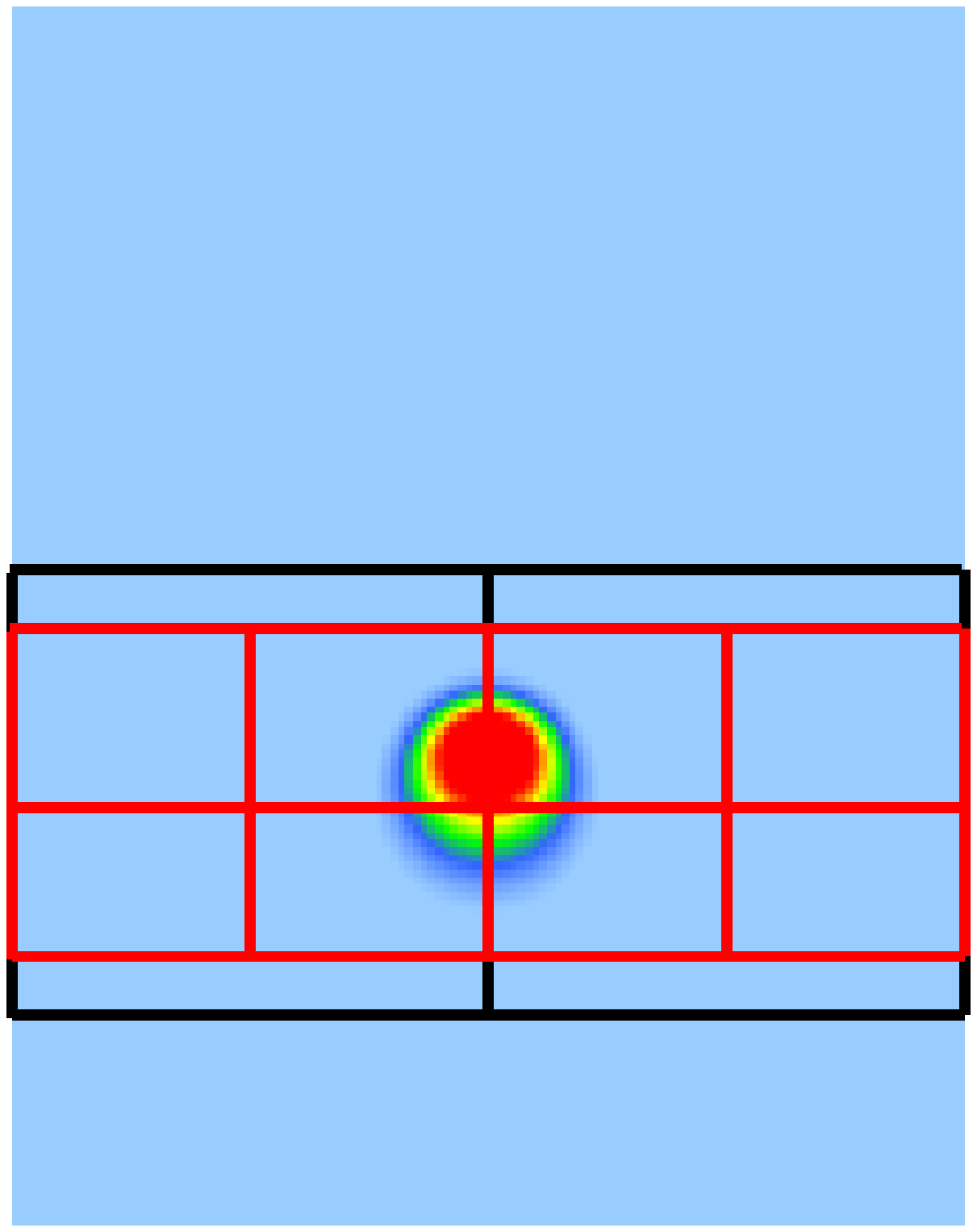}~
\includegraphics[width=1in]{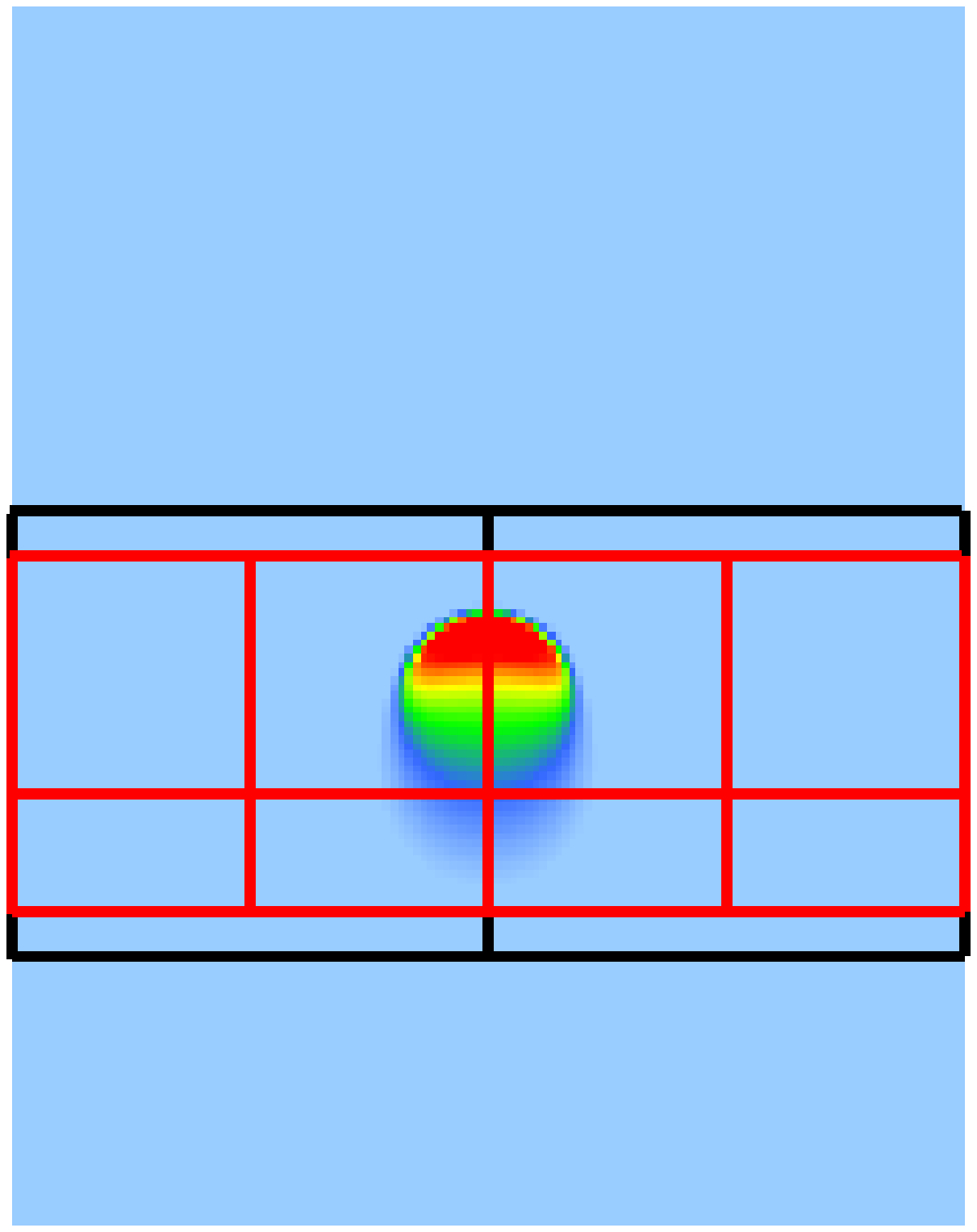}~
\includegraphics[width=1in]{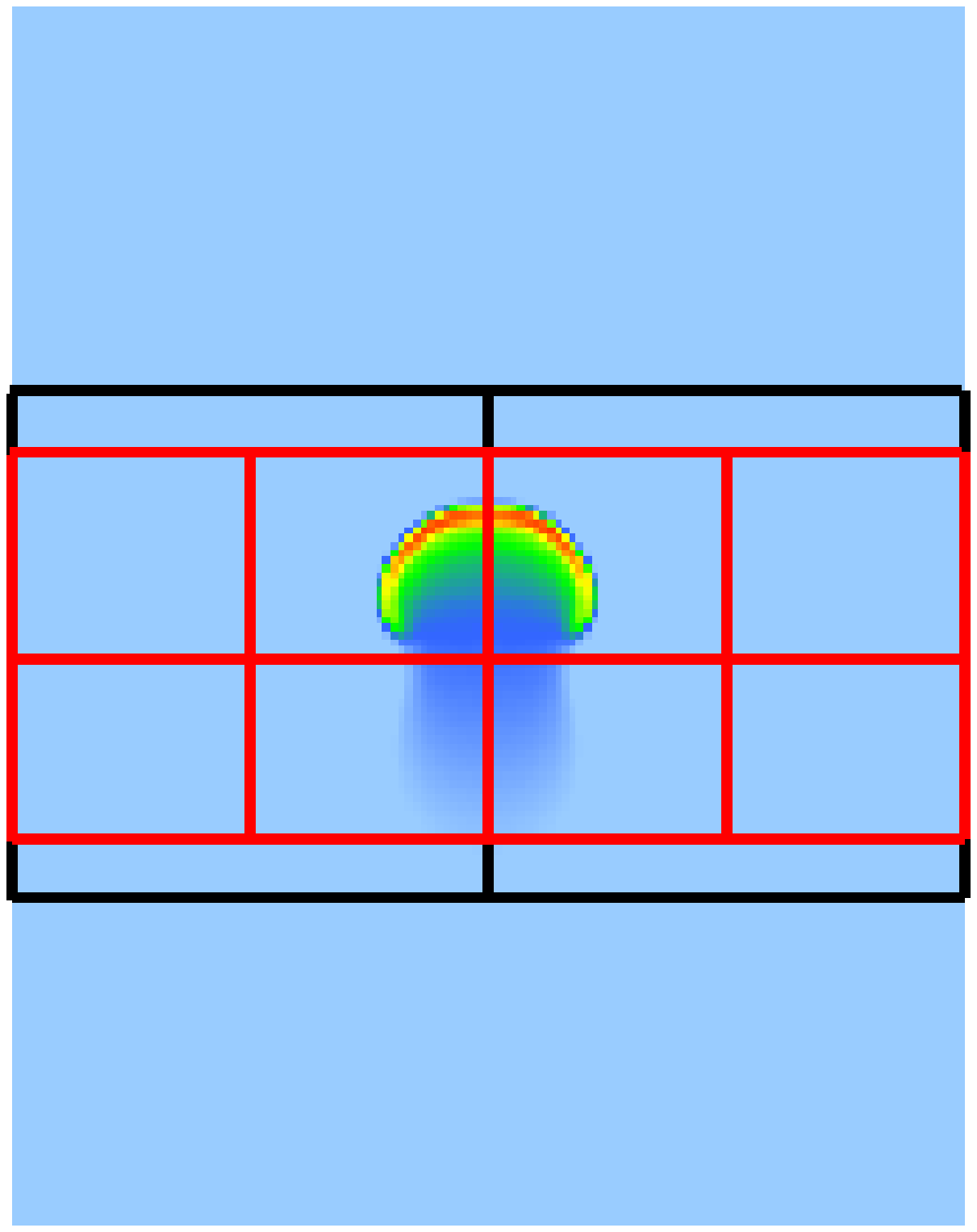}~
\includegraphics[width=1in]{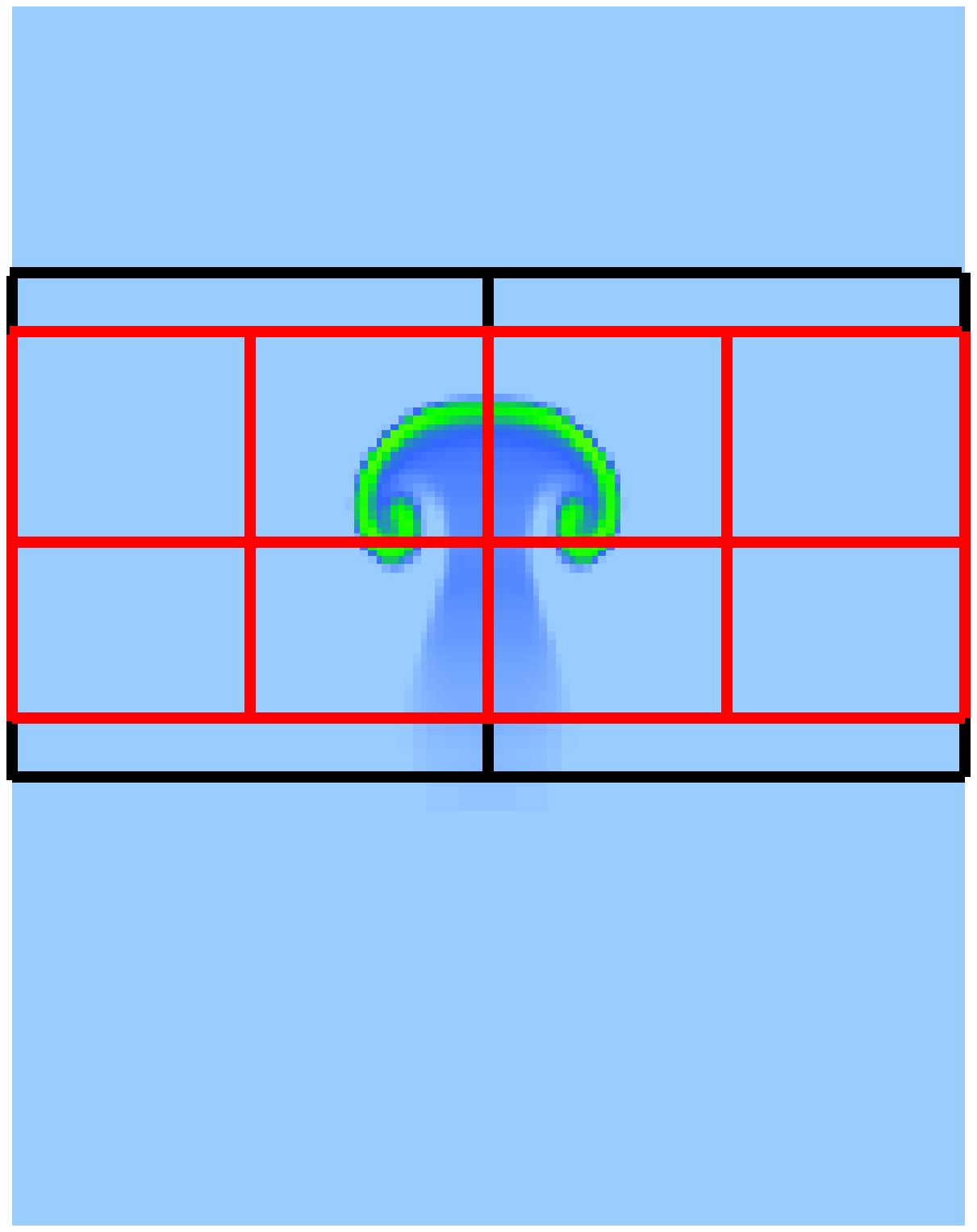}~
\includegraphics[width=1in]{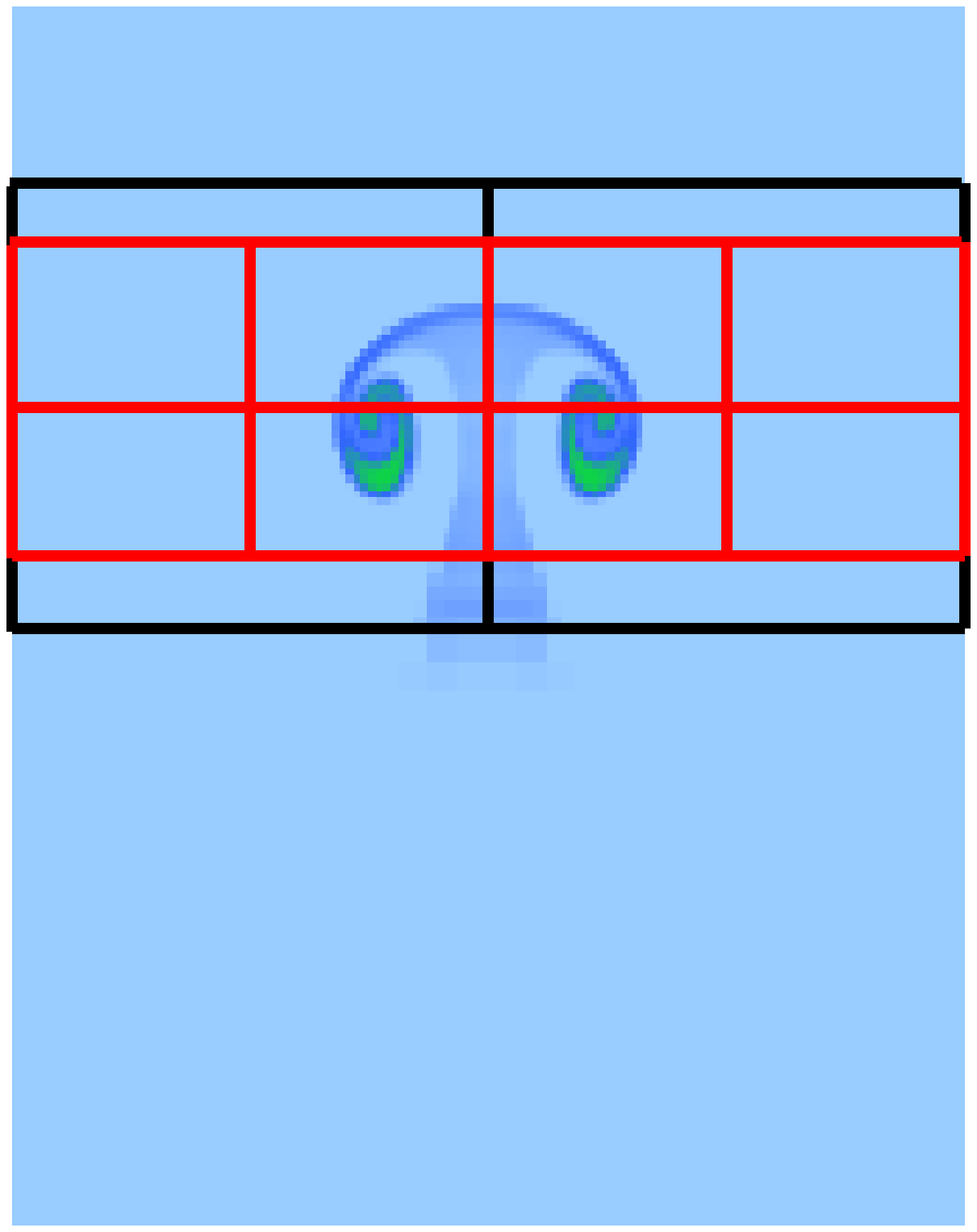}
\caption{\label{Fig:test2_2d}
Time-lapse cross section of Figure \ref{Fig:test2_3d} at $t=0,0.5,1.0,1.5,2.0$, and $2.5$~s
for a single-level simulation (above) an adaptive simulation with two levels of refinement
at the same effective resolution (below).}
\end{figure}

Figure \ref{Fig:test2_3d} shows the initial profile of $T-\overline{T}$ 
and the initial grid structure of the multi-level run.
The single-level simulation has 8,388,608 grid cells 
and takes approximately 32 seconds per time step.  The adaptive simulation initially 
has 131,072 grid cells at the coarsest level, 114,688 cells at the first level 
of refinement, and 688,128 cells at the finest level of refinement 
(the number of grid cells at the finer levels changes slightly with time
as the grid structure changes to track the bubble) 
and takes approximately 12 seconds per time step, for a factor of 2.7 speedup.
Both computations were performed 
using 32 processors on the Franklin XT4 machine at NERSC.
Figure \ref{Fig:test2_2d} shows a series of planar slices of the 
simulations at 0.5 second intervals in order to show that
the adaptive simulation captures the same dynamics as the single-level
simulation.  The vertical distance shown is from $z=0$ to 
$9.2\times 10^7$~cm.  
%%%%%%%%%%%%%%%%%%%%%%%%%%%%%%%%%
%%%%%%%%%%%%%%%%%%%%%%%%%%%%%%%%%
%%%%%%%%%%%%%%%%%%%%%%%%%%%%%%%%%
%%%%%%%%%%%%%%%%%%%%%%%%%%%%%%%%%

\subsection{Forced Convection}\label{Sec:Forced Convection}
To test the expansion of the base state in a multi-level, two-dimensional planar
simulation, we simulate a white-dwarf environment with an external heating layer.
We also show that the multi-level simulation captures the same fine-scale structure
as a single-level simulation at the same effective resolution for a short time.
We performed a similar test in Paper III, but without refinement.

We choose a domain size of $2.5\times 10^8$~cm by $4\times 10^8$~cm and for each 
simulation, we generate an initial model with $\dr$ equal to the effective $\Delta x$ 
for that simulation using the method described in Appendix 
\ref{Sec:Test Problem Initial Model} with $r_\mathrm{base} = 5\times 10^7$~cm.
The low entropy layer beneath our model atmosphere prevents the convective flow 
from reaching our lower boundary.

We apply an external heating layer of the form
\begin{equation}
H_{{\rm ext},ij} = H_0 e^{(r_j-r_{\rm layer})^2}\left[1+\sum_{m=1}^3 b_m\sin\left(\frac{k_m\pi x_i}{L_x} + \Psi_m\right)\right]
\end{equation}
with $r_{\rm layer} = 1.25\times 10^8$~cm, $H_0 = 2.5\times 10^{16}$~erg~g$^{-1}$~s$^{-1}$,
$L_x = 2.5\times 10^8$~cm, and $r_j$ and $x_i$ being the radial and horizontal
physical coordinates of cell $(i,j)$.  The perturbation parameters are
$b = (0.00625, 0.01875, 0.0125), k = (2,6,8)$, and $\Psi = (0,\pi/3,\pi/5)$.
We disable reactions for this test, since the heating layer was chosen to expand
the base state over a very short time period, rather than accurately model reactions.

We use cutoff densities of $\rho_{\rm cutoff}=\rho_{\rm anelastic}=3\times 10^6~\gcc$ 
and a sponge with $r_{\rm sp}$ equal to the radius where $\rho_0 = 10^8~\gcc$, 
$r_{\rm md}$ equal to the radius where $\rho_0 = \rho_{\rm cutoff}$, and 
$\kappa=100$~s$^{-1}$.  We specify periodic boundary conditions on the side walls, 
outflow at the top, and a solid wall at the bottom of the domain.

In this test, we use a CFL number of 0.9.  We perform two single-level simulations
using $80\times 128$ and $320\times 512$ cells, and a simulation with two levels
of refinement and an effective resolution of $320\times 512$ cells.  
For the multi-level simulation, we fix the refined grids, ensuring
that $r\in [9.375\times 10^7,1.5626\times 10^8]$~cm (which contains the external
heating layer) is at the finest level of refinement.  We run each
simulation to $t=30$~s.

\begin{figure}[tb]
\centering
\includegraphics{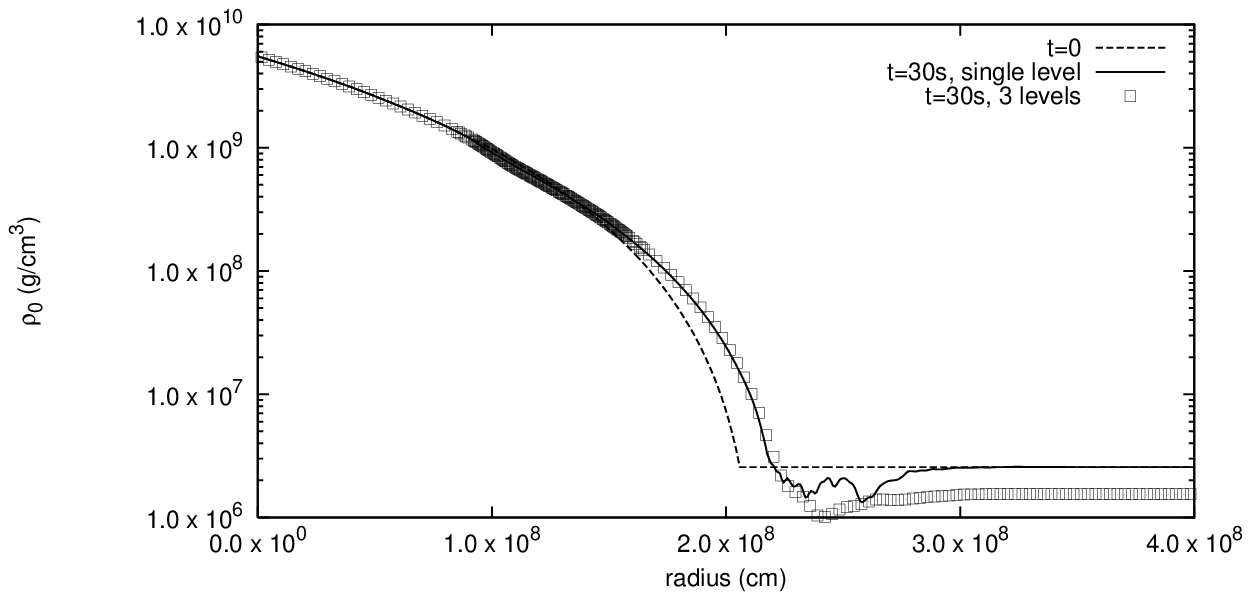}\\
\includegraphics{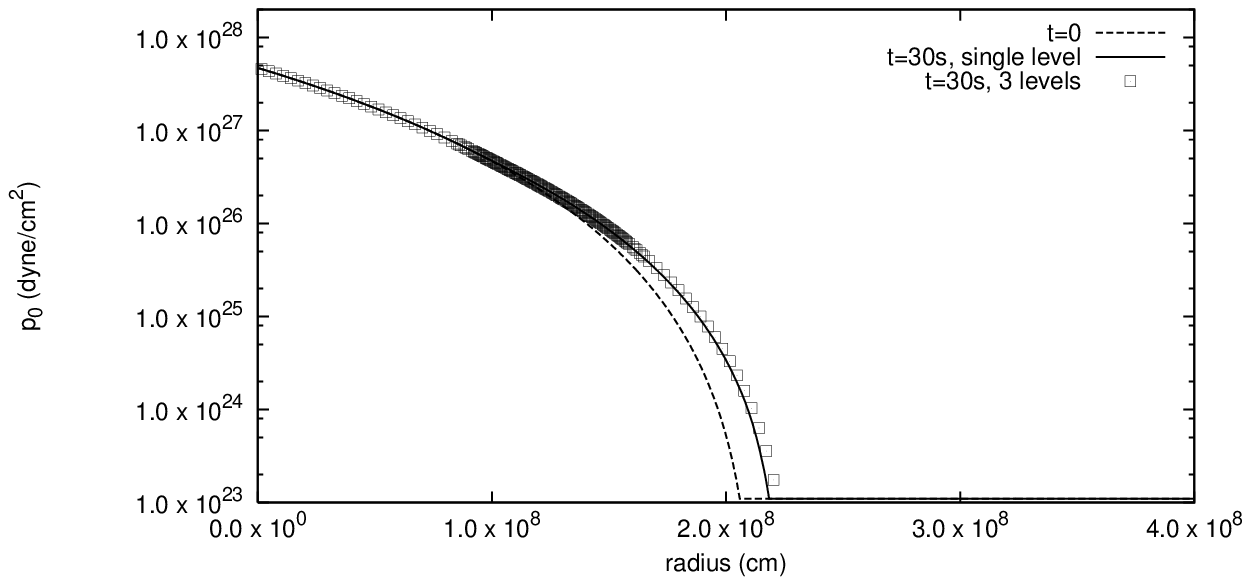}\\
\includegraphics{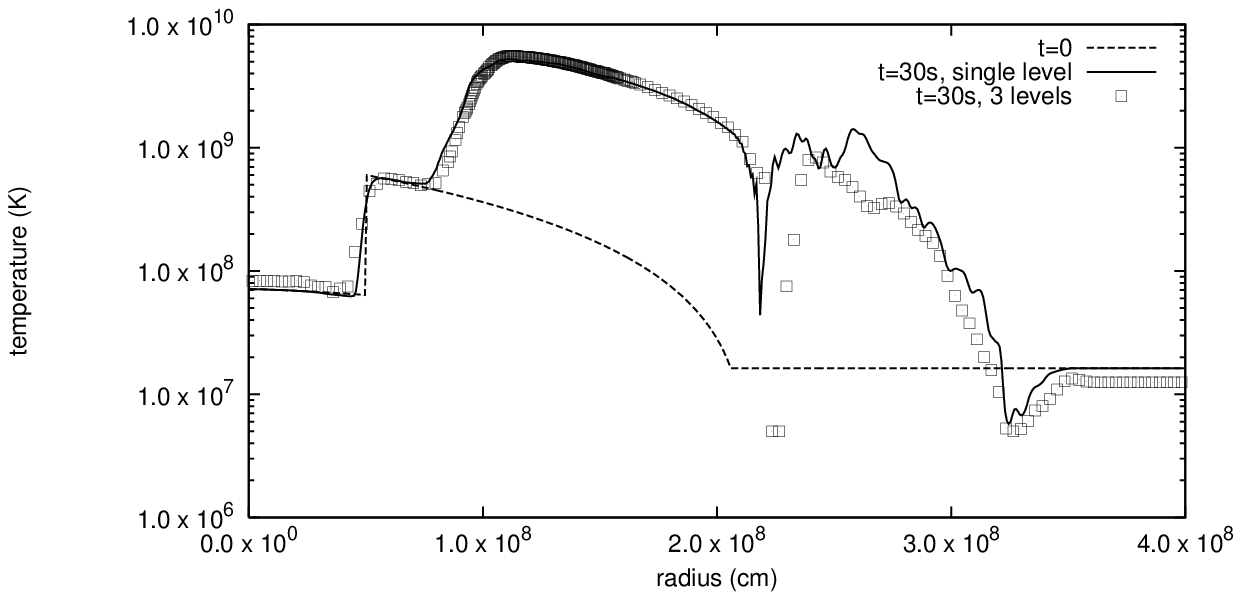}
\caption{\label{Fig:test_convect1} Comparison between $\rho_0$ (Top), 
  $p_0$ (Middle), and $\overline{T}$ (Bottom) at $t=0$, and $t=30$~s 
  in a white dwarf environment with a heating layer
  for the single-level fine grid and multi-level simulations.  The multi-level 
  simulation captures the same expansion of the base state as the single-level 
  simulation.}
\end{figure}

Figure \ref{Fig:test_convect1}
shows $\rho_0,p_0$, and $\overline{T}$ after 30 seconds of convection for the 
single-level fine grid simulation and the multi-level simulation.  There is an 
excellent agreement between these two simulations, except in the temperature
profiles at the top of the domain.  However, this corresponds to a region with
density below the cutoff densities where the temperature is extremely sensitive
to small density perturbations, and furthermore, is not fully refined in
this test, so this is an acceptable difference.  Both simulations were performed using 4 
processors; the single-level fine grid run required approximately 1.9 seconds 
per time step and the multi-level run required approximately 0.6 seconds per 
time step, for a factor of 3 speedup.
%%%%%%%%%%%%%%%%%%%%%%%%%%%%%%%%%
%%%%%%%%%%%%%%%%%%%%%%%%%%%%%%%%%
 
\begin{figure}[tb]
\raggedleft
\includegraphics[height=1.15in]{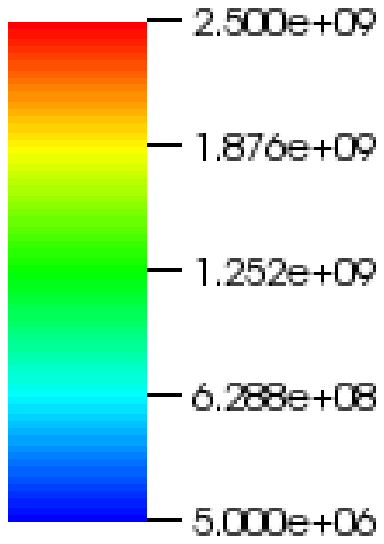}
\includegraphics[height=1.15in]{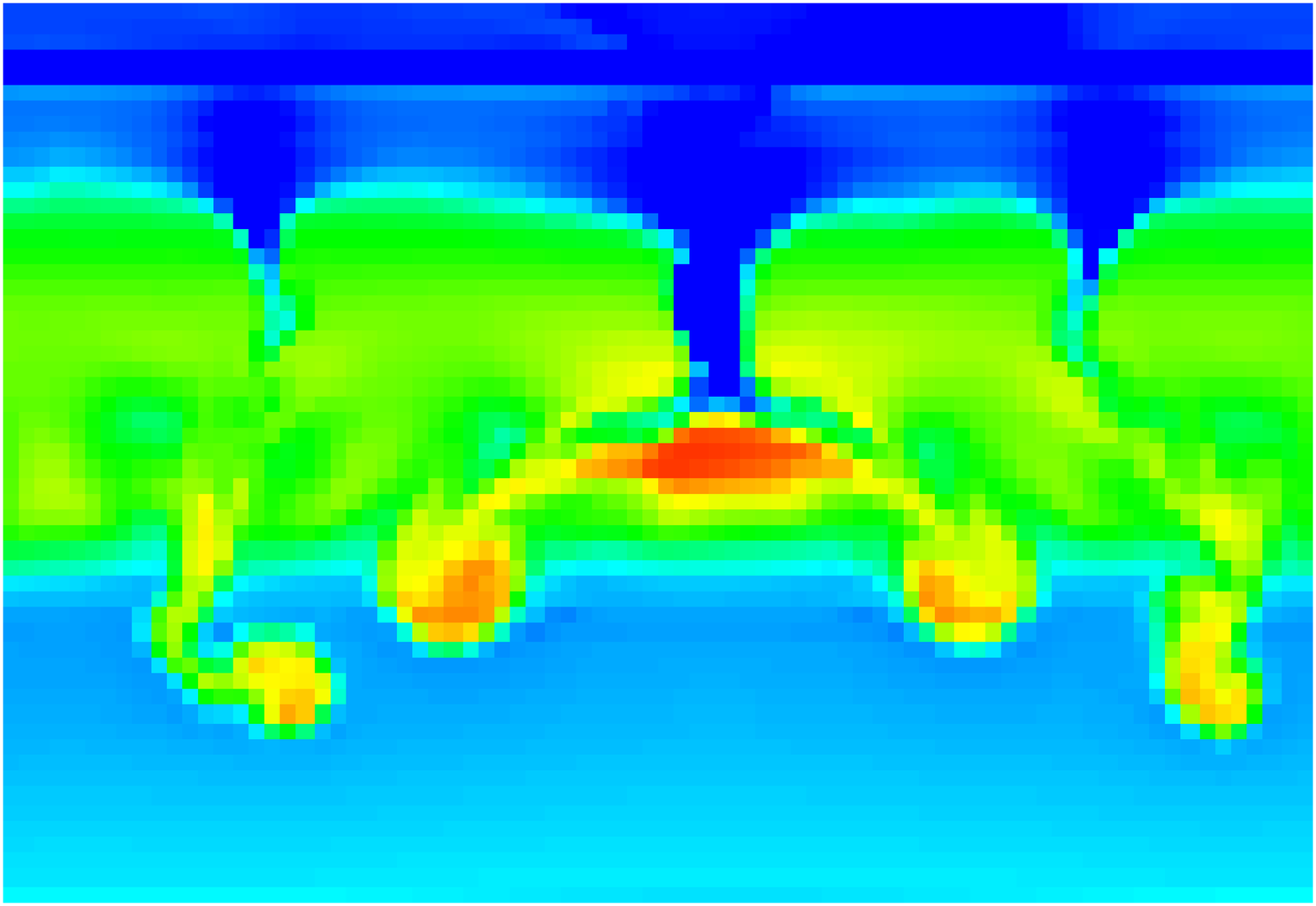}
\includegraphics[height=1.15in]{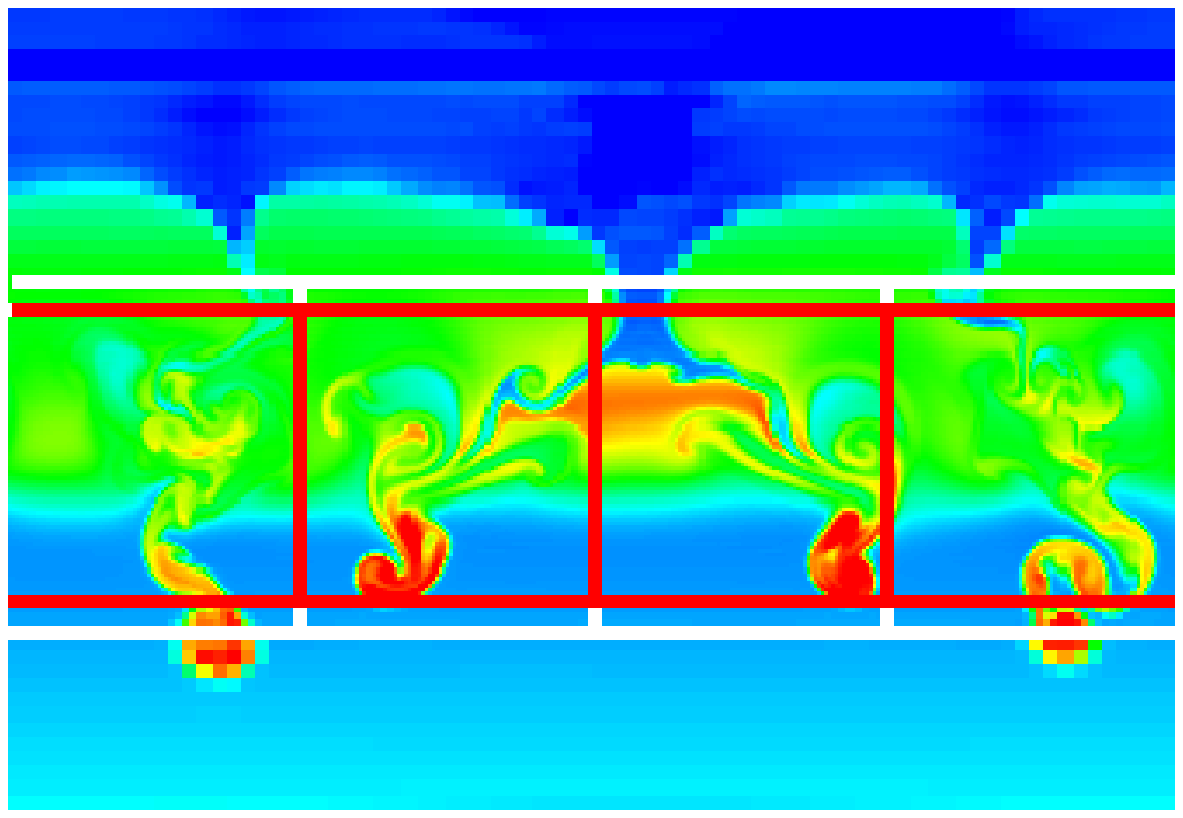}
\includegraphics[height=1.15in]{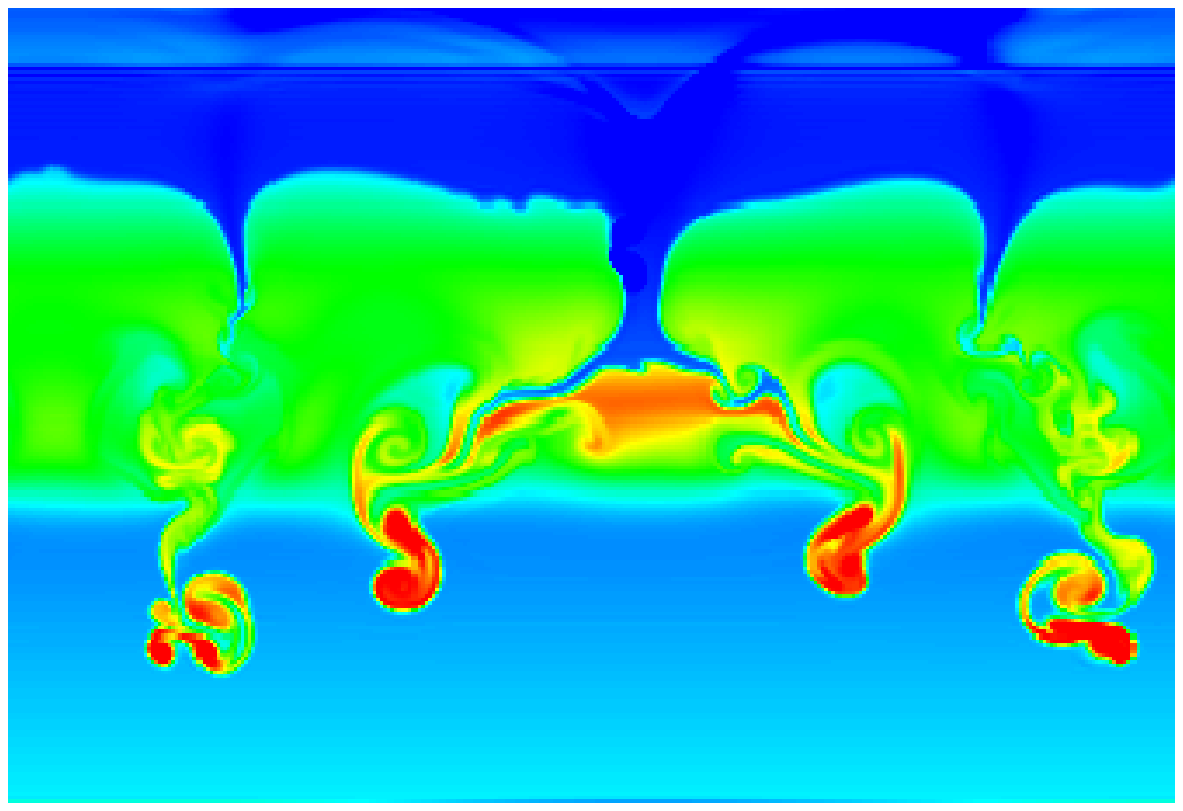}
\\
\includegraphics[height=1.15in]{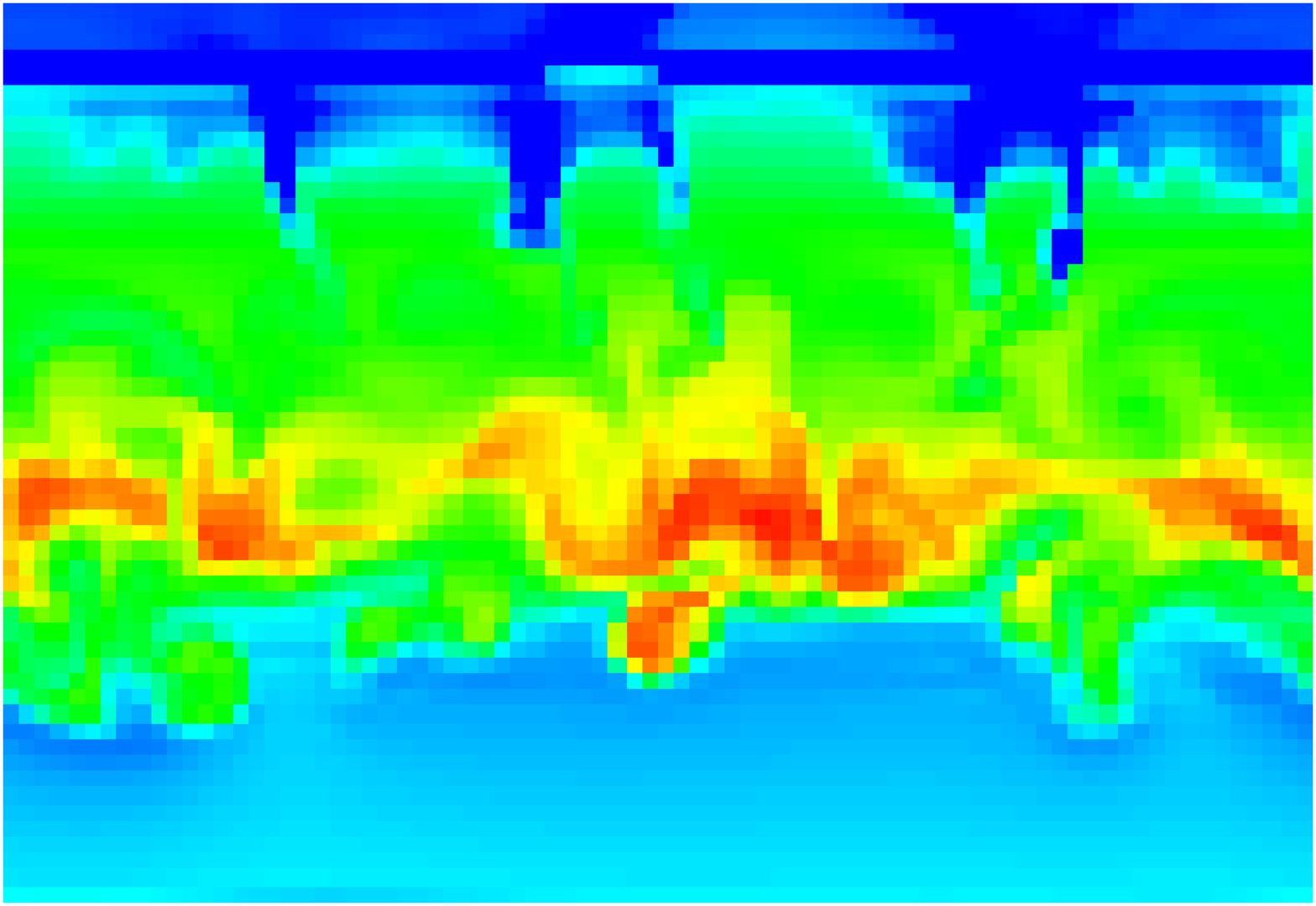}
\includegraphics[height=1.15in]{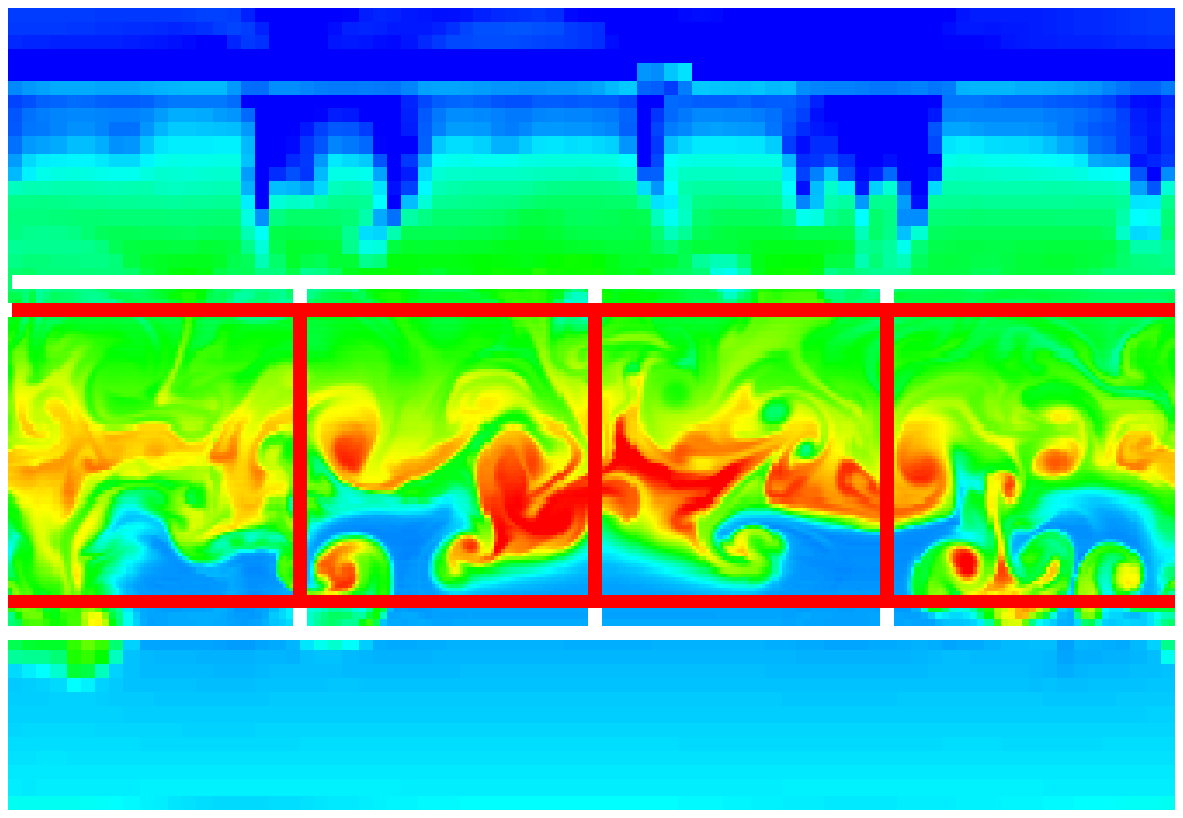}
\includegraphics[height=1.15in]{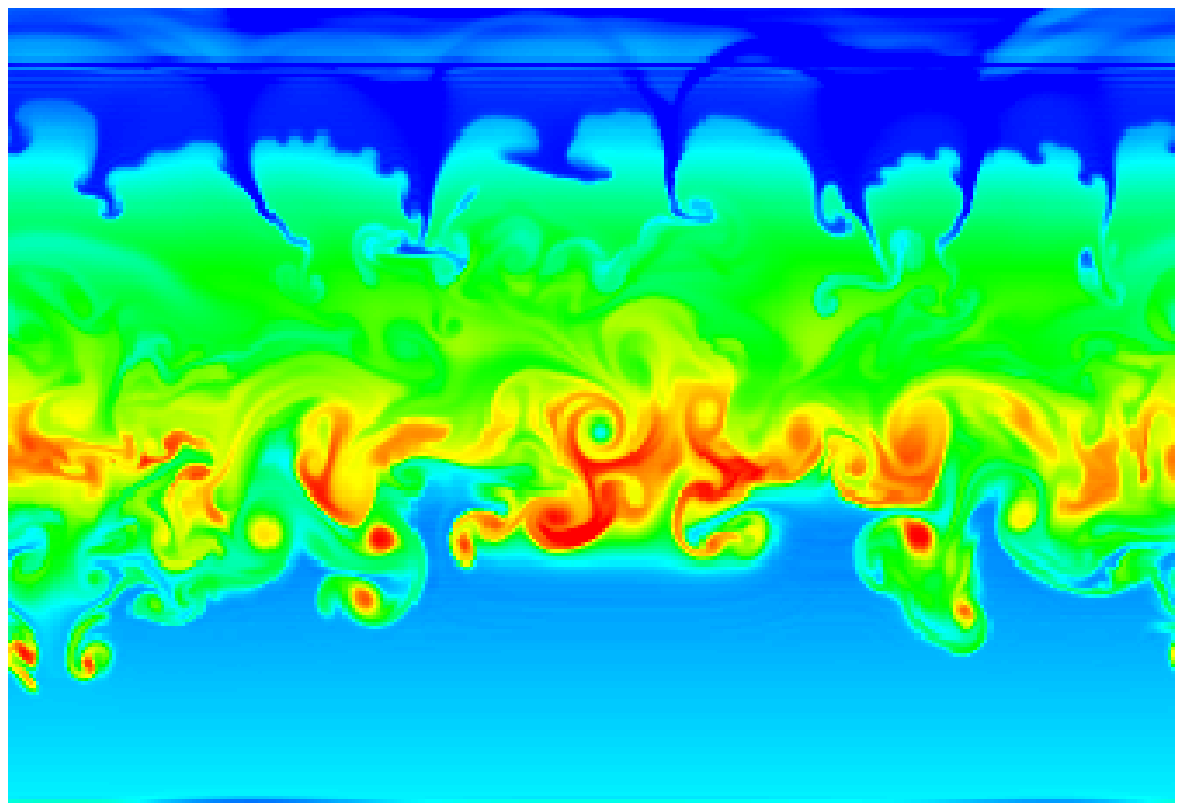}
\caption{\label{Fig:test_convect_3sec} Temperature plots at $t=3$~s (above) 
  and $t=4$~s (below) of a white dwarf environment with a heating layer.  
  The single-level coarse grid solution is on the left, the multi-level solution 
  is in the center, and the single-level fine grid solution as on the right. 
  The colored boxes indicate the grid structure at each level of refinement.  
  The vertical distance shown is from $z=5\times 10^7$~cm to $2.2\times 10^8$~cm. 
  At $t=3$~s, the multi-level simulation is able to capture the finer-scale structure 
  visible in the single-level fine grid simulation, which is not captured as well in 
  the single-level coarse grid simulation.  At $t=4$~s, a finer-scale structure is 
  still visible in the multi-level simulation, but the solution begins to drift from the 
  single-level fine grid simulation, which is expected since we are deliberately 
  refining only a part of the convective region.}
\end{figure}

Figure \ref{Fig:test_convect_3sec} shows 
the temperature profile after 3 and 4 seconds for each of the three simulations.  
The vertical distance shown is from $z=5\times 10^7$~cm to $2.2\times 10^8$~cm.  
At $t=3$~s, the multi-level simulation is able to capture the finer-scale structure 
visible in the single-level fine grid simulation, which is not captured in 
the single-level coarse grid simulation.  At $t=4$~s, a finer-scale structure is 
still visible in the multi-level simulation, but the solution begins to drift from the 
single-level fine grid simulation, which is expected since we are deliberately 
refining only a part of the convective region.
%%%%%%%%%%%%%%%%%%%%%%%%%%%%%%%%%
%%%%%%%%%%%%%%%%%%%%%%%%%%%%%%%%%

\subsection{Full-Star AMR}\label{Sec:Full-Star AMR}
We now compute three-dimensional, full-star calculations of convection in a white
dwarf.  This problem models the convection and energetics of a white dwarf that is a few
hours from reaching ignition.  We performed similar simulations using an earlier 
version of the algorithm in Paper IV.

We begin with the one-dimensional white dwarf model described in \S 2.4 in Paper IV.
We map this one-dimensional model into the center of a computational domain of 
$5\times 10^8$~cm on a side.  The first simulation is a single-level simulation with
384$^3$ cells.  The second simulation is adaptive with two levels of refinement and
an effective resolution of 384$^3$ cells.  We use the reaction network strategy 
from \citet{chamulak2008} to compute the energetics from the carbon burning.
This modified network differs from the one used in paper IV in the
ash composition (we now burn to an ash with $A = 18$ and $Z = 8.8$) and the
energy release (we use a quadratic fit to the  $q$-values tabulated on page 164 of 
\citealt{chamulak2008}).  Finally instead of destroying two carbon nuclei for each 
reaction, we use the $M_{12}$ value of 2.93 describe in that paper.
We initialize the simulation with a velocity perturbation described
exactly as in \S 2.4 in Paper IV.  We use the same cutoff densities,
sponge parameters, and boundary conditions as in \S \ref{Sec:Spherical Base State}.

\begin{figure}[tb]
\centering
\includegraphics[width=2.8in]{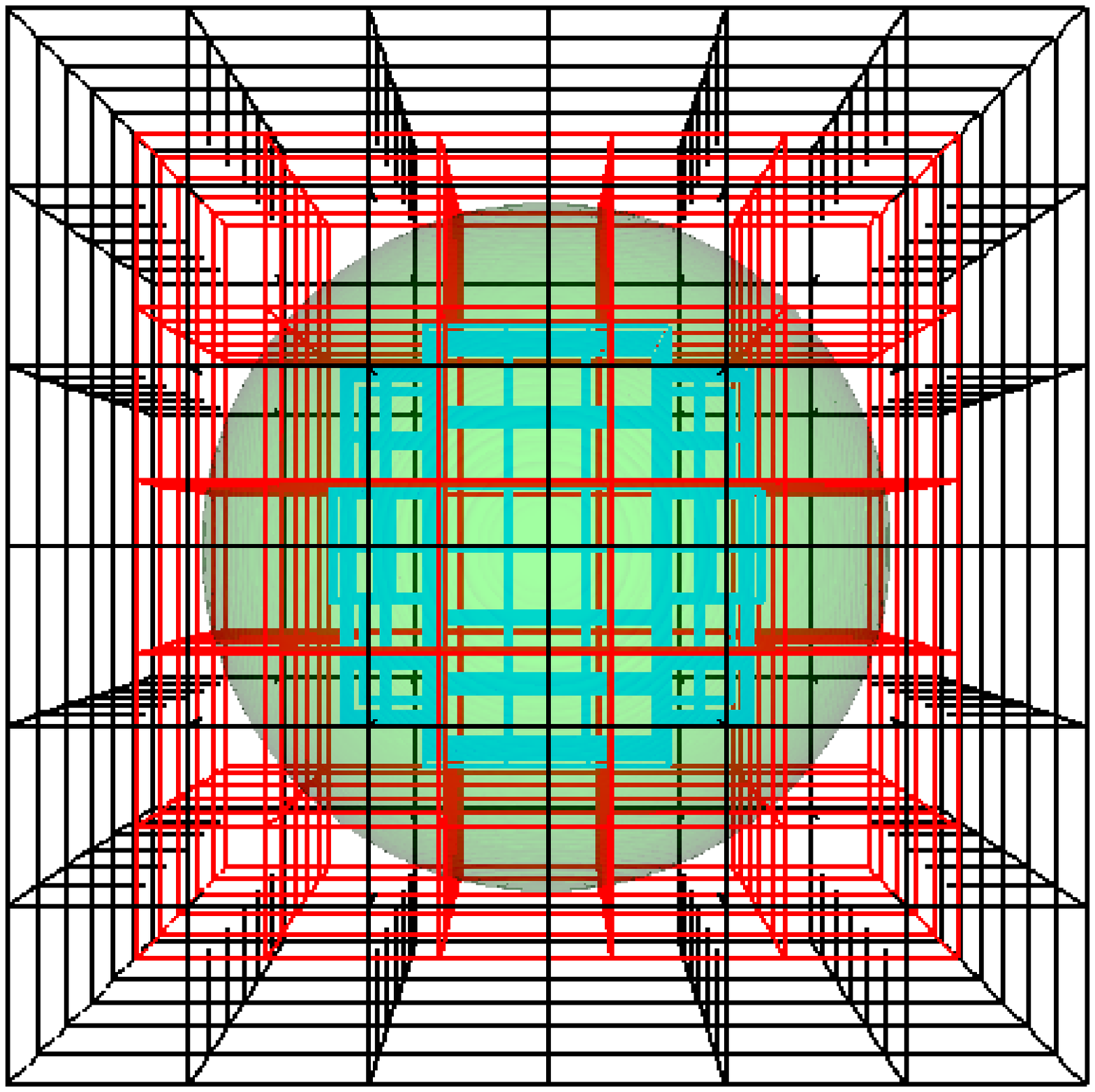}
\caption{\label{Fig:wdconvect_grid}
Initial grid structure for a full white dwarf star simulation with 2 levels of 
refinement.  The colored boxes indicated the grid structure at each level 
of refinement, each grid containing up to 32$^3$ cells.  The finest grids 
have effective resolution of 384$^3$.
The shaded sphere indicates where the density is 10$^5~\gcc$ or greater.  
In this simulation, we have chosen to include all points where $\rho > 10^5~\gcc$ 
at the first level of refinement
and all points were where $\rho > 10^8~\gcc$ at the second level of refinement.
There are 216 black grids at the coarse 
level, 125 red grids at the next level, and 212 blue grids at the finest level.}
\end{figure}

Figure \ref{Fig:wdconvect_grid} shows the initial grid structure of the adaptive
simulation.  Based on our work in Paper IV, we choose to fully refine all cells 
where $\rho>10^8~\gcc$, since at early times, the dynamics of the star are
driven by the reactions and convection in this inner region.
We wish to examine whether the adaptive simulation can give the 
same result as the single-level simulation, and the computational efficiency 
of each simulation.
%%%%%%%%%%%%%%%%%%%%%%%%%%%%%%%%%
%%%%%%%%%%%%%%%%%%%%%%%%%%%%%%%%%

We use a CFL number of 0.7 and compute to $t=900$~s.  We choose two
diagnostics used in Paper IV to compare the simulations.  Peak
temperature is a useful diagnostic since the reaction rates are
extremely sensitive to temperature, and thus peak temperature serves
as a good guide for observing the progression toward ignition.  Peak
radial velocity is another useful diagnostic as it is a simple measure of
the strength of the convection within the star.  Since the solution of
our equation is highly non-linear, and the reaction rates
scale with temperature as $\sim T^{23}$ \citep{Woosley:2004}, we
expect that errors from the coarse grid will perturb the solution at
the finest level, eventually causing significant differences in the
exact flow field.  However, as shown in Paper IV, when we run our
simulation to the point of ignition, we require upwards of hundreds of
thousands of time steps.  Therefore, in the comparison diagnostics in
this test, it is sufficient to compare the overall qualitative
solution.  An exact quantitative comparison is impossible over long
times.

\begin{figure}[tb]
\centering
\includegraphics{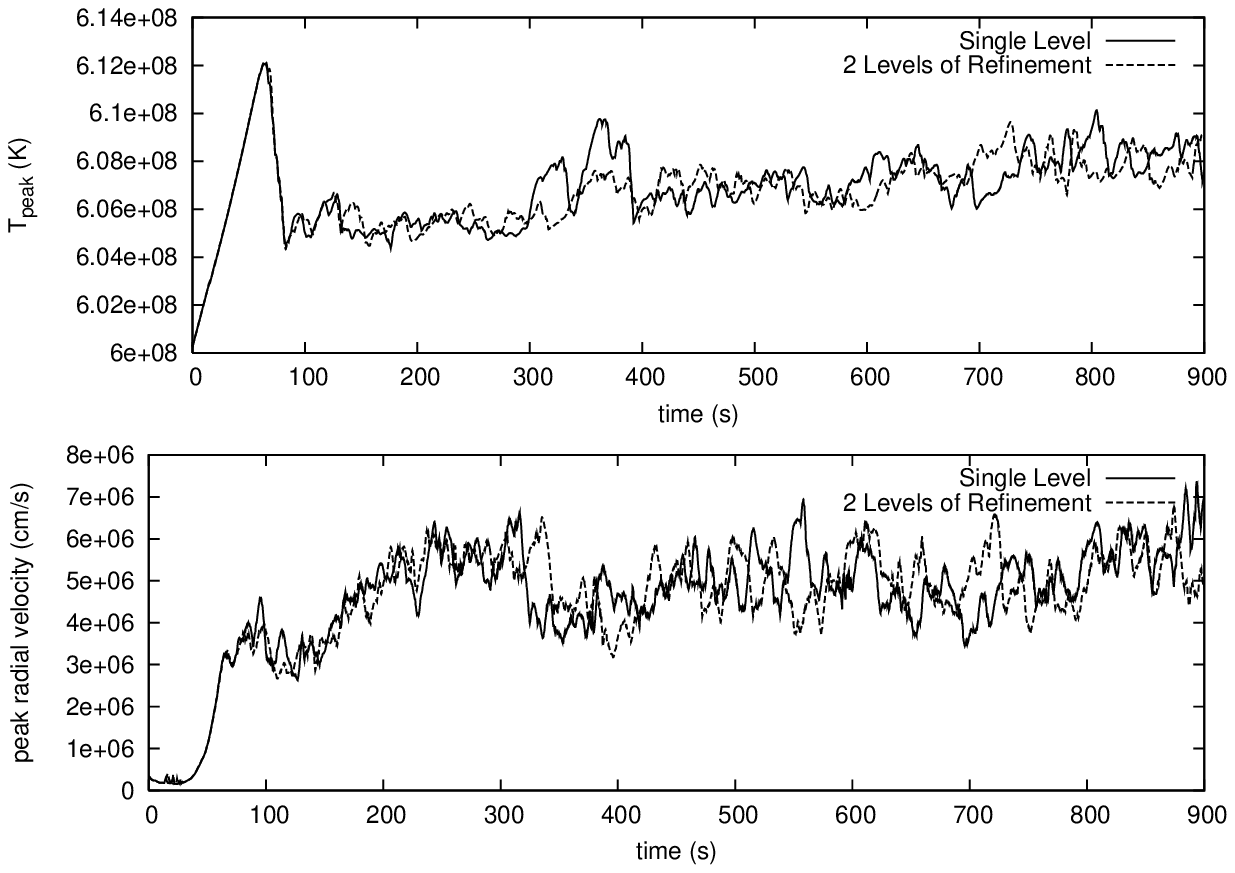}
\caption{\label{Fig:wdconvect1}
Peak temperature (Top) and peak radial velocity (Bottom) as a function of time 
for a single-level and adaptive simulation, each with an effective 384$^3$ resolution.}
\end{figure}

Figure \ref{Fig:wdconvect1} shows the evolution of the peak temperature 
and peak radial velocity over the first 900~s for both simulations.  
The adaptive simulation gives the same qualitative result as the single-level 
simulation.  As mentioned before, the curves do not match up more closely because
the equations are highly non-linear, and slight differences in the
solution caused by solver tolerance and discretization error 
change details of the results, but not the overall qualitative
results.  The single-level simulation has 56,623,760 grid cells 
and takes approximately 36 seconds per time step.  The adaptive simulation initially 
has 884,736 grid cells at the coarsest level, 3,511,808 cells at the first level 
of refinement, and 4,282,048 cells at the finest level of refinement 
(the number of grid cells at the finer levels changes slightly with time)
and takes approximately 18 seconds per time step, for a factor of 2 speedup.
Also note that the overall memory requirements are significantly less for the
adaptive simulation, as can be seen by the reduced total cell count.
Each simulation was run using the Franklin XT4 machine at NERSC with 216 
processors.
%%%%%%%%%%%%%%%%%%%%%%%%%%%%%%%%%
%%%%%%%%%%%%%%%%%%%%%%%%%%%%%%%%%

We note that in the future, when we perform longer calculations up
to the point of ignition, we may have to refine a greater portion of the
star in order to properly capture the overall dynamics as the
convective region expands.  In a forthcoming paper, we plan on performing higher
resolution studies, where AMR will save us both time and computational resources.

%==========================================================================
% Conclusions
%==========================================================================
\section{Conclusions and Future Work}\label{Sec:Conclusions}
We have developed a low Mach number hydrodynamics algorithm suitable for
full star flows and local atmospheric regions with a time-evolving base
state within an AMR framework.  In forthcoming papers, we will use {\tt MAESTRO}
to further our scientific investigation of the convective phase of SNe~Ia.  
Our previous simulations in Paper IV were at modest resolution 
and assumed a constant base state.  We are now performing simulations at higher effective
resolutions with the use of AMR along with a time-varying base state.
As part of this study, we are examining the tagging
conditions necessary to model a full-star up to the point of ignition.
We are also studying Type I X-ray bursts \citep{STRO_BILD06,Lin:2006}, 
which are believed to 
be caused by the thermonuclear explosive burning of hydrogen and/or helium gas 
accreted into a thin shell on the surface of neutron stars.  We pose this 
problem in planar geometry, model a small patch of the star, and refine near the base of 
the accreted layer.

%==========================================================================
% Acknowledgments
%==========================================================================
\acknowledgments

We thank Frank Timmes for making his equation of state routines
publicly available, and for useful discussions regarding
thermodynamics.  The work at LBNL was supported by the SciDAC Program
of the DOE Office of Mathematics, Information, and Computational
Sciences under the U.S. Department of Energy under contract
No.\ DE-AC02-05CH11231.  The
work at Stony Brook was supported by a DOE/Office of Nuclear Physics
Outstanding Junior Investigator award, grant No.\ DE-FG02-06ER41448,
to Stony Brook. 
This research used resources of the National Energy Research Scientific 
Computing Center (NERSC), which is supported by the Office of Science of the
U.S. Department of Energy under Contract No. DE-AC02-05CH11231. 

\appendix

%==========================================================================
% Appendix A: Time Advancement Algorithm
%==========================================================================
\section{Time Advancement Algorithm}\label{Sec:Time Advancement Algorithm}

\subsection{Single-Level Algorithm Changes From Papers III and IV}\label{Sec:Changes}
The single-level algorithm has gone through numerous changes since
Papers III and IV.  The current single-level algorithm is
presented in \S \ref{Sec:Gravity}--\ref{Sec:Cutoffs};
we summarize here the changes since Papers III and IV:
\begin{itemize}
\item We have extended the base state evolution to spherical problems
  by defining {\bf Advect Base Density Spherical} (\S \ref{Sec:rho0 advection}), 
  {\bf Advect Base Enthalpy Spherical} (\S \ref{Sec:rhoh0 advection}), and 
  {\bf Compute} {\boldmath $w_0$} {\bf Spherical} (\S \ref{Sec:Computing w0}).
  We have also defined spherical discretizations for $\psi$ and $\etarho$ 
  ({\bf Steps 4C, 4F, 8C} and {\bf 8F} in \S \ref{Sec:Main Algorithm Description}).
\item For spherical problems, we have improved the accuracy of the mapping of 
  data between the one-dimensional radial array and the three-dimensional Cartesian 
  grid (\S \ref{Sec:Mapping}).
\item We have upgraded our unsplit piecewise-linear Godunov method for 
  time-centered edge state prediction for the base state and Cartesian 
  grid data to use the unsplit piecewise-parabolic method (PPM) \citep{ppm} 
  with full corner coupling in three dimensions \citep{ppmunsplit,saltzman}.  We shall
  refer to this procedure as ``computing the time-centered edge states''
  (\S \ref{Sec:rho0 advection}, \S \ref{Sec:rhoh0 advection}, and 
  {\bf Steps 3, 4, 7, 8}, and {\bf 11} in \S \ref{Sec:Main Algorithm Description}).
\item As introduced in Paper~IV, we update $T$
  after the call to {\bf React State} (\S \ref{Sec:Reactions}).
\item Rather than evolve $p_0$ using an evolution equation, we simply
  update $p_0$ using the hydrostatic equilibrium equation (\S \ref{Sec:p0 Update}).
  These two methods are analytically equivalent, but
  in our experience, the numerical drift associated with evolving $p_0$ using
  an evolution equation causes the entire solution to drift from 
  thermodynamic equilibrium over time.
\item As explained in \S \ref{Sec:Governing Equations}, in the advection 
  step, we predict $(\rho h)'$ to edges instead of $T$ 
  ({\bf Steps 4Hi} and {\bf 8Hi} in \S \ref{Sec:Main Algorithm Description}).
  Thus, a base state enthalpy,
  $(\rho h)_0$ is required in order to construct the enthalpy fluxes for
  the conservative update.  Unlike $\rho_0$, we do not need to carry the
  complete evolution of $(\rho h)_0$.  In practice, we set $(\rho h)_0$
  equal to the lateral average of the full state enthalpy after the first call
  to {\bf React State}, i.e., $(\rho h)_0^n = \overline{(\rho h)^{(1)}}$.
  We then advect $(\rho h)_0$ without reactions or 
  heating to mirror the treatment of $(\rho h)$ in the advection step
  ({\bf Steps 4G} and {\bf 8G} in \S \ref{Sec:Main Algorithm Description}).
\item In the advection step, rather than simultaneously updating each of
  the base state quantities, followed by an update of all the full
  state quantities, we have interwoven these updates in order to
  obtain better accuracy.  In order: we advect $\rho_0$, advect
  $\rho$, correct $\rho_0$, advance $p_0$, advect $(\rho h)_0$, and
  advect $(\rho h)$ ({\bf Steps 4} and {\bf 8} in 
  \S \ref{Sec:Main Algorithm Description}).  
  This enables us to use a time-centered $\psi$ rather than a time-lagged
  $\psi$ for the enthalpy update.
\item Rather than compute $\nabla\cdot\etarho$ and use it to adjust
  $\rho_0$ to account for convecting overturning, as was done in {\bf
  Correct Base} in Paper III, we simply use the lateral average operator to
  enforce $\rho_0 = \overline{\rho}$, which is simpler and analytically
  equivalent ({\bf Steps 4D} and {\bf 8D} in 
  \S \ref{Sec:Main Algorithm Description}).  We use the
  improved averaging procedure described in \S \ref{Sec:Lateral Average},
  which greatly improves the accuracy of this mapping for spherical
  problems.
\item We have moved the first reaction step 
  (formerly {\bf Step 3} in \S \ref{Sec:Main Algorithm Description}) to occur 
  before {\bf Steps 1} and {\bf 2} in \S \ref{Sec:Main Algorithm Description}.
  This is a non-functional change; it was only made so that {\bf Steps 2-5} 
  mirror {\bf Steps 6-9} in the overall predictor-corrector scheme.
\item For planar problems, we evolve the enthalpy 
  for the sole purpose of updating the temperature, which is subsequently 
  fed into the reaction network and used in computing thermodynamic derivatives.
  For spherical problems, as described in Paper IV, we instead define 
  $T$ from $\rho$,$p_0$, and $X_k$.  This completely decouples the
  enthalpy from the rest of the solution.  Our experience has shown that, 
  with spherical geometry, the discretization errors are minimized by using 
  the hydrostatic, radial base state pressure to define temperature.  We 
  still evolve the enthalpy, but we use it as a diagnostic to 
  examine to what extent our numerical method keeps the solution in 
  thermodynamic equilibrium.
\end{itemize}

\subsection{Gravity}\label{Sec:Gravity}
For planar problems, we treat gravity as 
constant in space and time.  For spherical problems, gravity is computed directly 
from $\rho_0$ in the following manner.  First, we define the enclosed mass at 
radial edges and cell centers:
\begin{equation}
m_{{\rm encl},j-\myhalf} = \sum_{k=1}^j \frac{4}{3}\pi\left(r_{k-\myhalf}^3 - r_{k-\sfrac{3}{2}}^3\right)\rho_{0,k-1}; \quad m_{{\rm encl},j} = m_{{\rm encl},j-\myhalf} + \frac{4}{3}\pi \left(r_j^3 - r_{j-\myhalf}^3\right)\rho_{0,j}.
\end{equation}
In practice, we compute $(r_{k-\myhalf}^3
- r_{k-\sfrac{3}{2}}^3 )$ as $\Delta r (r^2_{k-\sfrac{1}{2}} +
r_{k-\sfrac{1}{2}} r_{k-\sfrac{3}{2}} + r^2_{k-\sfrac{3}{2}})$, 
in order to minimize roundoff error at large radii
and use a similar formula for the radial difference term in the 
equation for $m_{{\rm encl},j}$.
Next, we define the gravity at both radial edges and cell centers:
\begin{equation}
g_{j-\myhalf} = \frac{Gm_{{\rm encl},j-\myhalf}}{r_{j-\myhalf}^2}; \qquad g_j = \frac{Gm_{{\rm encl},j}}{r_j^2}.
\end{equation}
We indicate which base
state density is used to compute gravity by using a shorthand notation
with superscripts on $g$, e.g., $g^n \equiv g(\rho_0^n)$.

\subsection{Main Algorithm Notation}\label{Sec:Main Algorithm Notation}
We make use of the following shorthand notations in outlining the algorithm:

\subsubsection{Reactions}\label{Sec:Reactions}
{\bf React State}$[\rho^{\inp},(\rho
  h)^{\inp},X_k^{\inp},T^{\inp}, (\rho\Hext)^{\inp},
  p_0^{\inp}] \rightarrow [\rho^{\outp}, (\rho h)^{\outp}, X_k^{\outp},
  T^{\outp}, (\rho \omegadot_k)^{\outp}, (\rho\Hnuc)^{\outp}]$ 
  evolves the species and enthalpy due to reactions through
  $\Delta t/2$ according to:
\begin{equation}
\frac{dX_k}{dt} = \omegadot_k(\rho,X_k,T) ; \qquad
\frac{dT}{dt}   = \frac{1}{c_p} \left ( -\sum_k \xi_k  \omegadot_k  + \Hnuc \right ).
\end{equation}
%This system is solved using the stiff ordinary differential equation
%integration package, {\tt VODE} \citep{vode}.  
Full details of the 
solution procedure can be found in Paper III. We then define:
\begin{eqnarray}
(\rho\omegadot_k)^{\outp} &=& \frac{\rho^{\outp} ( X_k^{\outp} - X_k^{\inp})}{\dt/2}, \\
(\rho h)^{\outp} &=& (\rho h)^{\inp} + \frac{\dt}{2} (\rho\Hnuc)^{\outp} + \frac{\dt}{2} (\rho\Hext)^{\inp}.
\end{eqnarray}
where the enthalpy update includes external heat sources $(\rho\Hext)^{\inp}$.
As introduced in Paper IV, we update the temperature using $T^{\outp} =
T(\rho^\outp,h^\outp,X_k^\outp)$ for planar geometry or $T^{\outp} =
T(\rho^\outp,p_0^\inp,X_k^\outp)$ for spherical geometry.
Note that the density remains unchanged within {\bf React State}, i.e.,
$\rho^{\outp} = \rho^{\inp}$.

\subsubsection{$\rho_0$ Advection}\label{Sec:rho0 advection}
{\bf Advect Base Density}$[\rho_0^\inp,w_0^\inp] \rightarrow
  [\rho_0^\outp, \rho_0^{\outp,\nph}]$ is the process by which we
  update the base state density through $\dt$ in time.  We keep the
  time-centered edge states, $\rho_0^{\outp,\nph}$,
  since they are used later in discretization of $\etarho$ for planar problems.
\begin{description}
\item[planar:] We discretize equation (\ref{eq:Base State Density}) to
  compute the new base state density,
\begin{equation}
\rho_{0,j}^{\outp} = \rho_{0,j}^{\inp} - \frac{\dt}{\dr} \left [ \left( \rho_0^{\outp,\nph} w_0^{\inp}\right)_{j+\myhalf} - \left( \rho_0^{\outp,\nph} w_0^{\inp}\right)_{j-\myhalf} \right ].
\end{equation}
  We compute the time-centered edge states, $\rho_0^{\outp,\nph}$,
  by discretizing an expanded form of equation (\ref{eq:Base State Density}):
\begin{equation}
\frac{\partial \rho_0}{\partial t} + w_0 \frac{\partial \rho_0}{\partial r} = - \rho_0 \frac{\partial w_0}{\partial r},
\end{equation}
  where the right hand side is used as the force term.
\item[spherical:] The base state density update now includes the area factors in the 
  divergences:
\begin{equation}
\rho_{0,j}^{\outp} = \rho_{0,j}^{\inp} - \frac{1}{r_j^2} \frac{\dt}{\dr} \left [ \left( r^2 \rho_0^{\outp,\nph} w_0^{\inp}\right)_{j+\myhalf} - \left( r^2 \rho_0^{\outp,\nph} w_0^{\inp}\right)_{j-\myhalf} \right].
\end{equation}
  In order to compute the time-centered edge states, an additional geometric 
  term is added to the forcing, due to the spherical discretization of 
  (\ref{eq:Base State Density}):
\begin{equation}
\frac{\partial \rho_0}{\partial t} + w_0 \frac{\partial \rho_0}{\partial r} = - \rho_0 \frac{\partial w_0}{\partial r} - \frac{2 \rho_0 w_0}{r}.
\end{equation}
\end{description}

\subsubsection{$p_0$ Update}\label{Sec:p0 Update}
{\bf Enforce HSE}$[p_0^{\inp},\rho_0^{\inp}]
  \rightarrow [p_0^{\outp}]$ has replaced {\bf Advect Base Pressure}
  from Paper III as the process by which we update the base state
  pressure.  Rather than discretizing the evolution equation for
  $p_0$, we enforce hydrostatic equilibrium directly, which is numerically simpler
  and analytically equivalent.  We first set 
  $p_{0,j=0}^{\outp} = p_{0,j=0}^{\inp}$ and then update $p_0^\outp$ using:
\begin{equation}
p_{0,j+1}^{\outp} = p_{0,j}^{\outp} + \Delta r g_{j+\myhalf}\frac{\left(\rho_{0,j+1}^{\inp}+\rho_{0,j}^{\inp}\right)}{2},
\end{equation}
  where $g=g(\rho_0^\inp)$.  As soon as $\rho_{0,j}^\inp < \rho_{\rm cutoff}$, we set 
  $p_{0,j+1}^\outp = p_{0,j}^\outp$ for all remaining values of $j$.  
  Then we compare $p_{0,j_{\rm max}}^\outp$ with $p_{0,j_{\rm max}}^\inp$ and offset
  every element in $p_0^\outp$ so that $p_{0,j_{\rm max}}^\outp = p_{0,j_{\rm max}}^\inp$.  
  We are effectively using the location where the $\rho_0^\inp$ drops below 
  $\rho_{\rm cutoff}$ as the starting point for integration.

\subsubsection{$(\rho h)_0$ Advection}\label{Sec:rhoh0 advection}
{\bf Advect Base Enthalpy}$[(\rho h)_0^\inp,w_0^\inp,\psi^\inp] \rightarrow [(\rho h)_0^\outp]$
  is the process by which we update the base state enthalpy through $\dt$ in time.  
\begin{description}
\item[planar:] We discretize equation (\ref{eq:Base State Enthalpy}), neglecting reaction source terms,
  to compute the new base state enthalpy,
\begin{equation}
(\rho h)_{0,j}^{\outp} = (\rho h)_{0,j}^{\inp} - \frac{\dt}{\Delta r} \left\{ \left[ (\rho h)_0^{\nph} w_0^{\inp}\right]_{j+\myhalf} - \left[ (\rho h)_0^{\nph} w_0^{\inp}\right]_{j-\myhalf} \right\} + \dt\psi_j^{\inp}.
\end{equation}
  We compute the time-centered edge states, $(\rho h)_0^{\nph}$, by discretizing
  an expanded form of equation (\ref{eq:Base State Enthalpy}):
\begin{equation}
\frac{\partial (\rho h)_0}{\partial t} + w_0 \frac{\partial (\rho h)_0}{\partial r} = -(\rho h)_0 \frac{\partial w_0}{\partial r} + \psi.
\end{equation}
\item[spherical:]  The base state enthalpy update now includes the area factors 
  in the divergences:
\begin{eqnarray}
(\rho h)_{0,j}^{\outp} &=& (\rho h)_{0,j}^{\inp} \nonumber \\
&& - \frac{1}{r_j^2} \frac{\dt}{\dr} \left \{ \left[ r^2 (\rho h)_0^{\nph} w_0^{\inp}\right]_{j+\myhalf} - \left[ r^2 (\rho h)_0^{\nph} w_0^{\inp}\right]_{j-\myhalf} \right\} +\dt\psi^{\inp,\nph}.\nonumber\\
\end{eqnarray}
  In order to compute the time-centered edge states, an additional geometric 
  term is added to the forcing, due to the spherical discretization of 
  (\ref{eq:Base State Enthalpy}):
\begin{equation}
\frac{\partial (\rho h)_0}{\partial t} + w_0 \frac{\partial (\rho h)_0}{\partial r} = -(\rho h)_0 \frac{\partial w_0}{\partial r} - \frac{2 (\rho h)_0 w_0}{r} + \psi.
\end{equation}
\end{description}

\subsubsection{Computing $w_0$}\label{Sec:Computing w0}
Here we describe the process by which we compute $w_0$.  The arguments 
are different for planar and spherical geometries.
\begin{description}

\item {\bf Compute} {\boldmath $w_0$} {\bf Planar}
$[\Sbar^{\inp},\gammaonebar^{\inp}, p_0^{\inp},\psi^{\inp}]\rightarrow [w_0^{\outp}]$:

In Paper III, we showed that $\psi=\etarho g$ for planar geometries, and 
derived an alternate expression for equation (\ref{eq:w0 divergence}).
We compute $w_0$ using equation (35) in Paper III using the following 
discretization:
\begin{equation}
\frac{w_{0,j+\myhalf}^\outp-w_{0,j-\myhalf}^\outp}{\Delta r} = \left(\Sbar^{\inp} - \frac{1}{\gammaonebar^{\inp} p_0^{\inp}}\psi^{\inp}\right)_j,
\end{equation}
with $w_{0,-\myhalf}=0$.

\item {\bf Compute} {\boldmath $w_0$} {\bf Spherical}
  $[\Sbar^{\inp},\gammaonebar^{\inp},\rho_0^{\inp},p_0^{\inp},\etarho^{\inp}]
  \rightarrow[w_0^{\outp}]$:

We begin with equation (\ref{eq:w0 divergence}) written in spherical coordinates:
\begin{equation}
\frac{1}{r^2}\frac{\partial}{\partial r} \left (r^2 \beta_0 w_0 \right ) = \beta_0 \left ( \Sbar - \frac{1}{\gammaonebar p_0} \frac{\partial p_0}{\partial t} \right ).
\end{equation}
We expand the spatial derivative and recall from Paper I that
\begin{equation}
\frac{1}{\gammaonebar p_0} \frac{\partial p_0}{\partial r} = \frac{1}{\beta_0} \frac{\partial \beta_0}{\partial r},
\end{equation}
giving:
\begin{equation}
\frac{1}{r^2} \frac{\partial}{\partial r} \left (r^2 w_0 \right ) = \Sbar - \frac{1}{\gammaonebar p_0} \underbrace{\left( \frac{\partial p_0}{\partial t} + w_0 \frac{\partial p_0}{\partial r} \right)}_{\psi}.\label{eq:psi def}
\end{equation}
We solve this equation for $w_0$ as described in Appendix \ref{Sec:Computing w0 spherical}.

\end{description}

\subsection{Main Algorithm Description}\label{Sec:Main Algorithm Description}
\begin{figure}[tb]
\centering
\includegraphics[scale=0.6]{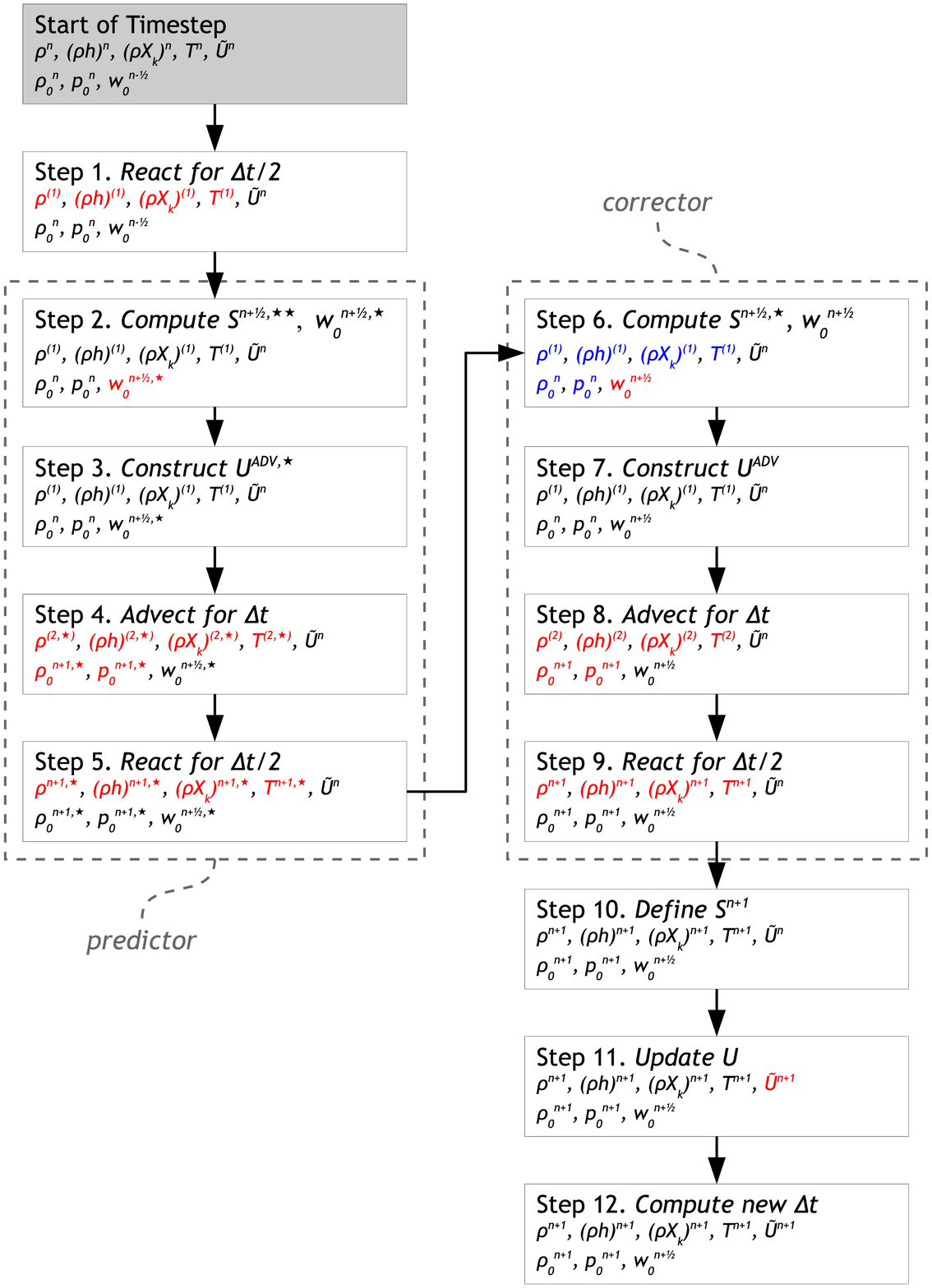}
\caption{\label{Fig:flowchart} A flowchart of the algorithm.  The
  thermodynamic state variables, base state variables, and local velocity are
  indicated in each step.  Red text indicates that quantity was
  updated during that step.  The predictor-corrector steps are
  outlined by the dotted box.  The blue text indicates state
  variables that are the same in {\bf Step 6} as they are in
  {\bf Step 2}, i.e., they are unchanged by the predictor steps.}
\end{figure}

\begin{figure}[tb]
\centering
\includegraphics[scale=0.6]{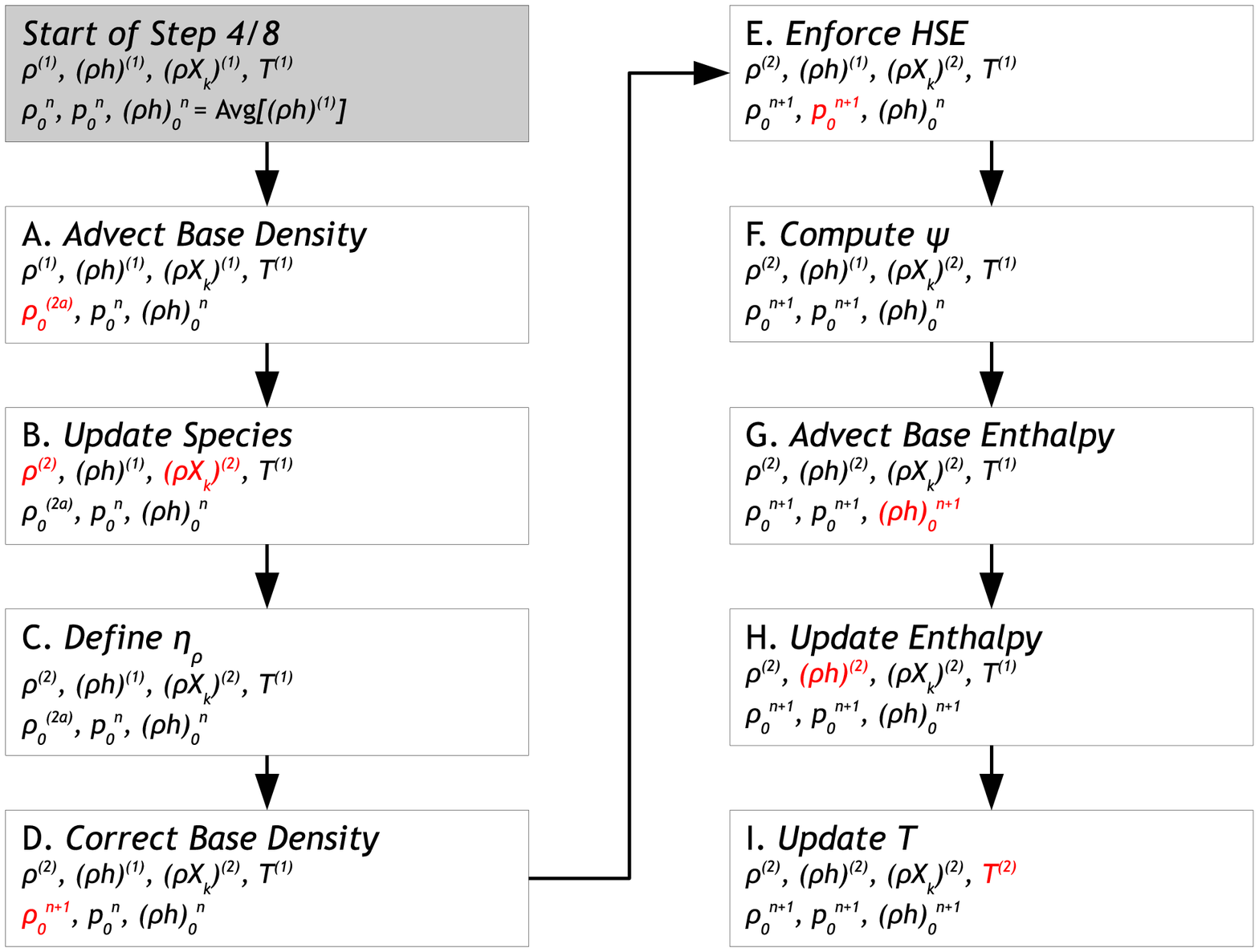}
\caption{\label{Fig:flowchart48}
A flowchart for {\bf Steps 4} and {\bf 8}.  The thermodynamic state variables 
and base state variables are indicated in each step.  Red text indicates that quantity
was updated during that step.  Note, for {\bf Step 4}, the updated
quantities should also have a $\star$ superscript, e.g., {\bf Step 8I} defines
$T^{(2)}$ while {\bf Step 4I} defines $T^{(2),\star}$ .}
\end{figure}

We now describe the main algorithm, making frequent use of the
shorthand developed above.  In summary, in the predictor step ({\bf
  Steps 2-5}) we use an estimate of the expansion term, $S$, to
compute a preliminary solution at the new time level, denoted with an
``$n+1,\star$'' superscript.  In the corrector step ({\bf Steps 6-9}),
we use the results from the predictor step to compute a more accurate
expansion term, and compute the final solution at the new time level,
denoted with an ``$n+1$'' superscript.  We use Strang-splitting to
achieve second-order accuracy in time.  See Figure \ref{Fig:flowchart}
for a flow chart of the algorithm, including the notation used as we
advance the solution by $\dt.$ Figure \ref{Fig:flowchart48} is a flow
chart of the advection steps ({\bf Steps 4} and {\bf 8}), which
includes the notation we use as we advect the solution through a time
interval of $\dt$.

The discussion that follows mirrors closely that in Paper III, but has been
updated to reflect all the changes throughout the algorithm.  The advance of 
the state through a single time step appears as:

\begin{description}

%--------------------------------------------------------------------------
% STEP 0
%--------------------------------------------------------------------------

\item[Step 0.] {\em Initialization}

This step remains unchanged from Paper III.
The initialization step only occurs at the beginning of the simulation.  
The initial values for $\Ub^0, \rho^0, (\rho h)^0, X_k^0, T^0, 
\rho_0^0, p_0^0$, and $\overline{\Gamma_1^0}$ are specified from the problem-dependent 
initial conditions.  The initial time step, $\dt^0$, is computed as in 
Paper III.  Finally, initial
values for $w_0^{-\myhalf}, \etarho^{-\myhalf}, \psi^{-\myhalf}, 
\pi^{-\myhalf}, S^0$, and $S^1$ come from a preliminary pass through
the algorithm.  

%--------------------------------------------------------------------------
% STEP 1
%--------------------------------------------------------------------------
\item[Step 1.] {\em React the full state through the first time interval of $\dt / 2.$}

Call {\bf React State}$[\rho^n, (\rho h)^n, X_k^n, T^n, (\rho\Hext)^n, p_0^n]$\\
\phantom{ }\hfill $\rightarrow [\rho^{(1)},(\rho h)^{(1)},X_k^{(1)},T^{(1)},(\rho \omegadot_k)^{(1)},(\rho \Hnuc)^{(1)}]$.

%--------------------------------------------------------------------------
% STEP 2
%--------------------------------------------------------------------------

\item[Step 2.] {\em Compute the provisional time-centered expansion,
  $S^{\nph,\star\star}$, provisional base state velocity,
  $w_0^{\nph,\star}$, and provisional base state velocity forcing.}

\begin{enumerate}
\renewcommand{\theenumi}{{\bf \Alph{enumi}}}

\item Compute $S^{\nph,\star\star}$.  We compute an estimate for the 
  time-centered expansion term in the velocity 
  divergence constraint (eq. [\ref{eq:utilde divergence}]).  For the  
  first time step ($n=0$), we set
\begin{equation}
S^{n+\myhalf,\star\star} = \frac{S^0 + S^1}{2},
\end{equation}
  where $S^1$ is found during initialization.  For other time steps
  $(n \ne 0)$, following \citet{Bell:2004}, we extrapolate
  to the half-time using $S$ at the previous and current
  time levels
\begin{equation}
S^{\nph,\star\star} = S^n + \frac{\dt^n}{2} \frac{S^n - S^{n-1}}{\dt^{n-1}}.
\end{equation}
  Next, compute
\begin{equation}
\overline{S^{\nph,\star\star}} = {\mathrm{\bf Avg}} \left(S^{\nph,\star\star}\right).
\end{equation}

\item Compute $w_0^{\nph,\star}$.
\begin{description}
\item For planar geometry, call\\
{\bf Compute} {\boldmath $w_0$} {\bf Planar}$[\overline{S^{\nph,\star\star}},\overline{\Gamma_1^n},p_0^n,\psi^{n-\myhalf}] \rightarrow [w_0^{\nph,\star}]$.
\item For spherical geometry, call\\
{\bf Compute} {\boldmath $w_0$} {\bf Spherical}$[\overline{S^{\nph,\star\star}},\overline{\Gamma_1^n},\rho_0^n,p_0^n,\etarho^{n-\myhalf}] \rightarrow [w_0^{\nph,\star}]$.
\end{description}

\item Compute the provisional base state velocity forcing.
  Rearrange equation (\ref{eq:w0 evolution}),
\begin{equation}
-\frac{1}{\rho_0} \frac{\partial \pi_0}{\partial r} = \frac{\partial w_0}{\partial t} + w_0 \frac{\partial w_0}{\partial r}, \label{eq:pizero}
\end{equation}
  with the following discretization:
\begin{equation}
\left ( \frac{1}{\rho_0} \frac{\partial \pi_0}{\partial r} \right )^{n,\star} = -\frac{w_0^{\nph,\star} - w_0^\nmh}{(\dt^n+\dt^{n-1})/2} - w_0^{n,\star} \left(\frac{\partial w_0}{\partial r}\right)^{n,\star},
\end{equation} 
  where $w_0^{n,\star}$ and $(\partial w_0 / \partial r)^{n,\star}$ are defined as
\begin{eqnarray}
w_0^{n,\star} &=& \frac{\dt^{n} w_0^{\nmh} + \dt^{n-1} w_0^{\nph,\star}}{\dt^n+\dt^{n-1}}, \\
\left(\frac{\partial w_0}{\partial r}\right)^{n,\star} &=& \frac{1}{\dt^n+\dt^{n-1}}\left [ \dt^{n} \left(\frac{\partial w_0 }{ \partial r}\right)^{\nmh} + \dt^{n-1} \left(\frac{\partial w_0 }{ \partial r}\right)^{\nph,\star} \right ].\nonumber \\
\end{eqnarray}
  If $n=0$, we use $\dt^{-1} = \dt^0$.
\end{enumerate}

%--------------------------------------------------------------------------
% STEP 3
%--------------------------------------------------------------------------
\item[Step 3.] {\em Construct the provisional time-centered advective velocity on 
edges, $\uadvone$.}

Using equation (\ref{eq:utildeupd}), we compute time-centered edge 
velocities, $\uadvonedag$, using 
$\Ub = \Ubt^n + w_0^{\nph,\star}$.  The $\dagger$ superscript refers to the 
fact that the predicted velocity field does not satisfy the divergence 
constraint.  We then construct $\uadvone$ from $\uadvonedag$
using a MAC projection, as described in detail in Appendix B of Paper III.  
We note that $\uadvone$ satisfies the discrete version of
$\overline{(\uadvone\cdot\eb_r)}=0$ as well as
\begin{eqnarray}
\nabla \cdot \left(\beta_0^n \uadvone\right) &=& \beta_0^n \left(S^{\nph,\star\star} - \overline{S^{\nph,\star\star}}\right),\\
 \beta_0^n &=& \beta_0 \left(\rho_0^n, p_0^n, \overline{\Gamma_1^n}\right),
\end{eqnarray}
where $\beta_0$ is computed as described in Appendix C of Paper III.

%--------------------------------------------------------------------------
% STEP 4
%--------------------------------------------------------------------------
\item[Step 4.] {\em Advect the base state and full state through a time interval of $\dt.$}

\begin{enumerate}
\renewcommand{\theenumi}{{\bf \Alph{enumi}}}

\item Update $\rho_0$, saving the time-centered density at radial edges by calling

{\bf Advect Base Density}$[\rho_0^{n},w_0^{\nph,\star}] \rightarrow [\rho_0^{(2a),\star}, \rho_0^{\nph,\star,\pred}]$.

\item Update $(\rho X_k)$ using a discretized version of equation 
(\ref{eq:species}) omitting the reaction terms, which were already 
accounted for in {\bf React State}.  The update consists of two steps:

\begin{enumerate}
\renewcommand{\labelenumii}{{\bf \roman{enumii}}.}

\item Compute the time-centered species edge states, $(\rho X_k)^{\nph,\star,\pred}$,
  for the conservative update of $(\rho X_k)^{(1)}$.  We use equations 
  (\ref{eq:Perturbational Density}) and (\ref{eq:Primitive Species}) to 
  predict $\rho^{'(1)} = \rho^{(1)} - \rho_0^n$ and 
  $X_k^{(1)} = (\rho  X_k)^{(1)} / \rho^{(1)}$ to time-centered edges using 
  $\Ub = \uadvone+w_0^{\nph,\star}\eb_r$, yielding $\rho^{'\nph,\star,\pred}$ 
  and $X_k^{\nph,\star,\pred}$.
  We convert the perturbational density full state density using
\begin{equation}
\rho^{\nph,\star,\pred} = \rho^{'\nph,\star,\pred} + \frac{\rho_0^n + \rho_0^{(2a),\star}}{2},
\end{equation}
  where the base state density terms are mapped to Cartesian edges.
  Then,\\
  $(\rho X_k)^{\nph,\star,\pred} = (\rho^{\nph,\star,\pred}X_k^{\nph,\star,\pred})$.

\item Evolve $(\rho X_k)^{(1)} \rightarrow (\rho X_k)^{(2),\star}$ using
\begin{eqnarray}
(\rho X_k)^{(2),\star} &=& (\rho X_k)^{(1)} \nonumber \\
&& - \dt \left\{ \nabla \cdot \left[ \left(\uadvone+w_0^{\nph,\star} \eb_r\right) (\rho X_k)^{\nph,\star,\pred} \right] \right\},\nonumber \\
\end{eqnarray}
\begin{equation}
\rho^{(2),\star} = \sum_k (\rho X_k)^{(2),\star},
\qquad
X_k^{(2),\star} = (\rho X_k)^{(2),\star} / \rho^{(2),\star}.
\end{equation}

\end{enumerate}

\item Define a radial edge-centered $\etarho^{\nph,\star}$.

\begin{description}
\item For planar geometry, since $\etarho = \overline{\rho'(\Ub\cdot\eb_r)} = \overline{\rho(\Ub\cdot\eb_r)}-\overline{\rho_0(\Ub\cdot\eb_r}) = \overline{\rho(\Ub\cdot\eb_r)} - \rho_0w_0$,
\begin{eqnarray}
 \etarho^{\nph,\star} &=&  {\rm {\bf Avg}} \sum_k \left[ \left(\uadvone \cdot \eb_r + w_0^{\nph,\star}\right) (\rho X_k)^{\nph,\star,\pred} \right]\nonumber\\
&& - w_0^{\nph,\star} \rho_0^{\nph,\star,\pred},
\end{eqnarray}
\item For spherical geometry, first construct 
$\etarho^{{\rm cart},\nph,\star} = [\rho'(\Ub\cdot\eb_r)]^{\nph,\star}$ on Cartesian cell centers using:
\begin{eqnarray}
\etarho^{{\rm cart},\nph,\star} &=& \left[\left(\frac{\rho^{(1)}+\rho^{(2),\star}}{2}\right)-\left(\frac{\rho_0^n+\rho_0^{(2a),\star}}{2}\right)\right] \nonumber \\
&&\cdot \left( \uadvone \cdot \eb_r  + w_0^{\nph,\star}\right).
\end{eqnarray}
Then,
\begin{equation}
\etarho^{\nph,\star} = {\rm {\bf Avg}}\left(\etarho^{{\rm cart},\nph,\star}\right).
\end{equation}
This gives a radial cell-centered $\etarho^{\nph,\star}$.  To get
$\etarho^{\nph,\star}$ at radial edges, average the two neighboring
radial cell-centered values.
\end{description}

\item Correct $\rho_0$ by setting $\rho_0^{n+1,\star} =$ {\bf Avg}$(\rho^{(2),\star})$.

\item Update $p_0$ by calling 
{\bf Enforce HSE}$[p_0^n,\rho_0^{n+1,\star}] \rightarrow [p_0^{n+1,\star}]$.

\item Compute $\psi^{\nph,\star}$.
\begin{description}
\item For planar geometry,
\begin{equation}
\psi_j^{\nph,\star} = \frac{1}{2} \left(\eta_{\rho,j-\myhalf}^{\nph,\star} 
+ \eta_{\rho,j+\myhalf}^{\nph,\star}\right) g.
\end{equation}
\item For spherical geometry, first compute:
\begin{eqnarray}
\overline{\Gamma_1^{(1)}} &=& {\rm{\bf Avg}} \left[ \Gamma_1\left(\rho^{(1)}, p_0^{n}, X_k^{(1)}\right) \right]  , \\
\overline{\Gamma_1^{(2),\star}} &=& {\rm{\bf Avg}} \left[ \Gamma_1\left(\rho^{(2),\star}, p_0^{n+1,\star}, X_k^{(2),\star}\right) \right].
\end{eqnarray}
Then, define $\psi^{\nph,\star}$ using equation (\ref{eq:psi def})
\begin{eqnarray}
\psi_j^{\nph,\star} 
&=& \left(\frac{\overline{\Gamma_1^{(1)}}+\overline{\Gamma_1^{(2),\star}}}{2}\right)_j
\left(\frac{p_0^n+p_0^{n+1,\star}}{2}\right)_j \nonumber \\
&& \left \{ \overline{S_j^{\nph,\star}} - \frac{1}{r_j^2} \left [ \left(r^2 w_0^{\nph,\star}\right)_{j+\myhalf} - \left(r^2 w_0^{\nph,\star}\right)_{j-\myhalf} \right ] \right \}.\nonumber \\
\end{eqnarray}
\end{description}

\item Update $(\rho h)_0$.  First, compute $(\rho h)_0^n = $ {\bf Avg}$[(\rho h)^{(1)}]$.
Then, call\\
{\bf Advect Base Enthalpy}$[(\rho h)_0^{n}, w_0^{\nph,\star}, \psi^{\nph,\star}] \rightarrow [(\rho h)_0^{n+1,\star}]$.

\item Update the enthalpy using a discretized version of equation
(\ref{eq:enthalpy}), again omitting the reaction and heating terms
since we already accounted for
them in {\bf React State}.  This equation takes the form:
\begin{equation}
\frac{\partial (\rho h)}{\partial t}  = - \nabla \cdot (\Ub \rho h) + \psi + (\Ubt \cdot \eb_r) \frac{\partial p_0}{\partial r}.
\end{equation}
For spherical geometry, we solve the
analytically equivalent form,
\begin{equation}
\frac{\partial (\rho h)}{\partial t}  = - \nabla \cdot (\Ub \rho h) + \psi + \nabla \cdot (\Ubt p_0) - p_0 \nabla \cdot \Ubt,
\end{equation}
which experience has shown to minimize the drift from thermodynamic equilibrium.  The update consists of two steps:

\begin{enumerate}
\renewcommand{\labelenumii}{{\bf \roman{enumii}}.}

\item Compute the time-centered enthalpy edge state, $(\rho h)^{\nph,\star,\pred},$
  for the conservative update of $(\rho h)^{(1)}$.  We use equation 
  (\ref{eq:Perturbational Enthalpy}) to predict
  $(\rho h)' = (\rho h)^{(1)} - (\rho h)_0^n$ to time-centered edges, 
  using $\Ub = \uadvone+w_0^{\nph,\star} \eb_r$,
  yielding $(\rho h)^{'\nph,\star,\pred}$.  We convert the perturbational 
  enthalpy to a full state enthalpy using
\begin{equation}
(\rho h)^{\nph,\star,\pred} = (\rho h)^{'\nph,\star,\pred} + \frac{(\rho h)_0^n + (\rho h)_0^{n+1,\star}}{2}.
\end{equation}
  For planar geometry, we map $(\rho h)_0$ directly to Cartesian edges.
  In spherical geometry, our experience has shown that a slightly different
  approach leads to reduced discretization errors.  We first map 
  $h_0 \equiv (\rho h)_0/\rho_0$ and $\rho_0$ to Cartesian edges separately, 
  and then multiply these terms to get $(\rho h)_0$.

\item Evolve $(\rho h)^{(1)} \rightarrow (\rho h)^{(2),\star}$.
\begin{description}
\item For planar geometry,
\begin{eqnarray}
(\rho h)^{(2),\star} 
&=& (\rho h)^{(1)} \nonumber \\
&&- \dt \left\{ \nabla \cdot \left[ \left(\uadvone+w_0^{\nph,\star} \eb_r\right) (\rho h)^{\nph,\star,\pred} \right] \right\} \nonumber \\ 
&& + \dt \left(\uadvone \cdot \eb_r\right) \left(\frac{\partial p_0}{\partial r} \right)^{n} + \dt \psi^{\nph,\star},
\end{eqnarray}

\item For spherical geometry,
\begin{eqnarray}
(\rho h)^{(2),\star} 
&=& (\rho h)^{(1)} \nonumber \\
&&- \dt \left\{ \nabla \cdot \left[ \left(\uadvone+w_0^{\nph,\star} \eb_r\right) (\rho h)^{\nph,\star,\pred} \right] \right\} \nonumber \\ 
&& + \dt \left \{ \nabla \cdot \left (\uadvone p_0^{n} \right ) - p_0^{n} \nabla \cdot \uadvone \right \} \nonumber \\
&&+ \dt \psi^{\nph,\star},
\end{eqnarray}
\end{description}

\end{enumerate}

Then, for each Cartesian cell where $\rho^{(2),\star} < \rho_\mathrm{cutoff}$,
we recompute enthalpy using
\begin{equation}
(\rho h)^{(2),\star} = \rho^{(2),\star}h\left(\rho^{(2),\star},p_0^{n+1,\star},X_k^{(2),\star}\right).
\end{equation}

\item Update the temperature using the equation of state:
$T^{(2),\star} = 
  T(\rho^{(2),\star}, h^{(2),\star}, X_k^{(2),\star})$ (planar geometry) or
$T^{(2),\star} = 
  T(\rho^{(2),\star}, p_0^{n+1,\star}, X_k^{(2),\star})$ (spherical geometry).
\end{enumerate}

%--------------------------------------------------------------------------
% STEP 5
%--------------------------------------------------------------------------
\item[Step 5.] {\em React the full state through a second time interval of $\dt / 2.$}

Call {\bf React State}$[ \rho^{(2),\star},(\rho h)^{(2),\star}, X_k^{(2),\star}, T^{(2),\star}, 
(\rho\Hext)^{(2),\star}, p_0^{n+1,\star}]$\\
\phantom{ }\hfill$\rightarrow [ \rho^{n+1,\star},(\rho h)^{n+1,\star}, X_k^{n+1,\star}, T^{n+1,\star}, (\rho \omegadot_k)^{(2),\star}, (\rho \Hnuc)^{(2),\star} ].$

%--------------------------------------------------------------------------
% STEP 6
%--------------------------------------------------------------------------
\item[Step 6.] {\em Compute the time-centered expansion, $S^{\nph,\star}$, base state
velocity, $w_0^{\nph}$, and base state velocity forcing.}

\begin{enumerate}
\renewcommand{\theenumi}{{\bf \Alph{enumi}}}

\item Compute $S^{\nph,\star}$.  First, compute $S^{n+1,\star}$ with
\begin{equation}
S^{n+1,\star} =  -\sigma  \sum_k  \xi_k  (\omegadot_k)^{(2),\star}  + \frac{1}{\rho^{n+1,\star} p_\rho} \sum_k p_{X_k}  ({\omegadot}_k)^{(2),\star} + \sigma \Hnuc^{(2),\star} + \sigma \Hext^{(2),\star},
\end{equation}
  where $(\omegadot_k)^{(2),\star} = (\rho \omegadot_k)^{(2),\star} /
  \rho^{(2),\star}$ and the thermodynamic quantities are defined using
  $\rho^{n+1,\star}, X_k^{n+1,\star},$ and $T^{n+1,\star}$ as inputs to
  the equation of state.  Then, define
\begin{equation}
\overline{S^{\nph,\star}} = {\mathrm{\bf Avg}} (S^{\nph,\star}),
\qquad
 S^{\nph.\star} = \frac{S^n + S^{n+1,\star}}{2},
\end{equation}
\item Compute $w_0^{\nph}$.  First, define
\begin{equation}
\overline{\Gamma_1^{\nph,\star}} = \frac{\overline{\Gamma_1^n} + \overline{\Gamma_1^{n+1,\star}}}{2}, 
\quad
\rho_0^{\nph,\star} = \frac{\rho_0^{n} + \rho_0^{n+1,\star}}{2},
\quad
p_0^{\nph,\star} = \frac{p_0^{n} + p_0^{n+1,\star}}{2},
\end{equation}
  with
\begin{equation}
 \overline{\Gamma_1^{n+1,\star}} = {\rm{\bf Avg}} \left[ \Gamma_1\left(\rho^{n+1,\star}, p_0^{n+1,\star}, X_k^{n+1,\star}\right) \right].
\end{equation}
\begin{description}
\item For planar geometry, call\\
{\bf Compute} {\boldmath $w_0$} {\bf Planar}$[\overline{S^{\nph,\star}},\overline{\Gamma_1^{\nph,\star}},p_0^{\nph,\star},\psi^{\nph,\star}]\rightarrow [w_0^{\nph}]$.
\item For spherical geometry, call\\
{\bf Compute} {\boldmath $w_0$} {\bf Spherical}$[\overline{S^{\nph,\star}},\overline{\Gamma_1^{\nph,\star}},\rho_0^{\nph,\star},p_0^{\nph,\star},\etarho^{\nph,\star}]$\\
\phantom{ }\hfill$\rightarrow [w_0^{\nph}]$.
\end{description}

\item Compute the base state velocity forcing.  Rearrange equation (\ref{eq:pizero}),
\begin{equation}
\left ( \frac{1}{\rho_0} \frac{\partial \pi_0}{\partial r} \right )^n = 
-\frac{w_0^{\nph} - w_0^\nmh}{\myhalf(\dt^n+\dt^{n-1})} 
- w_0^n \left(\frac{\partial w_0}{\partial r}\right)^n,
\end{equation}
  where $w_0^{n}$ and $(\partial w_0 / \partial r)^{n}$ are defined as
\begin{eqnarray}
w_0^n &=& \frac{\dt^{n} w_0^{\nmh} + \dt^{n-1} w_0^{\nph}}{\dt^n+\dt^{n-1}}, \\
\left(\frac{\partial w_0}{\partial r}\right)^{n} &=& \frac{1}{\dt^n+\dt^{n-1} } \left [ \dt^{n} \left(\frac{\partial w_0 }{ \partial r}\right)^{\nmh} + \dt^{n-1} \left(\frac{\partial w_0 }{ \partial r}\right)^{\nph} \right ].\nonumber \\
\end{eqnarray}
  If $n=0$, we use $\dt^{-1} = \dt^0$.

\end{enumerate}

%--------------------------------------------------------------------------
% STEP 7
%--------------------------------------------------------------------------
\item[Step 7.] {\em Construct the time-centered advective velocity on edges, $\uadvtwo$.}

The procedure to construct $\uadvtwodag$ is identical to the procedure
for computing $\uadvonedag$ in {\bf Step 3}, but uses 
the updated values $w_0^{\nph}$ and $\pi_0^n$ rather than $w_0^{\nph,\star}$ 
and $\pi_0^{n,\star}$.  We note that $\uadvtwo$ satisfies the discrete version of
$\overline{(\uadvtwo\cdot\eb_r)}=0$ as well as
\begin{equation}
\nabla \cdot \left(\beta_0^{\nph,\star} \uadvtwo\right) =
\beta_0^{\nph,\star}\left(S^{\nph,\star} - \overline{S^{\nph,\star}}\right),
\end{equation}
\begin{equation}
\beta_0^{\nph,\star} = \frac{ \beta_0^n +  \beta_0^{n+1,\star} }{2};
\qquad
 \beta_0^{n+1,\star} = \beta_0 \left(\rho_0^{n+1,\star}, p_0^{n+1,\star}, \overline{\Gamma_1^{n+1,\star}}\right).
\end{equation}

%--------------------------------------------------------------------------
% STEP 8
%--------------------------------------------------------------------------
\item[Step 8.] {\em Advect the base state and full state through a time interval of $\dt.$}

\begin{enumerate}
\renewcommand{\theenumi}{{\bf \Alph{enumi}}}

\item Update $\rho_0$, saving the time-centered density at radial edges by calling

{\bf Advect Base Density}$[\rho_0^{n},w_0^{\nph}] \rightarrow [\rho_0^{(2a)}, \rho_0^{\nph,\pred}]$.

\item Update $(\rho X_k)$.  This step is identical to {\bf Step 4B} except we use
  the updated values $w_0^{\nph}, \uadvtwo$, and $\rho_0^{(2a)}$ rather than 
  $w_0^{\nph,\star}, \uadvone$, and $\rho_0^{(2a),\star}$.  In particular:

\begin{enumerate}
\renewcommand{\labelenumii}{{\bf \roman{enumii}}.}

\item Compute the time-centered species edge states, $(\rho X_k)^{\nph,\pred}$,
  for the conservative update of $(\rho X_k)^{(1)}$.  We use equations 
  (\ref{eq:Perturbational Density}) and (\ref{eq:Primitive Species}) to 
  predict $\rho^{'(1)} = \rho^{(1)} - \rho_0^n$ and 
  $X_k^{(1)} = (\rho  X_k)^{(1)} / \rho^{(1)}$ to time-centered edges
  with $\Ub = \uadvtwo+w_0^{\nph} \eb_r$,
  yielding $\rho^{'\nph,\pred}$ and $X_k^{\nph,\pred}$.  
  We convert the perturbational density to a full state density using
\begin{equation}
\rho^{\nph,\pred} = \rho^{'\nph,\pred} + \frac{\rho_0^n + \rho_0^{(2a)}}{2}.
\end{equation}
  Then, $(\rho X_k)^{\nph,\pred} = (\rho^{\nph,\pred}X_k^{\nph,\pred})$.

\item Evolve $(\rho X_k)^{(1)} \rightarrow (\rho X_k)^{(2)}$ using
\begin{equation}
(\rho X_k)^{(2)} = (\rho X_k)^{(1)} 
- \dt \left\{ \nabla \cdot \left[\left(\uadvtwo+w_0^{\nph} \eb_r\right)  
(\rho X_k)^{\nph,\pred} \right] \right\},
\end{equation}
\begin{equation}
\rho^{(2)} = \sum_k (\rho X_k)^{(2)},
\qquad
X_k^{(2)} = (\rho X_k)^{(2)} / \rho^{(2)}.
\end{equation}

\end{enumerate}

\item Define a radial edge-centered $\etarho^{\nph}$.
\begin{description}
\item For planar geometry,
\begin{eqnarray}
 \etarho^{\nph} &=& {\rm {\bf Avg}} \sum_k \left [\left(\uadvtwo \cdot \eb_r + w_0^{\nph}\right) (\rho X_k)^{\nph,\pred} \right] \nonumber \\
&&- w_0^{\nph} \rho_0^{\nph,\pred},
\end{eqnarray}
\item For spherical geometry, first construct 
$\etarho^{{\rm cart},\nph} = [\rho'(\Ub\cdot\eb_r)]^{\nph}$ on Cartesian
  cell centers using:
\begin{equation}
\etarho^{{\rm cart},\nph} = \left[\left(\frac{\rho^{(1)}+\rho^{(2)}}{2}\right)-\left(\frac{\rho_0^n+\rho_0^{(2a)}}{2}\right)\right] \left(\uadvtwo \cdot \eb_r + w_0^{\nph}\right).
\end{equation}
Then,
\begin{equation}
\etarho^{\nph} = {\rm {\bf Avg}}\left(\etarho^{{\rm cart},\nph}\right).
\end{equation}
This gives a radial cell-centered $\etarho^{\nph}$.  To get
$\etarho^{\nph}$ at radial edges, average the two neighboring
cell-centered values.
\end{description}

\item Correct $\rho_0$ by setting $\rho_0^{n+1} =$ {\bf Avg}$(\rho^{(2)})$.

\item Update $p_0$ by calling
{\bf Enforce HSE}$[p_0^n,\rho_0^{n+1}] \rightarrow [p_0^{n+1}]$.

\item Compute $\psi^{\nph}$.
\begin{description}
\item For planar geometry, 
\begin{equation}
\psi_j^{\nph} = \frac{1}{2} \left(\eta_{\rho,j-\myhalf}^{\nph} 
+ \eta_{\rho,j+\myhalf}^{\nph}\right) g.
\end{equation}
\item For spherical geometry, first compute:
\begin{equation}
\overline{\Gamma_1^{(2)}} = {\rm{\bf Avg}} \left[ \Gamma_1\left(\rho^{(2)}, p_0^{n+1}, 
X_k^{(2)}\right) \right].
\end{equation}
Then, define $\psi^{\nph}$ using equation (\ref{eq:psi def}):
\begin{eqnarray}
\psi_j^{\nph} 
&=& \left(\frac{\overline{\Gamma_1^{(1)}}+\overline{\Gamma_1^{(2)}}}{2}\right)_j \left(\frac{p_0^n+p_0^{n+1}}{2}\right)_j \nonumber \\
&& \left \{ \overline{S_j^{\nph}} - \frac{1}{r_j^2} \left [ \left(r^2 w_0^{\nph}\right)_{j+\myhalf} - \left(r^2 w_0^{\nph}\right)_{j-\myhalf} \right ] \right \}.
\end{eqnarray}
\end{description}

\item Update $(\rho h)_0$ by calling 
{\bf Advect Base Enthalpy}$[(\rho h)_0^n, w_0^{\nph}, \psi^{\nph}] \rightarrow [(\rho h)_0^{n+1}]$.

\item Update the enthalpy.  This step is identical to {\bf Step 4H} except we use
  the updated values $w_0^{\nph}, \uadvtwo, \rho_0^{n+1}, (\rho h)_0^{n+1}, p_0^{n+\myhalf}$, 
  and $\psi^{n+\myhalf}$ rather than\\
  $w_0^{\nph,\star}, \uadvone, \rho_0^{n+1,\star}, (\rho h)_0^{n+1,\star}, p_0^n$, 
  and $\psi^{n+\myhalf,\star}$.  In particular:

\begin{enumerate}
\renewcommand{\labelenumii}{{\bf \roman{enumii}}.}

\item Compute the time-centered enthalpy edge state, $(\rho h)^{\nph,\pred},$
  for the conservative update of $(\rho h)^{(1)}$.  We use equation 
  (\ref{eq:Perturbational Enthalpy}) to predict
  $(\rho h)' = (\rho h)^{(1)} - (\rho h)_0^n$ to time-centered edges
  with $\Ub = \uadvtwo+w_0^{\nph} \eb_r$,
  yielding $(\rho h)^{'\nph,\pred}$.  
  We convert the perturbational enthalpy to a full state enthalpy using
\begin{equation}
(\rho h)^{\nph,\pred} = (\rho h)^{'\nph,\pred} + \frac{(\rho h)_0^n + (\rho h)_0^{n+1}}{2}.
\end{equation}

\item Evolve $(\rho h)^{(1)} \rightarrow (\rho h)^{(2)}$.
\begin{description}
\item For planar geometry,
\begin{eqnarray}
(\rho h)^{(2)} 
&=& (\rho h)^{(1)} - \dt \left\{ \nabla \cdot \left[ \left(\uadvtwo+w_0^{\nph} \eb_r\right)  (\rho h)^{\nph,\pred} \right] \right\} \nonumber \\
&& + \dt \left(\uadvtwo \cdot \eb_r\right) \left(\frac{\partial p_0}{\partial r} \right)^\nph + \dt \psi^{\nph},
\end{eqnarray}

\item For spherical geometry,
\begin{eqnarray}
(\rho h)^{(2)} 
&=& (\rho h)^{(1)} - \dt \left\{ \nabla \cdot \left[ \left(\uadvtwo+w_0^{\nph} \eb_r\right)  (\rho h)^{\nph,\pred} \right] \right\} \nonumber \\
&& + \dt \left[ \nabla \cdot \left (\uadvtwo p_0^{\nph} \right ) - p_0^{\nph} \nabla \cdot \uadvtwo \right] + \dt \psi^{\nph},\nonumber \\
\end{eqnarray}
\end{description}
where $p_0^\nph$ is defined as $p_0^\nph = (p_0^n+p_0^{n+1})/2$.

\end{enumerate}

Then, for each Cartesian cell where $\rho^{(2)} < \rho_\mathrm{cutoff}$, we recompute enthalpy using
\begin{equation}
(\rho h)^{(2)} = \rho^{(2)}h\left(\rho^{(2)},p_0^{n+1},X_k^{(2)}\right).
\end{equation}

\item Update the temperature using the equation of state:
$T^{(2)} = 
   T(\rho^{(2)}, h^{(2)}, X_k^{(2)})$ (planar geometry) or
$T^{(2)} = 
   T(\rho^{(2)}, p_0^{n+1}, X_k^{(2)})$ (spherical geometry).
\end{enumerate}

%--------------------------------------------------------------------------
% STEP 9
%--------------------------------------------------------------------------
\item[Step 9.] {\em React the full state through a second time interval of $\dt / 2.$}

Call {\bf React State}$[\rho^{(2)},(\rho h)^{(2)}, X_k^{(2)},T^{(2)}, (\rho\Hext)^{(2)}, p_0^{n+1}]$\\
\phantom{ }\hfill$\rightarrow [\rho^{n+1}, (\rho h)^{n+1}, X_k^{n+1}, T^{n+1}, (\rho \omegadot_k)^{(2)}, (\rho \Hnuc)^{(2)} ].$  

%--------------------------------------------------------------------------
% STEP 10
%--------------------------------------------------------------------------
\item[Step 10.] {\em Define the new time expansion, $S^{n+1}$, and $\overline{\Gamma_1^{n+1}}$.}

\begin{enumerate}
\renewcommand{\theenumi}{{\bf \Alph{enumi}}}
\item Define
\begin{equation}
  S^{n+1} =  -\sigma  \sum_k  \xi_k (\omegadot_k)^{(2)}  + \sigma \Hnuc^{(2)} +
  \frac{1}{\rho^{n+1} p_\rho} \sum_k p_{X_k}  ({\omegadot}_k)^{(2)}  
   + \sigma \Hext^{(2)},
\end{equation}
where $(\omegadot_k)^{(2)} = (\rho \omegadot_k)^{(2)} / \rho^{(2)}$
and the thermodynamic quantities are defined using $\rho^{n+1}$,
$X_k^{n+1}$, and $T^{n+1}$ as inputs to the equation of state.
Then, compute
\begin{equation}
\overline{S^{n+1}} = {\mathrm{\bf Avg}} (S^{n+1}).
\end{equation}

\item Define
\begin{equation}
\overline{\Gamma_1^{n+1}} = {\rm{\bf Avg}}\left[\Gamma_1\left(\rho^{n+1}, p_0^{n+1}, 
X_k^{n+1}\right) \right].
\end{equation}

\end{enumerate}

%--------------------------------------------------------------------------
% STEP 11
%--------------------------------------------------------------------------
\item[Step 11.] {\em Update the velocity}.  

First, we compute the time-centered edge velocities, $\Ubt^{\nph,\pred}$.
Then, we define
\begin{equation}
\rho^\nph = \frac{\rho^n + \rho^{n+1}}{2}, \qquad \rho_0^\nph = \frac{\rho_0^n + \rho_0^{n+1}}{2}.
\end{equation}
We update the velocity field $\Ubt^n$ to $\Ubt^{n+1,\dagger}$ by discretizing 
equation (\ref{eq:utildeupd}) as
\begin{eqnarray}
\Ubt^{n+1,\dagger} 
&=& \Ubt^n - \dt \left[\left(\uadvtwo+ w_0^{\nph} \eb_r\right) \cdot \nabla \Ubt^{\nph,\pred} \right] \nonumber \\
&&- \dt \left(\uadvtwo \cdot \eb_r\right)  \left(\frac{\partial w_0}{\partial r} \right)^\nph \eb_r \nonumber \\
&& + \dt \left[ - \frac{1}{\rho^\nph} \mathbf{G} \pi^\nmh + \left(\frac{1}{\rho_0}\frac{\partial\pi_0}{\partial r}\right)^n \eb_r - \frac{\left(\rho^\nph-\rho_0^\nph\right)}{\rho^\nph} g^{\nph} \eb_r \right],\nonumber \\
\end{eqnarray}
where $\mathbf{G}$ approximates a cell-centered gradient from nodal
data.  Again, the $\dagger$ superscript refers 
to the fact that the updated velocity does not satisfy the divergence 
constraint.

Finally, we use an approximate nodal projection to define $\Ubt^{n+1}$
from $\Ubt^{n+1,\dagger},$  such that $\Ubt^{n+1}$ approximately
satisfies 
\begin{equation}
\nabla \cdot \left(\beta_0^{\nph} \Ubt^{n+1} \right) 
= \beta_0^{\nph} \left(S^{n+1} - \overline{S^{n+1}} \right),
\end{equation}
where $\beta_0^{\nph}$ is defined as
\begin{equation}
\beta_0^{\nph} = \frac{\beta_0^n + \beta_0^{n+1}}{2}; \qquad
\beta_0^{n+1} = \beta \left(\rho_0^{n+1}, p_0^{n+1}, \overline{\Gamma_1^{n+1}}, g^{n+1}\right).
\end{equation}
As part of the projection we also define the new-time perturbational pressure,
$\pi^\nph.$  This projection necessarily differs from the MAC projection used in 
{\bf Step 3} and {\bf Step 7} because the velocities in those steps are defined
on edges and $\Ubt^{n+1}$ is defined at cell centers, requiring different divergence
and gradient operators.  Details of the approximate projection are given in Paper III.

%--------------------------------------------------------------------------
% STEP 12
%--------------------------------------------------------------------------
\item[Step 12.] {\em Compute a new $\dt.$}

Compute $\dt$ for the next time step with the procedure described in
\S 3.4 of Paper III using $w_0$ as computed in {\bf Step 6} and $\Ubt^{n+1}$
as computed in {\bf Step 11}.

\end{description}

\noindent This completes one step of the algorithm.

Figure \ref{Fig:flowchart} is a flow chart summarizing the 12 step algorithm, 
including the notation used as we advance the solution by $\dt.$  Figure 
\ref{Fig:flowchart48} is a flow chart of the advection steps ({\bf Steps 4} and 
{\bf 8}), which includes the notation we use as we advect the solution through 
a time interval of $\dt$.
%%%%%%%%%%%%%%%%%%%%%%%%%%%%%%%%%
%%%%%%%%%%%%%%%%%%%%%%%%%%%%%%%%%
%%%%%%%%%%%%%%%%%%%%%%%%%%%%%%%%%
%%%%%%%%%%%%%%%%%%%%%%%%%%%%%%%%%

\subsection{Numerical Cutoffs}\label{Sec:Cutoffs}
As discussed in Paper IV, in order to prevent the velocity from
becoming too large in low density regions far from the center of the
star, we impose a cutoff at a moderately small density, $\rho_{\rm
  cutoff}$, and hold the density at this constant value outside of the
star.  The cutoff affects the evolution in the following ways:

\begin{itemize}
\item After advancing enthalpy in the advection step ({\bf Steps 4Hii} 
  and {\bf 8Hii} in \S \ref{Sec:Main Algorithm Description}), we recompute the 
  enthalpy using the equation of state  if $\rho\le\rho_{\rm cutoff}$.

\item When computing gravity (\S \ref{Sec:Gravity}) we only add $\rho_0$ to 
  $m_{\rm encl}$ if $\rho_0 > \rho_{\rm cutoff}$ in order to prevent an unphysical amount
  of mass from contributing to the calculation.

\item When computing $p_0$ in {\bf Enforce HSE} (\S \ref{Sec:p0 Update}), 
  we hold $p_0$ constant once $\rho_0 \le \rho_{\rm cutoff}$.

\item In {\bf React State} (\S \ref{Sec:Reactions}), we set $\omegadot_k=0$ 
  and $\rho\Hnuc=0$ if $\rho \le \rho_{\rm cutoff}$.

\item When computing $\psi$, ({\bf Steps 4F} and {\bf 8F} in 
  \S \ref{Sec:Main Algorithm Description}) we set $\psi=0$ if 
  $\rho_0\le\rho_{\rm cutoff}$.

\item When we compute the velocity forcing in {\bf Steps 3}, {\bf 7}, 
  {\bf 11}, and {\bf 12}, we set the buoyancy term 
  (the term proportional to $\rho-\rho_0$) to zero if $\rho < 5\rho_{\rm cutoff}$.

\end{itemize}

Additionally, we use an anelastic cutoff density, $\rho_{\rm anelastic}$, in the 
computation of $\beta_0$ ({\bf Steps 3}, {\bf 7}, and {\bf 11} in 
\S \ref{Sec:Main Algorithm Description}).  When $\rho_{0,j}\le\rho_{\rm anelastic}$, 
we set $\beta_{0,j} = (\rho_{0,j}/\rho_{0,j-1})\beta_{0,j-1}$.

%==========================================================================
% Appendix B: Computing w_0 for spherical problems
%==========================================================================
\section{Computing $w_0$ for spherical problems}\label{Sec:Computing w0 spherical}

Recall that we want to solve
\begin{equation}
\frac{1}{r^2} \frac{\partial}{\partial r} \left (r^2 w_0 \right ) = \Sbar - \frac{1}{\gammaonebar p_0} \underbrace{\left( \frac{\partial p_0}{\partial t} + w_0 \frac{\partial p_0}{\partial r} \right)}_{\psi}.
\end{equation}
for the base state velocity, $w_0.$ We first decompose $w_0$ by setting $w_0 = \ow + \delta w_0$,
where the $\ow$ term is the contribution to $w_0$ due to the expansion term:
\begin{equation}
\frac{1}{r^2} \frac{\partial}{\partial r} \left (r^2 \ow \right ) = \Sbar^{\inp}.
\end{equation}
Then we can write an equation for the remaining term, $\delta w_0$:
\begin{equation}
\frac{1}{r^2} \frac{\partial}{\partial r} \left (r^2 \dw \right ) = - \frac{1}{\gammaonebar p_0} \left[\frac{\partial p_0}{\partial t} + (\ow + \dw)\frac{\partial p_0}{\partial r} \right].\label{eq:w01a}
\end{equation}
Multiplying equation (\ref{eq:w01a}) through by $\gammaonebar p_0$,
taking another derivative with respect to $r$, and switching the order
of temporal and spatial derivatives, we get:
\begin{equation}
\frac{\partial}{\partial r} \left [ \frac{\gammaonebar p_0 }{r^2} \frac{\partial}{\partial r} (r^2 \dw) \right ] = -\frac{\partial}{\partial t} \frac{\partial p_0}{\partial r} - \frac{\partial}{\partial r} \left [(\ow+\dw) \frac{\partial p_0}{\partial r} \right ].\label{eq:sphconstraint_2}
\end{equation}
To solve for $\dw$ we will need to substitute for the derivatives of
$p_0.$ To do so we start with the hydrostatic equilibrium equation,
\begin{equation}
\frac{\partial p_0}{\partial r} = -\rho_0 g;
\qquad
g = \frac{G m_{\mathrm encl}}{r^2},
\end{equation} 
where $m_{\mathrm encl}(r)$ is the mass enclosed at radius $r$ and $G$
is the gravitational constant.  Using this, we can then write equation
(\ref{eq:sphconstraint_2}) as:
\begin{eqnarray}
\frac{\partial}{\partial r} \left[\frac{\gammaonebar p_0}{r^2} \frac{\partial}{\partial r}(r^2 \dw)\right] 
&=& \frac{\partial}{\partial t} \left (\rho_0 g \right ) +\frac{\partial}{\partial r} \left (w_0 \rho_0 g \right)\nonumber\\
&=& g \left[ \frac{\partial \rho_0}{\partial t} + \frac{\partial}{\partial r} (w_0 \rho_0) \right] + \rho_0 \left( \frac{\partial g}{\partial t} + w_0 \frac{\partial g}{\partial r} \right).\label{eq:sphconstraint_3}
\end{eqnarray}
The mass enclosed inside any radius, $r$, is $m_{\mathrm encl}(r) = 4
\pi \int_0^r \rho_0(s) s^2 ds$, or alternately, $\partial
m_{\mathrm encl}/\partial r = 4\pi r^2 \rho_0$.  The Lagrangian
derivative of the enclosed mass is then:
\begin{eqnarray}
\frac{D_0 m_{\mathrm encl}}{D t} &=& \frac{\partial m_{\mathrm encl}}{\partial t} + w_0 \frac{\partial m_{\mathrm encl}}{\partial r} \nonumber \\
&=& 4\pi \left( \frac{\partial}{\partial t} \int_0^r \rho_0(s) s^2 ds + w_0 r^2 \rho_0 \right) \nonumber 
    = 4\pi \left( \int_0^r \frac{\partial \rho_0}{\partial t} s^2 ds + w_0 r^2 \rho_0 \right) \nonumber \\
&=& 4\pi \left\{ -\int_0^r \left [ \frac{1}{s^2}\frac{\partial(s^2 \rho_0 w_0)}{\partial s} + \frac{1}{s^2}\frac{\partial (s^2 \etarho)}{\partial s} \right ]s^2 ds + w_0 r^2 \rho_0 \right\} \nonumber \\
&=& \left . 4\pi \left(-s^2 \rho_0 w_0 \right |_0^r - \left . s^2 \etarho \right |_0^r + w_0 r^2 \rho_0 \right) 
   = -4\pi r^2 \etarho,\label{eq:dmdt}
\end{eqnarray}
where we used the spherical form of equation (29) in Paper III, 
\begin{equation}
\frac{\partial \rho_0}{\partial t} + \frac{1}{r^2} \frac{\partial (r^2 \rho_0 w_0)}{\partial r} + \frac{1}{r^2} \frac{\partial (r^2 \etarho )}{\partial r}  = 0,\label{eq:sph_continuity}
\end{equation}
to eliminate $\partial \rho_0/\partial t$.  We note that in the
absence of any mixing, $\etarho=0$, and $D_0 m_{\mathrm encl}/Dt = 0$.
Equation (\ref{eq:dmdt}) allows us to write the Lagrangian change in
the gravitational acceleration as:
\begin{eqnarray}
\frac{D_0 g}{D t} = \frac{\partial g}{\partial t} + w_0 \frac{\partial
  g}{\partial r} &=& \frac{D_0}{Dt}\left(\frac{G
  m_{\mathrm encl}}{r^2}\right) 
= G m_{\mathrm encl}\frac{D_0}{D t}\left(\frac{1}{r^2}\right) +
  \frac{G}{r^2}\frac{D_0 m_{\mathrm encl}}{Dt} \nonumber \\ 
&=& -\frac{2 w_0 G m_{\mathrm encl}}{r^3} - 4 \pi G \etarho  =
-\frac{2 w_0 g}{r} - 4 \pi G \etarho.
\end{eqnarray}
Putting it all together, equation (\ref{eq:sphconstraint_3}) becomes:
\begin{equation}
 \frac{\partial}{\partial r} \left[ \frac{\gammaonebar p_0}{r^2} \frac{\partial}{\partial r} (r^2 \dw) \right] = g \left[\frac{\partial \rho_0}{\partial t} + \frac{\partial}{\partial r} (w_0 \rho_0) \right] + \rho_0 \left(\frac{-2 w_0 g}{r} - 4 \pi G \etarho \right).\label{eq:B5}
\end{equation}
Finally, we can use equation (\ref{eq:sph_continuity}) to write
equation (\ref{eq:B5}) as:
\begin{eqnarray}
\frac{\partial}{\partial r} \left[ \frac{\gammaonebar p_0}{r^2} \frac{\partial}{\partial r} (r^2 \dw) \right] 
&=& g \left [ -\frac{1}{r^2} \frac{\partial}{\partial r} (r^2 w_0 \rho_0) - \frac{1}{r^2} \frac{\partial}{\partial r} (r^2 \etarho) + \frac{\partial}{\partial r} (w_0 \rho_0) \right ]
 \nonumber \\
&&+ \rho_0 \left(\frac{-2 w_0 g}{r} - 4 \pi G \etarho \right) \nonumber \\
&=& - \frac{g}{r^2} \frac{\partial (r^2 \etarho)}{\partial r} - \frac{4 (\ow + \dw) \rho_0 g}{r} - 4 \pi G \rho_0 \etarho.
\end{eqnarray}
We discretize this elliptic equation in the radial dimension as:
\begin{eqnarray}
&& \frac{1}{\Delta r} \left\{\left[ \frac{\gammaonebar p_0}{r^2} \frac{\partial (r^2 \dw)}{\partial r} \right]_{j} - \left[ \frac{\gammaonebar p_0}{r^2} \frac{\partial (r^2 \dw)}{\partial r} \right]_{j-1} \right\} + \left[ \frac{4 (r^2 \dw) \rho_0 g}{r^3} \right]_{j-\myhalf} \nonumber\\
&=& - \frac{g_{j-\myhalf}}{r_{j-\myhalf}^2 \Delta r} \left[ \left( r^2 \etarho \right)_{j} - \left( r^2 \etarho  \right)_{j-1} \right] - \left( \frac{4 \ow \rho_0 g}{r} \right)_{j-\myhalf} - \left(  4 \pi G \rho_0 \etarho \right)_{j-\myhalf},
\end{eqnarray}
where we choose to solve for $(r^2 \dw)$ rather than $\dw$ so that we can easily enforce
$\partial (r^2 \dw) / \partial r = 0$ at the the upper boundary.
Then, using hydrostatic equilibrium, we expand this to 
\begin{eqnarray}
&& \frac{1}{\Delta r} \left\{ \left( \frac{\gammaonebar p_0}{r^2}\right)_{j  } \frac{\left[ (r^2 \dw)_{j+\myhalf} - (r^2 \dw)_{j-\myhalf} \right]}{\Delta r} \right. \nonumber \\
&& \left.
- \left( \frac{\gammaonebar p_0}{r^2}\right)_{j-1} \frac{\left[ (r^2 \dw)_{j-\myhalf} - (r^2 \dw)_{j-\thalf}\right]}{\Delta r} \right\} \nonumber \\
&&  - \left( \frac{4}{r_{j-\myhalf} ^3} \frac{p_{0,j} - p_{0,j-1}}{\dr} \right) (r^2 \dw)_{j-\myhalf} = \left( \frac{4}{r_{j-\myhalf} ^3} \frac{p_{0,j} - p_{0,j-1}}{\dr} \right) (r^2 \ow)_{j-\myhalf} \nonumber \\
&& - \frac{g_{j-\myhalf}}{r_{j-\myhalf}^2 \Delta r} \left[ \left( r^2 \etarho \right)_{j} - \left( r^2 \etarho  \right)_{j-1} \right] - \left(  4 \pi G \rho_0 \etarho   \right)_{j-\myhalf}.
\end{eqnarray}
If we write this in matrix form, so that:
\begin{equation}
A_j (r^2 \dw)_{j-\thalf} + B_j (r^2 \dw)_{j-\myhalf} + C_j (r^2 \dw)_{j+\myhalf} = F_j,
\end{equation}
then:
\begin{eqnarray}
A_j &=& \frac{1}{\Delta r^2} \left( \frac{\gammaonebar^{\inp} p_0^{\inp}}{r^2}\right)_{j-1}, \\
B_j &=& -\frac{1}{\Delta r^2} \left[ \left( \frac{\gammaonebar^{\inp} p_0^{\inp}}{r^2}\right)_{j}  + \left( \frac{\gammaonebar^{\inp} p_0^{\inp}}{r^2}\right)_{j-1} \right] - \left( \frac{4}{r_{j-\myhalf} ^3} \frac{p_{0,j}^{\inp} - p_{0,j-1}^{\inp}}{\dr} \right), \\
C_j &=& \frac{1}{\Delta r^2} \left( \frac{\gammaonebar^{\inp} p_0^{\inp}}{r^2}\right)_{j}  , \\
F_j &=&  \left( \frac{4}{r_{j-\myhalf} ^3} \frac{p_{0,j}^{\inp} - p_{0,j-1}^{\inp}}{\dr} \right) (r^2 \ow)_{j-\myhalf} - \frac{g_{j-\myhalf}^\inp}{r_{j-\myhalf}^2 \Delta r} \left[ \left( r^2 \etarho^{\inp} \right)_{j} - \left( r^2 \etarho^{\inp} \right)_{j-1} \right] \nonumber \\
&& - \left(  4 \pi G \rho_0^{\inp} \etarho^{\inp} \right)_{j-\myhalf},
\end{eqnarray}
We define the lower boundary condition, 
$\dw = 0$ at $r=0$, which corresponds to $j=0,$ by setting:
\begin{equation}
A_0 = C_0 = F_0 = 0; \quad B_0 = 1.
\end{equation}
We also specify $\partial (r^2 \dw) / \partial r = 0$ at the the upper
boundary, which corresponds to the location where $\rho_0$ falls below
$\rho_{\rm cutoff}$, by setting:
\begin{eqnarray}
A_N = -1; \quad
B_N =  1; \quad
C_N = F_N = 0.
\end{eqnarray}
Finally, $w_0^{\outp} = \ow + \delta w_0$.  Once $\rho_0$ falls below $\rho_{\rm cutoff}$, 
we hold $r^2 w_0^{\outp}$ constant.

%==========================================================================
% Appendix C: Test Problem Initial Model
%==========================================================================
\section{Test Problem Initial Model}\label{Sec:Test Problem Initial Model}

We use the same general initial model for the convergence
test (\S \ref{Sec:Convergence Test}), adaptive bubble rise test (\S
\ref{Sec:Bubble Rise}) and forced convection test (\S \ref{Sec:Forced
  Convection}).  We define a base temperature, $T_\mathrm{base} =
6\times 10^8$~K, and density, $\rho_\mathrm{base} = 2.6\times
10^9~\gcc$, at some height $r_\mathrm{base}$ above the bottom of the
domain ($r_\mathrm{base}$ varies in each problem, and can be equal to
zero).  This defines a base entropy, $s_\mathrm{base} =
s(\rho_\mathrm{base}, T_\mathrm{base})$.  The gravitational
acceleration, $g = -1.5\times 10^{10}~\mathrm{cm~s^{-2}}$, is
constant.  The composition is uniform everywhere with
$X(^{12}\mathrm{C}) = 0.3$ and $X(^{16}\mathrm{O}) = 0.7$.  For $r >
r_\mathrm{base}$, we integrate the equation of hydrostatic equilibrium
along with the constraint that entropy is constant:
\begin{eqnarray}
p_{0,j+1} &=& p_{0,j} + \frac{1}{2} \Delta r ( \rho_{0,j} +
\rho_{0,j+1} ) g. 
\label{eq:hse_constantg} \\
s_{0,j+1} &=& s_\mathrm{base} \label{eq:entropy_constraint_app}
\end{eqnarray}
upwards from
$r_\mathrm{base}$.  We define a temperature
cutoff, $T_\mathrm{cutoff} = 10^7$~K, and when $T_{0,j+1} <
T_\mathrm{cutoff}$, we replace equation (\ref{eq:entropy_constraint_app})
with $T_{0,j+1} = T_\mathrm{cutoff}$.  If we choose $r_\mathrm{base} >
0$, then we must also define the model for $0 < r < r_\mathrm{base}$.
In this case, we define a desired linear entropy profile with a
discontinuous jump from $s_\mathrm{base}$ as
\begin{equation}
s_\mathrm{want} = \frac{1}{3}s_\mathrm{base} + \frac{r - r_\mathrm{base}}{r_\mathrm{base}}\frac{s_\mathrm{base}}{12}.
\end{equation}
This entropy profile creates a convectively stable layer below the
atmosphere to prevent any motions generated from heating above from
interfering with the lower boundary.  The initial model for this
region is then computed by integrating:
\begin{eqnarray}
p_{0,j} &=& p_{0,j+1} - \frac{1}{2} \Delta r ( \rho_{0,j} +
\rho_{0,j+1} ) g,
\label{eq:hse below base} \\
s_{0,j} &=& s_\mathrm{want}(r_j), \label{eq:entropy_constraint below base}
\end{eqnarray}
downward from $r = r_\mathrm{base}$.

\clearpage

%==========================================================================
% References
%==========================================================================

%\bibliographystyle{apj}
%\bibliography{ws} 

\begin{thebibliography}{27}
\expandafter\ifx\csname natexlab\endcsname\relax\def\natexlab#1{#1}\fi

\bibitem[{{Almgren} {et~al.}(2010){Almgren}, {Beckner}, {Bell}, {Day},
  {Howell}, {Joggerst}, {Lijewski}, {Nonaka}, {Singer}, \&
  {Zingale}}]{castro:2010}
{Almgren}, A.~S. {et~al.} 2010, accepted for publication in ApJS

\bibitem[{Almgren {et~al.}(1998)Almgren, Bell, Colella, Howell, \&
  Welcome}]{AlmBelColHowWel98}
Almgren, A.~S., Bell, J.~B., Colella, P., Howell, L.~H., \& Welcome, M. 1998,
  Journal of Computational Physics, 142, 1

\bibitem[{{Almgren} {et~al.}(2008){Almgren}, {Bell}, {Nonaka}, \&
  {Zingale}}]{ABNZ:III}
{Almgren}, A.~S., {Bell}, J.~B., {Nonaka}, A., \& {Zingale}, M. 2008,
  Astrophysical Journal, 684, 449

\bibitem[{Almgren {et~al.}(2006{\natexlab{a}})Almgren, Bell, Rendleman, \&
  Zingale}]{ABRZ:I}
Almgren, A.~S., Bell, J.~B., Rendleman, C.~A., \& Zingale, M.
  2006{\natexlab{a}}, Astrophysical Journal, 637, 922

\bibitem[{Almgren {et~al.}(2006{\natexlab{b}})Almgren, Bell, Rendleman, \&
  Zingale}]{ABRZ:II}
---. 2006{\natexlab{b}}, Astrophysical Journal, 649, 927

\bibitem[{{Almgren} {et~al.}(2007){Almgren}, {Bell}, \&
  {Zingale}}]{maestro-scidac2007}
{Almgren}, A.~S., {Bell}, J.~B., \& {Zingale}, M. 2007, Journal of Physics
  Conference Series, 78, 2085

\bibitem[{Bell {et~al.}(1994)Bell, Berger, Saltzman, \& Welcome}]{Bell:1994}
Bell, J.~B., Berger, M.~J., Saltzman, J.~S., \& Welcome, M. 1994, SIAM J. Sci.
  Statist. Comput., 15, 127

\bibitem[{Bell {et~al.}(2004)Bell, Day, Rendleman, Woosley, \&
  Zingale}]{Bell:2004}
Bell, J.~B., Day, M.~S., Rendleman, C.~A., Woosley, S.~E., \& Zingale, M.~A.
  2004, Journal of Computational Physics, 195, 677

\bibitem[{Berger \& Colella(1989)}]{berger-colella}
Berger, M.~J., \& Colella, P. 1989, Journal of Computational Physics, 82, 64

\bibitem[{Berger \& Rigoutsos(1991)}]{bergerRigoutsos:1991}
Berger, M.~J., \& Rigoutsos, J. 1991, IEEESMC, 21, 1278

\bibitem[{{Chamulak} {et~al.}(2008){Chamulak}, {Brown}, {Timmes}, \&
  {Dupczak}}]{chamulak2008}
{Chamulak}, D.~A., {Brown}, E.~F., {Timmes}, F.~X., \& {Dupczak}, K. 2008,
  Astrophysical Journal, 677, 160

\bibitem[{Colella \& Woodward(1984)}]{ppm}
Colella, P., \& Woodward, P.~R. 1984, Journal of Computational Physics, 54, 174

\bibitem[{Day \& Bell(2000)}]{DayBell:2000}
Day, M.~S., \& Bell, J.~B. 2000, Combust. Theory Modelling, 4, 535

\bibitem[{Durran(1990)}]{Durran_mono}
Durran, D. 1990, Meteorol. Monographs, 23, 59

\bibitem[{{Fryxell} {et~al.}(2000){Fryxell}, {Olson}, {Ricker}, {Timmes},
  {Zingale}, {Lamb}, {MacNeice}, {Rosner}, {Truran}, \& {Tufo}}]{flash}
{Fryxell}, B. {et~al.} 2000, Astrophysical Journal Supplement, 131, 273

\bibitem[{{Glasner} {et~al.}(2007){Glasner}, {Livne}, \&
  {Truran}}]{glasner:2007}
{Glasner}, S.~A., {Livne}, E., \& {Truran}, J.~W. 2007, Astrophysical Journal,
  665, 1321

\bibitem[{{H{\" o}flich} \& {Stein}(2002)}]{hoflichstein:2002}
{H{\" o}flich}, P., \& {Stein}, J. 2002, Astrophysical Journal, 568, 779

\bibitem[{{Kuhlen} {et~al.}(2006){Kuhlen}, {Woosley}, \&
  {Glatzmaier}}]{kuhlen-ignition:2005}
{Kuhlen}, M., {Woosley}, S.~E., \& {Glatzmaier}, G.~A. 2006, Astrophysical
  Journal, 640, 407

\bibitem[{{Lin} {et~al.}(2006){Lin}, {Bayliss}, \& {Taam}}]{Lin:2006}
{Lin}, D.~J., {Bayliss}, A., \& {Taam}, R.~E. 2006, Astrophysical Journal, 653,
  545

\bibitem[{{Meakin} \& {Arnett}(2007)}]{meakin:2007}
{Meakin}, C.~A., \& {Arnett}, D. 2007, Astrophysical Journal, 665, 690

\bibitem[{{Miller} \& {Colella}(2002)}]{ppmunsplit}
{Miller}, G.~H., \& {Colella}, P. 2002, Journal of Computational Physics, 183,
  26

\bibitem[{Pember {et~al.}(1998)Pember, Howell, Bell, Colella, Crutchfield,
  Fiveland, \& Jessee}]{pember-flame}
Pember, R.~B., Howell, L.~H., Bell, J.~B., Colella, P., Crutchfield, W.~Y.,
  Fiveland, W.~A., \& Jessee, J.~P. 1998, Comb. Sci. Tech., 140, 123

\bibitem[{Saltzman(1994)}]{saltzman}
Saltzman, J. 1994, Journal of Computational Physics, 115, 153

\bibitem[{{Strohmayer} \& {Bildsten}(2006)}]{STRO_BILD06}
{Strohmayer}, T., \& {Bildsten}, L. 2006, Compact Stellar X-Ray Sources, ed.
  W.~H.~G. {Lewin} \& M.~{van der Klis} (Cambridge: Cambridge Univ. Press),
  113--+

\bibitem[{Timmes \& Swesty(2000)}]{timmes_swesty:2000}
Timmes, F.~X., \& Swesty, F.~D. 2000, Astrophysical Journal Supplement, 126,
  501

\bibitem[{Woosley {et~al.}(2004)Woosley, Wunsch, \& Kuhlen}]{Woosley:2004}
Woosley, S.~E., Wunsch, S., \& Kuhlen, M. 2004, Astrophysical Journal, 607, 921

\bibitem[{Zingale {et~al.}(2009)Zingale, Almgren, Bell, Nonaka, \&
  Woosley}]{ZABNW:IV}
Zingale, M., Almgren, A.~S., Bell, J.~B., Nonaka, A., \& Woosley, S.~E. 2009,
  Astrophysical Journal, 704, 196

\end{thebibliography}

\end{document}